\documentclass[12pt]{article}
\usepackage{hyperref,natbib,amsmath,amssymb, amsthm}
\usepackage{import, color,graphicx}

\newcommand{\blu}[1]{\textcolor{black}{#1}} 
\newcommand{\blunew}[1]{\textcolor{black}{#1}} 

\usepackage{epsf}
\usepackage{epsfig}
\usepackage{amsfonts}
\usepackage{setspace}
\usepackage[boxed]{algorithm}
\usepackage{algpseudocode}
\usepackage[]{algorithm}
\usepackage{algorithmicx}

\makeatother

\usepackage{caption}
\usepackage{subcaption}
\usepackage[normalem]{ulem}

\DeclareMathOperator{\Diag}{Diag}

\newcommand{\blind}{0}

\newcommand{\bm}[1]{\mbox{\boldmath $#1$}}

\def\spacingset#1{\renewcommand{\baselinestretch}%
{#1}\small\normalsize} \spacingset{1}

\begin{document}

\if0\blind
{
\title{\vspace{-1cm} Multi-output calibration of a honeycomb seal\\via on-site surrogates}
\author{
 Jiangeng Huang\thanks{Genentech, Inc., South San Francisco, CA 94080 \href{mailto:huangj@vt.edu}{\tt huang.jiangeng@gene.com}; corresponding author}\
\and 
Robert B.~Gramacy\thanks{Department of Statistics, Virginia Tech}\
}
\date{}
\maketitle
}\fi

\if1\blind
{
  \bigskip
  \bigskip
  \bigskip
  \begin{center}
    {\LARGE Multi-output calibration of a honeycomb seal\\via on-site surrogates}
\end{center}
  \medskip
  \bigskip
} \fi

\begin{abstract}

We consider large-scale industrial computer model calibration, combining
multi-output simulation with limited physical observation, involved in the
development of a honeycomb seal. Toward that end, we adopt a localized
sampling and emulation strategy called ``on-site surrogates (OSSs)'', designed
to cope with the amalgamated challenges of high-dimensional inputs,
large-scale simulation campaigns, and nonstationary response surfaces.  In
previous applications, OSSs were one-at-a-time affairs for multiple outputs
leading to dissonance in calibration efforts for a common parameter set across
outputs for the honeycomb.  \blunew{We demonstrate that} a
principal-components representation, adapted from ordinary Gaussian process
surrogate modeling to the OSS setting, can resolve this tension.   With a
two-pronged -- optimization and fully Bayesian -- approach, we show how
pooled information across outputs can reduce uncertainty and enhance
efficiency in calibrated parameters \blu{and
prediction} for the honeycomb relative to the previous, ``data-poor''
univariate analog.

\end{abstract}

\if0\blind
{
\bigskip
\noindent {\bf Keywords:}
Bayesian calibration,  multivariate analysis, computer experiment,
hierarchical model, big data, surrogate modeling, uncertainty
quantification}\fi

 \doublespacing

\section{Introduction}
\label{sec:intro}


Computer simulation experiments play an increasingly pivotal role in
scientific inquiry. STEM training, cheap hardware, and robust numerical
libraries have democratized simulation as a means of exploration of complex
physical \citep[e.g.,][]{mehta2014modeling}, biological
\citep[e.g.,][]{johnson:2008}, engineering/materials
\cite[e.g.,][]{zhang2015microstructure}, and economic phenomena
\citep[e.g.,][]{kita2016realistic}, to provide a few representative examples.
We are interested in the use of simulation in the development of a so-called
{\em honeycomb seal}, a device integral to oil and gas recovery, together with
colleagues at Baker Hughes, a General Electric company (BHGE).   The honeycomb
experiment mirrors a common setup in engineering applications, leveraging a
general purpose simulator called {\tt ISOTSEAL}, a commercial spin-off of
tools developed first at Texas A\&M \citep{isotseal}.

Adapting a solver like {\tt ISOTSEAL} to a particular
application like BHGE's honeycomb is a two-stage process.  One first
\blunew{tailors} configuration files to reflect the design and operational
parameters/conditions of the particular instance under study. Inevitably some
of these computer model settings are unknown (precisely enough) in the actual
system, so there is interest in {\em calibrating} these parameters based on a
limited physical/field experimentation campaign. Simulations, collected under
a range of settings of both unknown {\em calibration parameters} and known
{\em design inputs} characterizing operating conditions, can then be
``matched'' with analogs observed in the field, and their residuals can drive
inference for the unknown settings, comprising the second stage.

Although there are several ways to operationalize that idea, the canonical
setup in the computer experiments literature is due to \citet[][KOH
hereafter]{Kennedy:O'Hagan:2001}.  At a high level, with details in Section
\ref{sec:review}, KOH couples two Gaussian processes (GPs) together -- one as
a surrogate for the computer model (e.g., {\tt ISOTSEAL}), and another to
capture bias and noise between simulation and field observation, to form a
multivariate normal (MVN) marginal likelihood which can drive inference
for all unknowns, including the calibration parameters.  KOH offers a nice
synthesis of information sources, but is susceptible to confounding
\citep[e.g.,][]{bryn2014learning, tuo2016,plumlee2017bayesian,gu2018jointly}.
More important for us, however, is that it is computationally daunting.  Cubic
matrix decompositions for MVNs involved in GP-based inference will severely
limit the size of simulation campaigns that can be entertained 
(See \cite{Santner2018, gramacy2020surrogates}).  Historically, computer
simulation was cumbersome, limiting the size of campaigns.  
With increasing computing power, the speed and size of computer simulation are evolving. 

The {\tt ISOTSEAL} simulator is fast, so a big campaign is reasonable
computationally, but not so fast nor steady enough that simulations can be
used directly (forgoing the surrogate) in an otherwise KOH-style
setting \citep{Higdon:2004}.  Meanwhile a large simulation campaign is
essential to capture stark changes in dynamics, simulation artifacts, and
other ``features'' common to modern numerical solvers.  A
flexible meta-modeling apparatus is essential in order to cope with large
training data sizes, and to adapt to and/or smooth over (i.e., separate signal
from) ``noise''/artifacts in such settings.  \citet{Huang:2018} proposed one
such approach based on {\em on-site surrogates} (OSSs), motivated specifically
by {\tt ISOTSEAL} and the BHGE honeycomb, but also vetted generically in
benchmark exercises.  \citeauthor{Huang:2018} showed how the KOH apparatus,
and variations based on modularization \citep{Liu:2009} and maximization
\citep{gra:etal:2015}, could be adapted to work with OSSs.

\blunew{\citeauthor{Huang:2018} explored a single output feature} and
honeycomb/{\tt ISOTSEAL} output is multi-dimensional. Our BHGE collaborators
are interested in at least four frequencies of outputs on four properties of
the system (16 total). Section \ref{sec:honeycomb} demonstrates \blunew{a
not-uncommon situation when attempting to apply KOH separately for
multiple outputs:} inconsistent inference about calibration parameters 
\blunew{across analyses}, challenging downstream data synthesis, and
ultimately \blunew{poor prediction}.  \blunew{Specifically to the honeycomb},
the physical dynamics in play are complex, and the imposition of complete
independence across property and frequency, \blunew{together to with a paucity
of field data observation}, throws too much information away.

KOH has subsequently been extended to multi-output simulation and field
experimentation \cite[e.g.][]{Higdon2008}, but with ordinary GP surrogates. In
this paper we detail the application of OSSs in that context, which is easier
said than done.  Although the basic ingredients are similar to 
\citeauthor{Higdon:2004}, via principal component decompositions, 
their application is non-trivial in this setting and requires care in
methodological development as well as implementation, as we provide. Following
\citet{Huang:2018}, we take a two-pronged approach: first via thrifty
marginalize--maximize calculations, rather than faithful KOH, then full Bayes
with MCMC. The result is a multi-output, large-scale, computer model
calibration framework that -- at least in the case of the honeycomb -- is able
to resolve stark multi-output calibration inconsistencies by making effective
use of the (joint) information in all of the available simulation and field
data.  \blunew{We wish to clarify that we do not offer a remedy to confounding
issues sometimes faced by KOH, whether for single or multiple outputs. However
for the honeycomb, we show that these may not be a first-order concern.
Our efforts target computation: building a high-fidelity multi-output KOH
apparatus via approximation that can cope with large simulation campaigns.}

The remainder of paper is organized as follows.  Section \ref{sec:review}
begins with review of basics: KOH, GPs, etc. Honeycomb specifics and
univariate OSS calibration in Section \ref{sec:honeycomb}. Section
\ref{sec:muloss} combines these building blocks for calibration with
multivariate outputs via OSSs leveraging linear dependence across output
frequency. Thrifty modular and fully Bayesian KOH are enumerated in turn.
Section \ref{sec:allcali} combines across output property,
\blunew{and Section \ref{sec:mpred}} demonstrates the impact on 
prediction down-stream. Section \ref{sec:discuss} concludes with a discussion.

\section{Review of elements}
\label{sec:review}

We introduce KOH calibration via GPs, and extensions for scale and
multivariate output.

\subsection{Basics: calibration and surrogate modeling}
\label{sec:basics}

\citet{Kennedy:O'Hagan:2001} described a univariate Bayesian calibration
framework, combining field experimental observations $y^F(\mathbf{x})$, at a
vector of input design variables $\mathbf{x}$, with computer simulations
$y^M(\mathbf{x}, \mathbf{u}^\star)$, under ideal or ``true'' calibration/tuning
parameter(s) $\mathbf{u}^\star$, through a discrepancy or bias correction
$b(\mathbf{x})$, between simulation and field:
\begin{equation}
y^F(\mathbf{x}) = y^M(\mathbf{x}, \mathbf{u}^\star) + b(\mathbf{x}) +\mathbf{\epsilon},
\quad \epsilon\stackrel{\mathrm{iid}}{\sim} \mathcal N(0,
\sigma^2_\epsilon).
 \label{eq:koh}
\end{equation}
Modeling and inference commences via Gaussian processes (GP) priors on
$y^M(\mathbf{x}, \mathbf{u}^\star)$ and  $b(\mathbf{x})$. GPs provide a flexible
nonparametric (Bayesian) structure for smooth functional relationships between
inputs $\mathbf{x}$ and output $f(\mathbf{x})$, where any finite number $N$ of
evaluations follow a multivariate normal (MVN) distribution:\blunew{
$f(\mathbf{x}) \sim
\mathcal{N}_N ( \mathbf{\mu}(\mathbf{x}), \mathbf{\Sigma}(\mathbf{x},\mathbf{x}'))$}. Inference
for any aspect of \blunew{$\mathbf{\mu}$ and $\mathbf{\Sigma}$} given training data $\mathbf{D}_N =
(\mathbf{X}_N,\mathbf{y}_N)$ may be facilitated by likelihoods, i.e., MVN
densities.  Often in practice \blunew{$\mathbf{\mu}(\cdot)=0$}.
%
%
MVN density/likelihood evaluation for aspects of $\mathbf{\Sigma}$, which is usually
based on inverse Euclidean distance up to several unknown hyperparameters
$\bm{\phi}$, involve cubic-in-$N$ matrix decomposition for inverses and
determinants, which can be a bottleneck in large-scale applications.

This is exacerbated in the KOH setting  where Eq.~(\ref{eq:koh})
implies a joint distribution for computer model training data
\blunew{
$\mathbf{D}_{N}= (\mathbf{X}_{N}, \mathbf{y}_{N})$ and field data
$\mathbf{D}_{F}= (\mathbf{X}_{F}, \mathbf{y}_{F})$:
 \begin{align}
\begin{bmatrix} \mathbf{y}_{N} \\ \mathbf{y}_{F} \end{bmatrix}
\sim \mathcal{N}_{N + F}(\mathbf{0}, \mathbf{\Sigma}(\mathbf{u})),
\quad \text{where} \quad 
\mathbf{\Sigma}(\mathbf{u}) 
\equiv \begin{bmatrix} \mathbf{\Sigma}_{N}& 
\mathbf{\Sigma}^\top_{F, N}(\mathbf{u})\\ 
\mathbf{\Sigma}_{F, N}(\mathbf{u}) &
\mathbf{\Sigma}_{F}(\mathbf{u})  + \mathbf{\Sigma}^b_{F} \end{bmatrix}.   
\label{eq:mvn}
\end{align}}
In (\ref{eq:mvn}) above,
\blunew{$\mathbf{\Sigma}_{N} \equiv \mathbf{\Sigma}([\mathbf{X}_{N},  \mathbf{U}_{N} ])$} 
is an \blunew{$N \times N$} covariance matrix for simulations $y^M(\cdot)$,
capturing pairwise covariance between 
$p_x + p_u$ dimensional inputs 
$( \mathbf{x},  \mathbf{u})$, and 
\blunew{$\mathbf{U}_{N} \equiv [ \mathbf{u}^\top_1, \dots,  \mathbf{u}^\top_{N}]$}
stacks \blunew{$N$} length $p_u$ row vectors.
The off-diganal \blunew{
$\mathbf{\Sigma}_{F, N}(\mathbf{u})$} is an \blunew{$F \times N$}
matrix capturing covariance between simulation inputs
\blunew{$[\mathbf{X}_{N},  \mathbf{U}_{N} ]$} 
and field inputs \blunew{$[\mathbf{X}_{F},  \mathbf{U}_{F}]$}, again under $y^M(\cdot)$, where
\blunew{$\mathbf{U}_{F} \equiv [ \mathbf{u}^\top, \dots,  \mathbf{u}^\top]$} 
stacks \blunew{$F$}  identical row vectors of (unknown) parameters  $\mathbf{u}$.  
Similarly, \blunew{$\mathbf{\Sigma}_{F}(\mathbf{u}) \equiv \mathbf{\Sigma}([\mathbf{X}_{F},  \mathbf{U}_{F}])$}
is \blunew{$F \times F$ for $[\mathbf{X}_{F},  \mathbf{U}_{F}]$}, 
$y^M(\cdot)$ dynamics between 
field data observations.
Lastly, \blunew{$\mathbf{\Sigma}^b_{F}$} is  an \blunew{$F \times F$}
 matrix of covariances specified by the bias correction 
GP $\blunew{b(\mathbf{x})}$ acting only field inputs $\mathbf{X}_{F}$. 
  
Fully Bayesian inference for unknown $\mathbf{u}$
under prior $p(\mathbf{u})$ hinges on \blunew{$p(
\mathbf{y}_{N}, \mathbf{y}_{F} \mid \mathbf{u})$}, the MVN (marginal)
likelihood implied by Eq.~(\ref{eq:mvn}). Computational challenges
are evident in inverse \blunew{$\mathbf{\Sigma}^{-1}(\mathbf{u})$} and determinant
\blunew{$|\mathbf{\Sigma}(\mathbf{u})|$} evaluations involved in such density evaluations. A
single Cholesky decomposition could furnish both in \blunew{$\mathcal{O}((N+F)^3)$}
flops. However, notice that not all blocks of
\blunew{$\mathbf{\Sigma}(\mathbf{u})$} change as $\mathbf{u}$ varies. While this implies
potential for economies, the computational demands are still
substantial (e.g., cubic in \blunew{$N \gg F$}) and nonetheless requires
\blunew{$\mathcal{O}((N+F)^2)$} for storage. For more details see \citet[][Section
8.1]{gramacy2020surrogates}.

If computational hurdles can be surmounted, synthesis of information at
varying fidelity (i.e., over field and simulation) can be a highly lucrative
affair in spite of notorious identifiability challenges 
with remedies of coming in Bayesian
\citep{Higdon:2004, bryn2014learning, plumlee2017bayesian, Gu:2018} and
frequentist \citep{tuo2015,tuo2016,wong2017,plumlee2019} flavors. \blunew{We
do not, ourselves, wade into these waters, in part because these remedies can
be at odds with computational tractibility in modern big-data application
\citep{marmin2022deep}.} However {\em modularization} \cite{Liu:2009},
\blunew{imposing} partial independence on the joint MVN structure, separating
surrogate $\hat{y}(\mathbf{x},
\mathbf{u})$ and bias $\hat{b}(\mathbf{x})$ \blunew{fitting}, 
\blunew{can be both simple and effective.}

\subsection{Scaling up}

\blunew{For our purposes}, a modular setup allows for thriftier surrogate
modeling, and thus more tractable posterior inference for $\mathbf{u}$ in
large-data settings. \cite{gra:etal:2015} leveraged modularization for a very
large radiative shock hydrodynamics experiment via local approximate Gaussian
processes \citep[LAGP;][]{gramacy:apley:2015}.  LAGP uses a neighborhood of
\blunew{$n \ll N$} nearby data subsets for much faster inference.
  \begin{align}
  \begin{array}{ccc}
 \blunew{\hat{\mathbf{y}}^M_{N} (\mathbf{X}^M_{N}, \mathbf{U}_{N}) } &
\longrightarrow  &
\blunew{\hat{\mathbf{y}}^M_{n}  (\mathbf{X}^M_{n}, \mathbf{U}_{n}) }\\
  \text{Global GP} &
\longrightarrow  &
\text{laGP}    
\label{eq:lagp}
  \end{array}
\end{align} 
Inference for $\mathbf{u}$ proceeds
via maximization of the posterior for $b(\cdot)$
 through the observed discrepancy 
\blunew{$\mathbf{D}^{B}_{F}(\mathbf{u}) =( \mathbf{X}_{F},  
 \mathbf{y}_{F} - \hat{\mathbf{y}}^M_{F}({\mathbf{u}}))$}: 
\begin{equation}
\hat{\mathbf{u}} = \mathrm{arg}\max_\mathbf{u} \left\{ p(\mathbf{u}) \left[ \max_{\bm{\phi}_b} 
p_b(\bm{\phi}_b \mid \mathbf{D}^{B}_{\blunew{F}}(\mathbf{u}))\right] \right\},
\label{eq:opt1}
\end{equation}
where $p_b$ is the \blunew{MVN} (marginal) likelihood for the GP prior on
$b(\cdot)$, and $\bm{\phi}_b$ are any hyperparameters  involved in
the bias covariance structure, e.g., lengthscales.

\cite{Huang:2018} developed on-site surrogates (OSSs), essentially
pairing a design strategy with local GP surrogate modeling for KOH.
Foreshadowing somewhat, as details are coming shortly in Section
\ref{sec:honeycomb}, they were motivated by the unique characteristics of
(cheap/fast but erratic and high-dimensional) {\tt ISOTSEAL} for the
honeycomb. Compared to conventional simulators (slow but smooth and
low-dimensional) the honeycomb demanded a large {\tt ISOTSEAL} campaign of
\blunew{$N = 292{,}000$} runs to fully map out the response surface. Yet only a small
handful \blunew{$F = 292$} of field-data observations were available.

To resolve this ``too big, too small" dilemma, \citeauthor{Huang:2018}~proposed to design/fit  {\tt ISOTSEAL} simulations/surrogates ``on-site'' as
follows: focus simulation designs, separately, on each of the physical
experimental input sites $\mathbf{x}_i$, for $i = 1, 2, \dots, \blunew{F},$ paired
with space-filling designs at calibration inputs $\mathbf{u}$ of size $n_i =
1000$. Extending chart (\ref{eq:lagp})\blunew{:}
 \begin{align}
 \begin{array}{ccccc}
\blunew{ \hat{\mathbf{y}}^M_{N}(\mathbf{X}^M_{N}, \mathbf{U}_{N})}  &
\longrightarrow  &
 \blunew{\hat{\mathbf{y}}^M_{n}  (\mathbf{X}^M_{n}, \mathbf{U}_{n})}  &
\longrightarrow & 
 \hat{\mathbf{y}}^M_{n_i}(\mathbf{U}_{n_i}),  i = 1, 2, \dots, \blunew{F} \\
 \text{Global GP} &
\longrightarrow  &
\text{laGP}  &
\longrightarrow & 
\text{OSSs} 
 \end{array}
\label{eq:oss}
\end{align} 
where $n_i$ denotes the number of {\tt ISOTSEAL} runs at the $i^\mathrm{th}$
site. In this way, the heavy $p_x + p_u =$ 17d simulation/emulation cargo is
decomposed onto a handful $(\blunew{F})$ of lighter, individually focused
$p_u=$ 4d sites.  Consequently,  OSSs address computational bottlenecks,
provide non-stationary flexibility, automatically smooth over artifacts in
some cases, and interpolates dynamics in others. Modularized calibration via
maximization (\ref{eq:opt1}) for $\hat{\mathbf{u}}$ is straightforward.   OSSs
also create a highly sparse kernel for \blunew{$\mathbf{\Sigma}(\mathbf{u})$}
in (\ref{eq:mvn}), so fully Bayesian KOH inference $\mathbf{u}$ is tractable
even with \blunew{$N$} in the millions.

\subsection{Calibration with multivariate outputs}

Although ideal for  {\tt ISOTSEAL} in many respects, this OSS strategy was for single-outputs.  In Section \ref{sec:multioutputs} we show that
separate application on honeycomb's multiple outputs (varying frequencies and
stiffness/damping coefficients) is problematic, motivating our main methodological
contribution [Section \ref{sec:muloss}].  Generally speaking, high dimensional
simulation output can manifest in many ways: functional \citep{Bayarri2007b,
Higdon2008}, time series \citep{conti2010, fadikar2018}, spatial
\citep{Bayarri2009},  spatial-temporal \citep{Gu2016}, spectral
\citep{Guinness2019}, with derivatives \citep{mcfarland2008calibration}.
Besides being more complex, multivariate output naturally implies
larger data size, aggravating computational challenges. Yet combining highly
multivariate, physically meaningful information, offers the potential for
improved posterior concentration and identification in calibration
\citep[see, e.g.][]{Arendt2012, Jiang2016}.

Dimension reduction, utilized appropriately, can help.  For example,
\cite{Higdon2008} extended univariate KOH into highly multivariate settings
through principal components,
\begin{equation}
\mathbf{y}^F(\mathbf{x}) = \mathbf{K}^M \mathbf{w}^M(\mathbf{x}, \mathbf{u}^\star)
 + \mathbf{K}^B\mathbf{w}^B(\mathbf{x}) +\mathbf{\epsilon}. 
  \label{eq:pccali}
\end{equation}
In this framework, high dimensional field observations
$\mathbf{y}^F(\mathbf{x})$ are modeled through $\mathbf{w}^M(\mathbf{x},
\mathbf{u}^\star)$ via orthogonal basis matrix $\mathbf{K}^M$, and discrepancies
$\mathbf{w}^B(\mathbf{x})$ via $\mathbf{K}^B$. Crucially, inference remains
tractable via MCMC, at least compared to the single-output analog, and so long
as training data sizes \blunew{$(N, F)$} are moderate. We aim to port this into the
OSS framework [Section \ref{sec:muloss}]. Other multi-output calibration
approaches, which seem less well-matched to our honeycomb setting, include
wavelet bases for functional outputs \citep{Bayarri2007b}, and the linear
model of co-regionalization \citep[LMC;][]{Paulo:2012}.

\section{Honeycomb specifics}
\label{sec:honeycomb}

Centrifugal compressors employ seals to minimize leakage in gas compression
phases,  preventing back flow and consequently performance decay. Conventional
annular gas seals, such as labyrinth and abradable seals, cause gas
recirculation around the shaft and produce destabilizing vibration effects.
Honeycomb seals are used in high performance turbomachinery to
promote stability via damping \citep[see, e.g.][]{childs}. Here we consider a
honeycomb rotor stabilizing gas seal under development at BHGE.

The system under study is characterized by input--output relationships between
variables representing seal geometry and flow dynamics. These include $p_x=13$
controllable physical design inputs $\mathbf{x}$, including rotational speed,
cell depth, seal diameter and length, inlet swirl, gas viscosity, gas
temperature, compressibility factor, specific heat, inlet/outlet pressure, and
inlet/outlet clearance. The field experiment, from BHGE's component-level
honeycomb seal test campaign, comprises \blunew{$F = 292$} runs varying a subset of
those conditions, \blunew{$\mathbf{X}_{F}$}, believed to have greatest variability
during turbomachinery operation: clearance, swirl, cell depth, seal length,
and seal diameter. Measured output features include direct/cross stiffness and
damping properties at multiple frequencies.

A general-purpose rotordynamic simulator called {\tt ISOTSEAL}, built upon
bulk-flow theory, virtually stress seals like the honeycomb. First developed
at Texas A\&M University \citep{isotseal}, it offers
fast evaluation (usually about one second) of gas seal force coefficients. Our
BHGE colleagues developed an {\sf R} interface mapping the seventeen scalar
inputs for the honeycomb into the format required for {\tt
ISOTSEAL}. Thirteen of those inputs match up with the columns of
\blunew{$\mathbf{X}_{F}$} (i.e., they are $\mathbf{x}$'s); four are calibration
parameters $\mathbf{u}$, which could not be controlled in the
field. These comprise statoric and rotoric friction coefficients $n_s, n_r$
and exponents $m_s, m_r$. We work with friction
factors coded to the unit cube: $(n_s, m_s, n_r, m_r)^\top \rightarrow (u_1,
u_2, u_3, u_4)^\top \in [0,1]^4$, primarily to protect BHGE's intellectual
property.  \blu{Throughout, we follow \citet{Huang:2018} and use independent
$\text{Beta}(2,  2)$ priors on $(u_1, u_2, u_3, u_4)$ to nudge the posterior
toward the interior of the space.  However, we comment briefly on a limited
sensitivity analysis in Section \ref{sec:baysresults}.}

\subsection{Multivariate outputs}
\label{sec:multioutputs}

The potential set of output features that could be monitored for the honeycomb
seal are many.  Here we focus on four rotordynamic coefficients, or
properties: direct stiffness ($K_d$),  cross stiffness ($k_c$), direct damping
($C_d$), and cross damping ($c_c$), measured at the following frequencies: 28,
70, 126, and 154 Hz; so 16 outputs in total. These are our
$\mathbf{y}$-values, measured either in the field as $\mathbf{y}^F_i$,
collected as $\mathbf{Y}_{\blunew{F}}$, or as $\mathbf{y}^M_i$ simulated via {\tt
ISOTSEAL}, collected as $\mathbf{Y}_{\blunew{N}}$.  \cite{Huang:2018} only considered
one of these: $K_d$ at 28 Hz.

The turbomachinery literature and bulk-flow theory \citep{Hirs:1973} provides
some insight into the relationship between these four properties. For example,
\cite{D'Souza:Childs:2002} demonstrate that classical transfer for a honeycomb
gas seal process can be expressed in a conventional linear
motion/reaction-force model
\begin{align}
- \begin{bmatrix} 
F_x  \\ 
F_y  
\end{bmatrix} = \begin{bmatrix} 
K_d & k_c \\ 
-k_c & K_d 
\end{bmatrix} \begin{bmatrix} 
x  \\ 
y 
\end{bmatrix} +
\begin{bmatrix} 
C_d & c_c  \\ 
-c_c & C_d 
\end{bmatrix} \begin{bmatrix} 
\dot x  \\ 
\dot y 
\end{bmatrix}
\label{eq:trans}
\quad \quad \mbox{(units omitted).}
\end{align}
Direct stiffness $K_d$ and damping $C_d$ account for orthogonal reaction
forces in $x$ and $y$ axes. Cross-coupled stiffness $k_c$ and damping $c_c$
describe reaction orthogonal to directions of motion. Other mechanical
engineering studies of rotordynamic coefficiets include
\citet{childs,isotseal,delgado2012}.

To access simulated versions of these outputs we augmented BHGE's {\sf R}
interface for {\tt ISOTSEAL} and then re-ran the campaign
of \cite{Huang:2018}, collecting following:
\begin{align}
\hat{\mathbf{Y}}^M(\mathbf{x}, \mathbf{u}) &\equiv  \hat{y}_{ijk}^M(\mathbf{u}), \quad \mbox{for } \left\{
\begin{array}{rl}
  i &= 1, 2, \dots,\blunew{F} \mbox{ (i.e., each field data pair)} \\
  j &\in \{1,2,3,4\} \mbox{ coding properties } \{K_d, k_c, C_d, c_c \} \\
  k &\in \{1,2,3,4\} \mbox{ coding frequencies.}
\end{array}
\right.
\label{eq:inds}
\end{align}
In total this involved $292 \times 1000 \times 4 \approx 1{,}168{,}000$  {\tt ISOTSEAL}
runs. Each run at inputs $(\mathbf{x}, \mathbf{u})$ is for a single output
frequency, producing all four rotordynamic coefficients simultaneously.
 In about 2\% of cases a convergence
issue is detected, terminating with an {\tt NA}-coded missing value after
about three seconds.  Collecting all $4{,}672{,}000$ measurements, over the four
frequencies, took about three days when divvied up across several multi-core
compute \blu{nodes}.  Our experiment resulted in \blunew{$N=\sum_{i=1}^{F}n_i= 286,282$}
successfully terminated runs, with most sites (241 out of 292) having a full $n_i =
1{,}000$. Of the 51 with missing responses of varying multitudes, the smallest
was $n_{238} = 574$. In total, we collected $16
\times 286{,}282 = 4{,}580{,}512$ on-site multivariate {\tt ISOTSEAL} runs.
The missingness pattern is similar to that reported in
\citet{Huang:2018}, even across output coefficients and frequencies because a
failed run at a particular input affects all outputs equally.  
One of the aims
of OSS calibration is to extrapolate to these unknown/missing parameter
regions after calibrating to field data.  See, e.g., \citet{marcy:2020},
albeit on a somewhat smaller scale.

A 16-fold increase in data, exhibiting all of the features of the
single-output case (nonstationarity, missingness, etc.), demands a scale-up of
OSS calibration in several directions.  We describe that in Section
\ref{sec:muloss}, but it shares pre-processing with a separate analysis of
each output.  We therefore turn first to a description of that simpler
process, which ultimately serves as a straw man against our fully multivariate
analysis.
 
\subsection{Separate univariate analysis}   
\label{sec:unioss}

Consider OSSs built separately for each of the outputs. Each OSS
comprises a fitted GP between successful on-site {\tt ISOTSEAL} run
outputs $y^M_{ijk}$ at $\mathbf{x}_i$ and with novel $1{,}000$-element maximin
Latin hypercube sample \citep[LHS;][]{morris:1995} $\mathbf{U}_{ijk}$, for
fixed $j$ and $k$. Specifically, $\hat{y}^M_{ijk}$ is built by fitting a
stationary zero-mean GP using a scaled and nugget-augmented separable Gaussian
kernel \blu{trained only on the space of parameter $\mathbf{u}$, }
\begin{align}
\blu{\blunew{\mathbf{\Sigma}}_{ijk}(\mathbf{u}_{jk}, \mathbf{u}_{jk}') = \tau_{ijk}^2 \left\{ \exp \left[  - \sum_{l=1}^{p_u} 
\frac{||\mathbf{u}_{ijkl} - \mathbf{u}'_{ijkl}||^2}{\theta_{ijkl}}  \right]   + \delta_{u,u'} \eta_{ijk} \right\},  }
 \label{eq:kernel}
\end{align}
where $\delta_{u,u'}$ is the
 Kronecker delta, $\tau_{ijk}^2$ is a scale parameter, $\bm{\theta}_{ijk}=(\theta_{ijk1},
\theta_{ijk2},  \dots, \theta_{ijkp_u} )^\top$ a vector of 
lengthscales, and $\eta_{ijk}$ is a nugget -- all \blu{being specific to the
$i^\mathrm{th}$ site, $j^\mathrm{th}$ output, and $k^\mathrm{th}$ frequency}.
Denote the set of hyperparameters of the $ijk^\mathrm{th}$ OSS as
$\bm{\phi}_{ijk}=\{\tau^2_{ijk},
\bm{\theta}_{ijk}, \eta_{ijk}\}$, with indices following Eq.~(\ref{eq:inds}).

\begin{figure}[ht!]
\centering
\includegraphics[scale=0.415, trim=0 45 25 0,clip=TRUE]{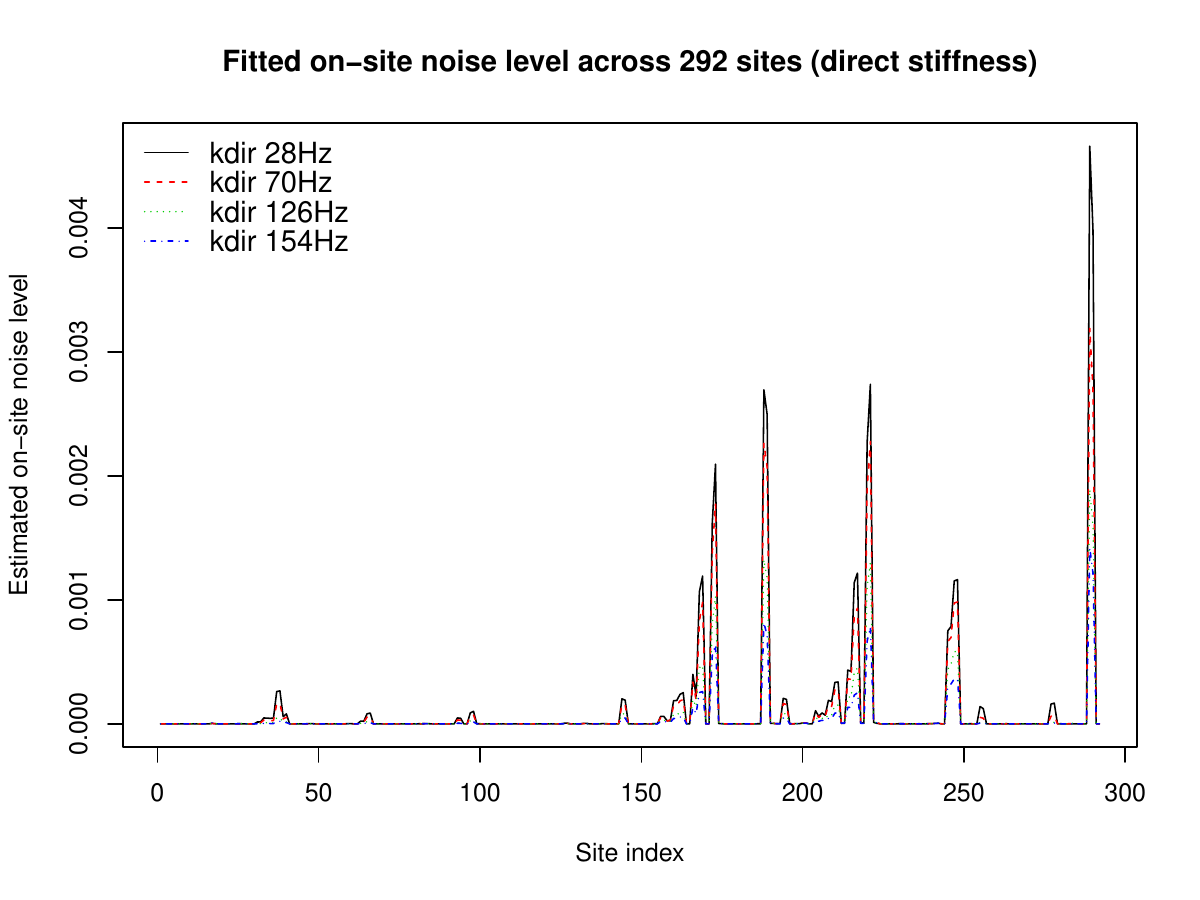}
\includegraphics[scale=0.415, trim=30 45 0 0,clip=TRUE]{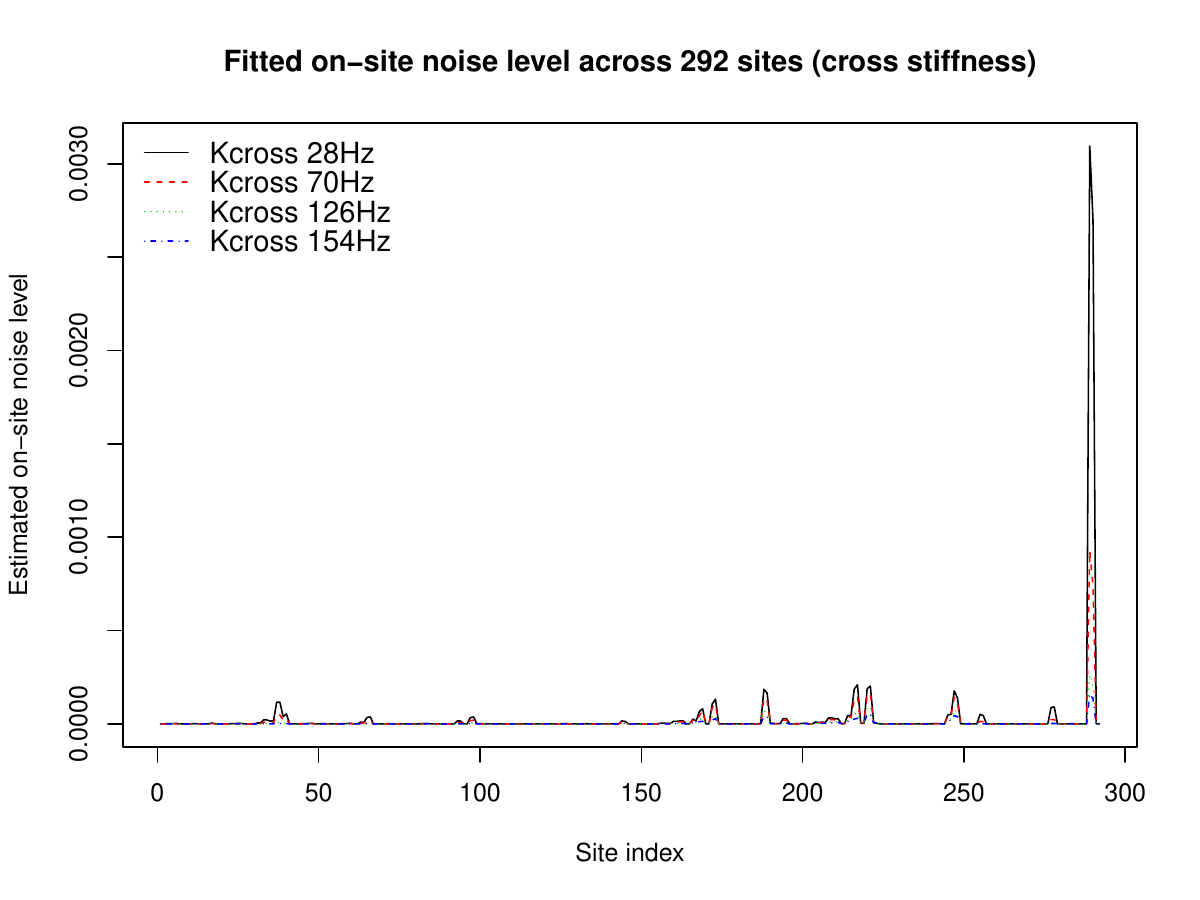}
\includegraphics[scale=0.415, trim=0 0 25 0,clip=TRUE]{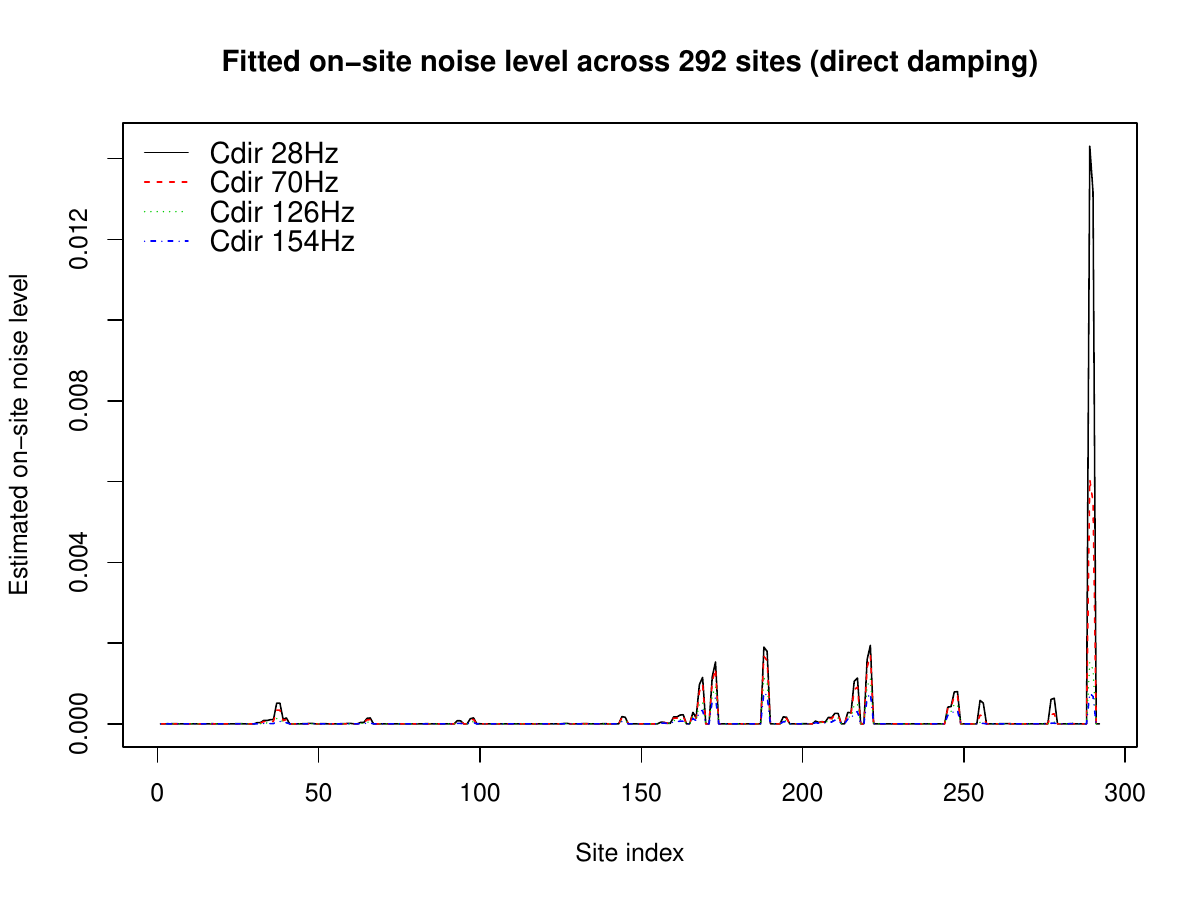}
\includegraphics[scale=0.415, trim=30 0 0 0,clip=TRUE]{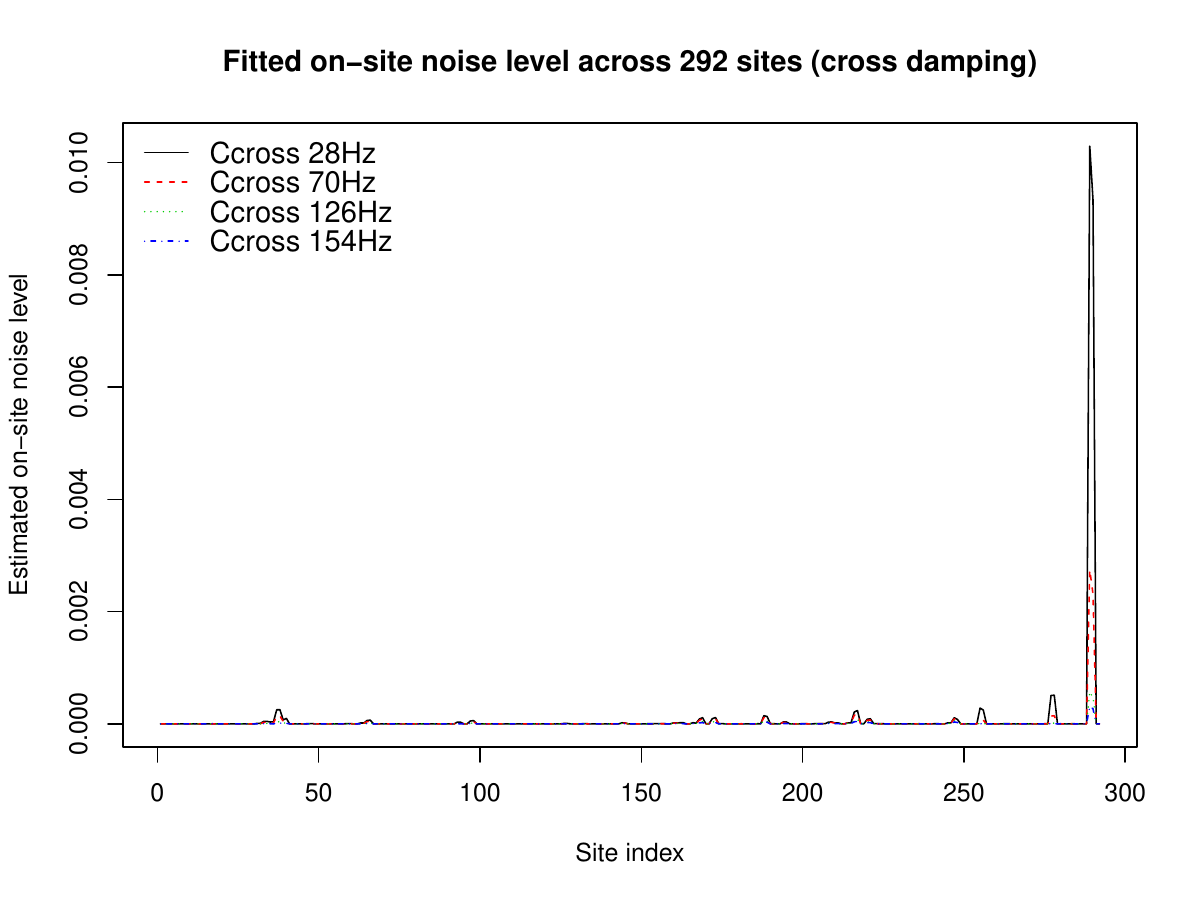}
\caption{Univariate OSS noise levels 
represented by the trained $\hat{\tau}_{ijk}^2 \hat{\eta}_{ijk}$ 
for direct/cross stiffness and damping at each of the four frequencies.}
\label{fig:noise1}%
\end{figure}

To offer some visual contrast between these fits, Figure \ref{fig:noise1}
shows the site-wise noise level $\hat{\tau}_{ijk}^2 \hat{\eta}_{ijk}$ across
the $\blunew{F}=292$ sites for the sixteen outputs. These are clearly non-constant, a
testament to the nonstationary nature of OSSs across the input space
represented by the 292 $\mathbf{x}_i$ settings, but also highly consistent
across outputs at each frequency level. This suggests that a certain amount of
information in each frequency is redundant. Although somewhat less obvious at
first glance, notice that the overall noise level decreases as the frequency
increases.  The amount by which noise level drops depends on the input
location, suggesting there is novel information in each output frequency,
which might manifest as spatial dependency, albeit in a large ($p_x =
17$d) space.

\begin{figure}[ht!]
\centering
\includegraphics[width=1\linewidth,trim= 0 25 0 50,clip=TRUE]{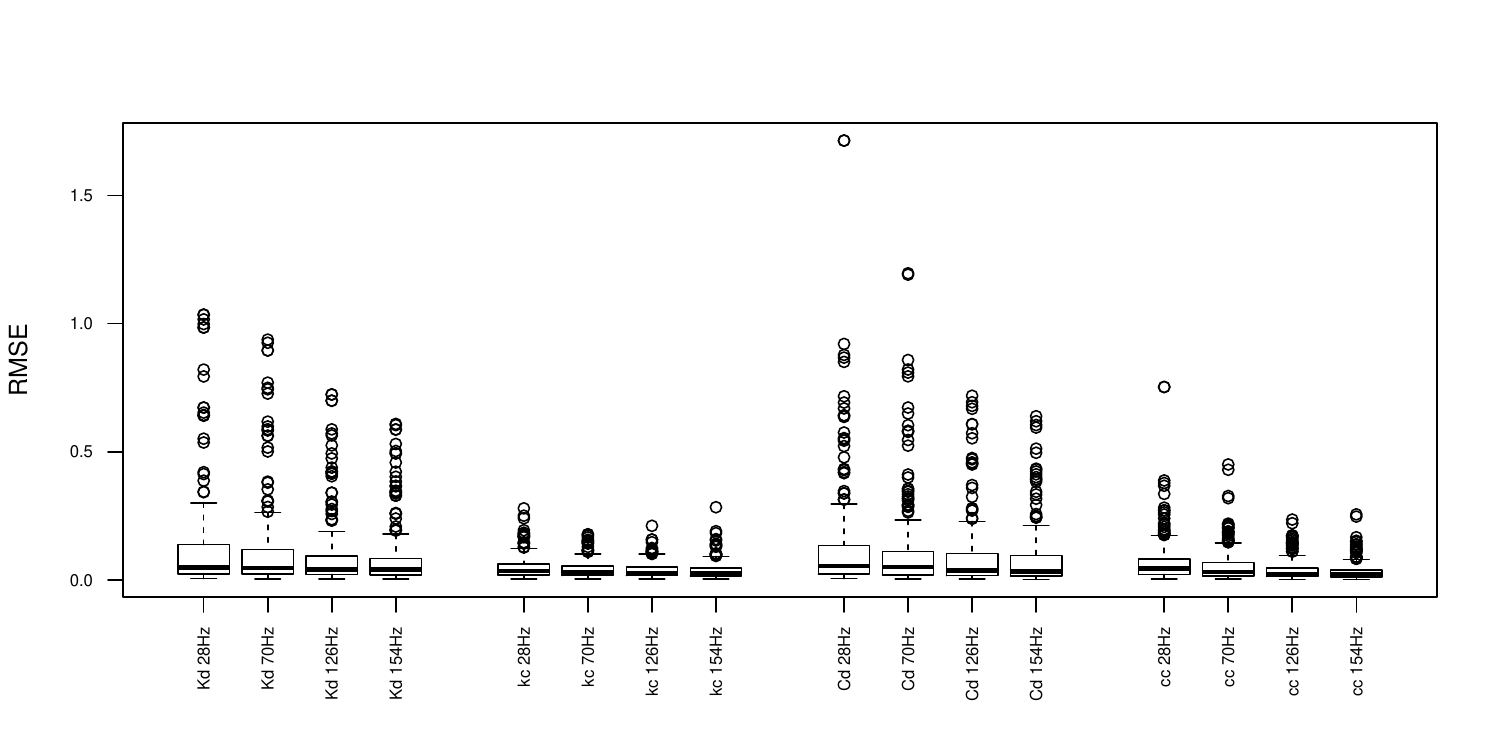}
\caption{Boxplots of 292 out-of-sample root mean-squared errors (RMSEs).}
\label{fig:rmse}%
\end{figure}

Before inserting these $292 \times 16 = 4{,}672$ OSSs into the KOH apparatus for
calibration, we performed a hyperparameter stability and predictive accuracy
check on a commensurately sized out-of-sample on-site testing data set, again
sixteen-fold across $4{,}672{,}000$ {\tt ISOTSEAL} runs.  The
hyperparameters looked similar to Figure \ref{fig:noise1}, and are not shown
here to save space. Prediction accuracies are summarized in Figure
\ref{fig:rmse}.   Each boxplot collects the 292 on-site root mean-squared
errors (RMSEs) from the $n_{ijk}
\approx 1{,}000$ converged runs on site $i$, separately for output $j$ and
frequency $k$.  Observe that most OSSs ($>50\%$ via the median lines in the
boxplots) act as interpolators on the sites, giving RMSEs are close to zero.
The rest act as extrapolators or smoothers, with a small handful ($\approx
5\%$ outside the whiskers) extremely so. Also notice that the scales of RMSEs
consistently decrease as frequency increases, coinciding with the in-sample
pattern(s) observed in Figure \ref{fig:noise1}.

\subsection{Challenges from univariate OSSs calibration}
\label{sec:challenge}

\begin{figure}[ht!]
\centering
\includegraphics[width=1\linewidth, trim=0 20 0 0,clip=TRUE]{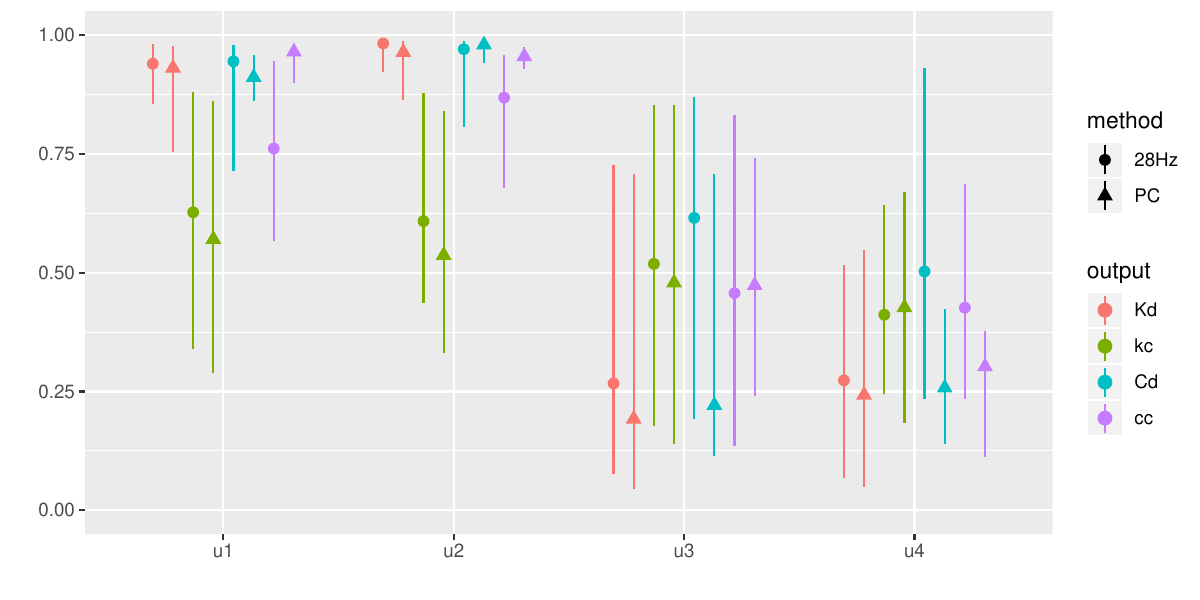}
\caption{Marginal posterior distributions of $\mathbf{u}$ 
 from outputs $K_d, k_c, C_d$, and $c_c$  at 28Hz (circles) and
via principal components combining all output frequencies (triangles, Section
\ref{sec:pcresult}). Dots indicate MAP values and error bars form 90\% credible
intervals.}
\label{fig:unibayes}
\end{figure}

\begin{figure}[ht!]
\centering
\includegraphics[width=1\linewidth, trim=0 5 0 0,clip=TRUE]{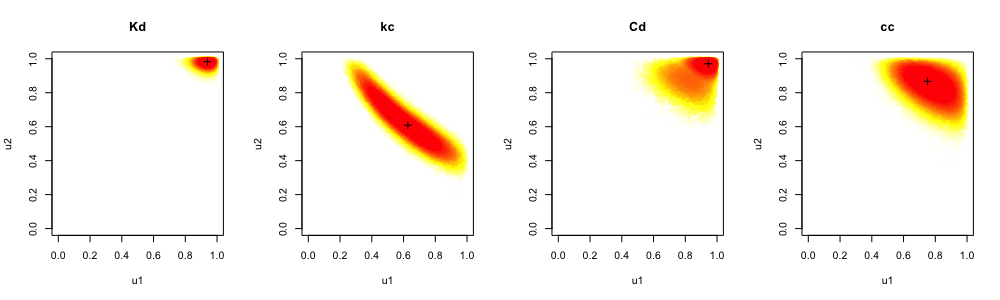}
\includegraphics[width=1\linewidth, trim=0 15 0 50,clip=TRUE]{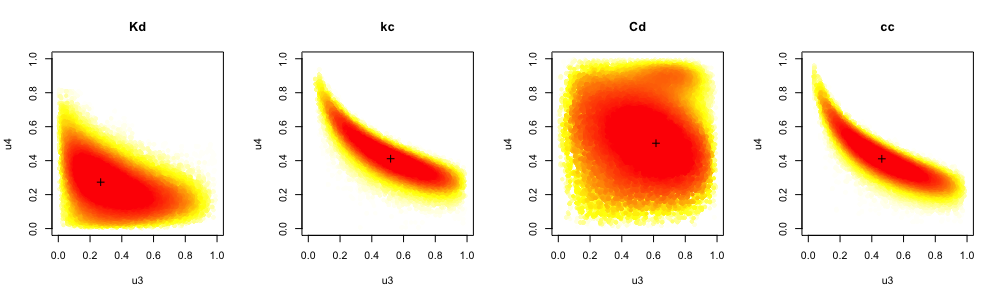}
\caption{Bivariate marginal plots of single-output posterior samples of $\mathbf{u}$ 
 from outputs $K_d, k_c, C_d$, and $c_c$  at 28 Hz. 
Dots are $100{,}000$ MCMC samples of $\mathbf{u}$, heat 
colored according to the rank of (log)
posterior probability. The ``+" indicates the MAP values. }
\label{fig:unibayes_2}
\end{figure}

Next, combining with field data, we performed separate OSS-based fully
Bayesian calibrations for $\mathbf{u}$, repeating the
\citeauthor{Huang:2018}~univariate approach for each output and frequency.
That is, we performed $4 \times 4 = 16$ MCMCs, $100{,}000$ samples each, and
discarded $5{,}000$ as burn-in. Each MCMC run took about three days to finish, 
after a warm start from modular optimization \blu{(\ref{eq:opt1})}. 
To demonstrate the drawbacks from this  univariate approach, Figures
\ref{fig:unibayes} and \ref{fig:unibayes_2} present two views into the
posteriors for $\mathbf{u}_{jk}$.

Figure \ref{fig:unibayes} shows marginal posteriors for the components of
$\mathbf{u}_{jk}$ for each output at 28Hz ($k=1$).   The other frequencies,
provided in Supplement \ref{sec:uniap}, look similar. To support a comparison
coming later in Section \ref{sec:pcresult}, the plot also shows results
combining all frequencies which will be discussed in due course.  Focusing on
the 28Hz results, notice that while some coordinates, like $u_3$ and $u_4$,
exhibit consistency in the high overlap of their credible intervals, others
like $u_1$ and $u_2$ do not.  There would appear to be complimentary (unused)
information in these independent analyses.
For a slightly higher resolution view, Figure \ref{fig:unibayes_2} shows
posteriors for pairs $u_1 \times u_2$ for each output (at 28Hz) across the
top, and $u_3 \times u_4$ across the bottom.  Again, the set of similar
visuals is completed in the supplemental material.  In the top row, $k_c$
stands out as the exception.  Despite wider credible regions in the bottom
row, only one pair ($k_c, c_c$) among the six resemble one another.

Taken together -- and even for just these four sets of $\mathbf{u}_{jk}$
posteriors, ignoring the other output frequencies -- there is no way to stack
them together in the 4-dimensional space to generate one single, concentrated
solution with appropriate evaluation of individual parameter uncertainty from
different outputs. These separately sampled  $\mathbf{u}_{jk}$ posteriors indicate
heterogeneity in parameter learning with varied amount of uncertainty,
creating more uncertainty downstream to any engineering decision-making. A
joint analysis not only holds the potential to reconcile (at times)
contradictory information, but would also drastically simplify downstream
decision-making with far fewer posterior views to scrutinize.

\section{On-site surrogates with basis representation}
\label{sec:muloss}

The essential insight behind OSS-based calibration is that learning for
$\mathbf{u}$ is primarily driven by model discrepancy $b(\mathbf{x})$, which
can only be observed at field input sites $\mathbf{X}_{\blunew{F}}$, through
residuals.  At a conceptual level, extending that to multivariate output
\blunew{means evaluating vectors of outputs}:
$\mathbf{y}_{ijk}^F({\mathbf{X}_{\blunew{F}}})$ in the field and
$\hat{\mathbf{y}}_{ijk}^{M}({\mathbf{X}_{\blunew{F}},  \mathbf{U}_i})$ for the
surrogate, and then calculating discrepancies with extra indices.
\begin{align}
\mathbf{y}_{ijk}^F({\mathbf{X}_{\blunew{F}}})-\hat{\mathbf{y}}_{ijk}^{M}({\mathbf{X}_{\blunew{F}}, \mathbf{U}_i}), \quad i = 1, 2, \dots, \blunew{F}, \quad j=1,2, ..., J, \quad k=1,2, ..., K,
\label{eq:onsite}
\end{align}
\blu{where $ \mathbf{U}_i$ is the design matrix (maximim LHS) 
of  calibration parameter $\mathbf{u}$
for converged $n_i$ on-site runs for the $i^\mathrm{th}$ site.}
As with univariate outputs, as long as the surrogate is
good at those locations, which OSSs facilitate, further considerations can be
pushed downstream.  For example, when new locations are of interest, say for
prediction, new OSSs/runs can be built/performed there, which is trivial if
(like {\tt ISOTSEAL}) the simulations are fast.

The devil is, however, in the details for how these residuals
(\ref{eq:onsite}) are modeled.  Section \ref{sec:challenge} demonstrates that
separating over $j$ and $k$ for the honeycomb leads to pathologies.  Here we
begin the description for a joint analysis, building a fitted bias for residuals
all at once, to resolve those inconsistencies while ensuring that all relevant
information is incorporated in posterior inference for $\mathbf{u}$, thus
filtering to downstream tasks like prediction.  What we propose below is
customized, to a degree, to the honeycomb setting, e.g., via fixed $j$ (output
classes) and pooling over $k$ (output frequencies) in Eq.~(\ref{eq:onsite}).
Completing the description by combining over output classes (pooling over $j$)
is deferred to Section \ref{sec:allcali}.  
We begin here with some notational setup that applies for
all $j$ and $k$, and then fix $j$ for the remainder of the section. Although
we believe other applications -- perhaps with more or fewer indices -- could
be set similarly, further speculation is left to Section \ref{sec:discuss}.

Let $\mathbf{Y}_i^M(\mathbf{U}) = [\mathbf{y}_{i11}^M(\mathbf{U}_i),
\mathbf{y}_{i12}^M(\mathbf{U}_i), \dots, \mathbf{y}_{i44}^M(\mathbf{U}_i)]$,
for $i=1, 2, \dots,\blunew{F}$, be a 16-column matrix (four outputs at four
frequencies) holding the $n_i \approx 1000$ rows of converged {\tt ISOTSEAL} runs
for the $i^\mathrm{th}$ site.  Now collect $\blunew{F}=292$ 
of these $\mathbf{Y}_i^M(\mathbf{U}) = [\mathbf{Y}_1^M(\mathbf{U})^\top, \dots, 
\mathbf{Y}_{\blunew{F}}^M(\mathbf{U})^\top ]^\top$ 
\begin{align}
\mathbf{Y}^M(\mathbf{U}) &= \label{eq:ym}
\begin{bmatrix} \mathbf{y}^M_{111}(\mathbf{U}_1) 
& \dots & \mathbf{y}^M_{144}(\mathbf{U}_1)  \\
 \vdots 
 & \ddots & \vdots \\
\mathbf{y}^M_{\blunew{F}11}(\mathbf{U}_{\blunew{F}}) 
& \dots & \mathbf{y}^M_{\blunew{F}44}(\mathbf{U}_{\blunew{F}}) \end{bmatrix}_{\blunew{N} \times 16} 
\end{align}
whose row dimension is  $\blunew{N} = \sum_{i=1}^{\blunew{F}} n_i = 286{,}282$.
Recall that $\mathbf{U}_i$ are $n_i \times p_u$ on-site (maximin LHS) design
matrices.

\begin{figure}[ht!]
\centering
\includegraphics[scale=0.65,trim=20 80 0 35,clip=TRUE]{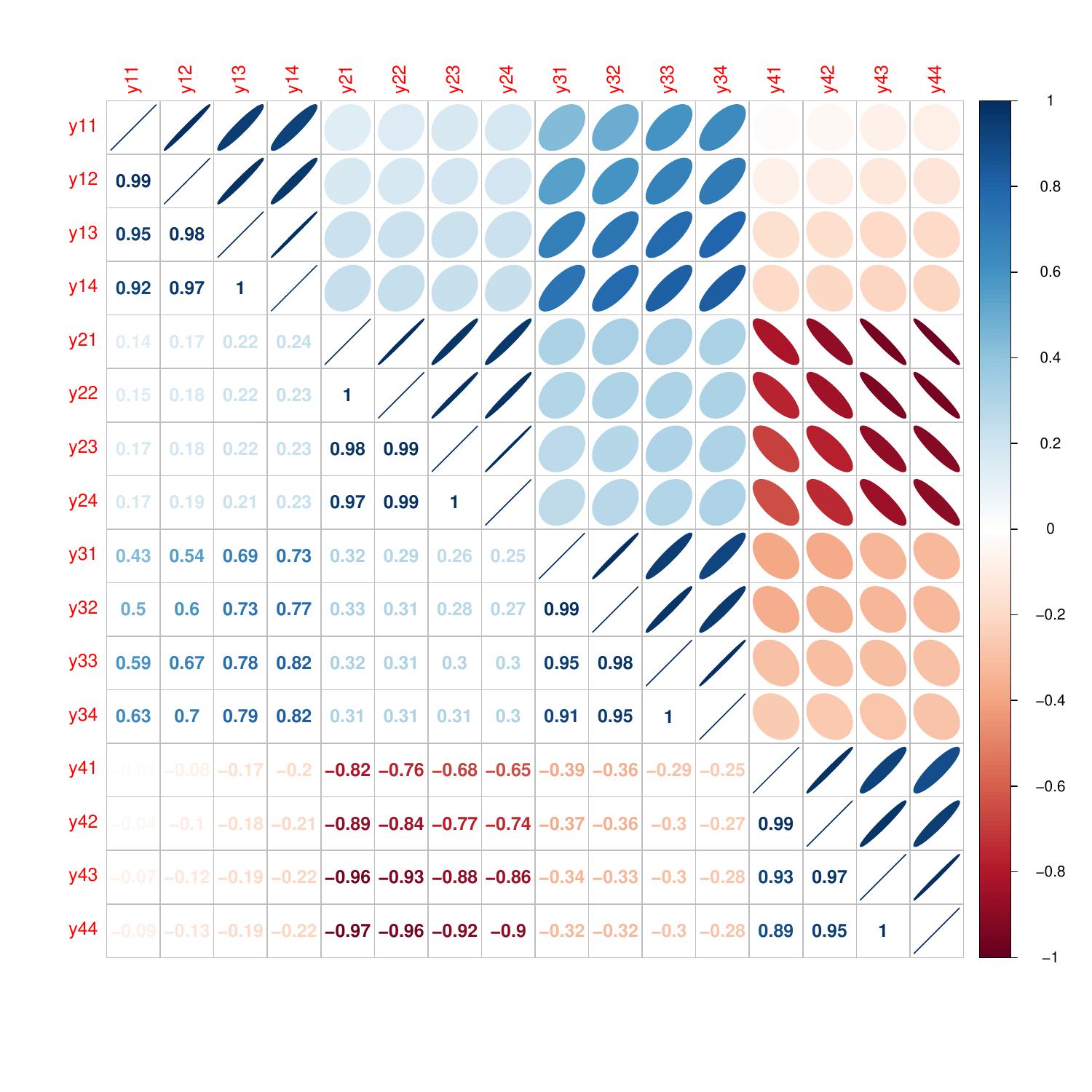}
\caption{Correlation matrix (number in lower triangle and ellipse in upper triangle) 
of $\blunew{N} = 286{,}282$ multivariate {\tt ISOTSEAL} simulation of direct/cross
stiffness and damping at frequencies 28, 70, 126, and 154 Hz. In the output
labels $y_{jk}$,  $j = 1, \dots, 4$ for output property and $k = 1, \dots, 4$ for
frequency, as defined in Eq.~(\ref{eq:inds}).}
\label{fig:cor_ym}
\end{figure}

We performed an initial exploratory data analysis (EDA) with these data,  one
aspect of which is $\mathbf{C}\mathrm{or}\{\mathbf{Y}^M(\mathbf{U})\}$,
visualized in Figure \ref{fig:cor_ym}.  Correlation strength and association
is indicated numerically in the lower-triangle, by elliptical shape and
direction in the upper half, and by color and shading in both. Four-by-four
blocks are clearly evident in this view, indicating strong linear correlation
between different frequency levels within the same property (block of) output,
but weaker correlation between types.
Our EDA also revealed similar correlations exhibited by the output-combined
field data $\mathbf{Y}^F$, defined similarly.  This is not shown here for
brevity.  Taken together, we conclude that parsimonious representation of
across-frequency information could be beneficial to a joint modeling
enterprise.

\subsection{Principal component OSSs}
\label{sec:osspca}

We propose performing principal component analysis (PCA) across frequency
outputs in an ``on-site'' fashion, i.e., fit a PC basis in the subspace spanned
by $\mathbf{U}_i, i=1,\dots, \blunew{F}$ on the $\blunew{F}$ observed physical sites
$\mathbf{X}_{\blunew{F}}$ for each type (block, indexed by $j$) of outputs. Fixing
$j$, we first center and standardize the correlated $K=4$ frequencies, perform
PCA on the \blunew{$N \times K$} dimensional ``on-site'' matrix,
\begin{align}
\mathbb{PC} \{ \mathbf{Y}_j^F - \mathbf{Y}_j^M \},  \quad \mbox{yielding eigenvectors} \quad \mathbf{W}_j, \quad 
\text{for} \quad j = 1, \dots, J.
\label{eq:pca}
\end{align}
In Eq.~(\ref{eq:pca}),  $\mathbf{Y}_j^F$ collects $K=4$ columns of
$n_i$-row  replicated field outputs across frequencies, 
 \begin{align} 
 \mathbf{Y}_j^F =
\begin{bmatrix} \mathbf{y}^F_{1j1} & \cdots 
& \mathbf{y}^F_{1j4} \\
 \vdots 
 & \ddots & \vdots \\
\mathbf{y}^F_{{\blunew{F}} j1} & \cdots 
& \mathbf{y}^F_{{\blunew{F}}j4} \\ \end{bmatrix}_{\blunew{N} \times 4,}
\label{eq:YF}
\end{align}
where $ \mathbf{y}^F_{ijk}\equiv (y^F_{ijk}(\mathbf{x}_i), \dots,
y^F_{ijk}(\mathbf{x}_i) )^\top$ is the $n_i$-time \blu{duplicated} field output
$y^F_{ijk}(\mathbf{x}_i)$ on $i^\mathrm{th}$ site for output $j$ at frequency $k$.
 Similarly, $\mathbf{Y}_j^M$ collects $K=4$  columns of on-site {\tt ISOTSEAL}
 simulations on $\blunew{F}$ sites at  multiple frequencies provided in Eq.~(\ref{eq:ym}).

\begin{figure}[ht!]
\centering
\includegraphics[width=0.243\linewidth, trim=10 10 10 20]{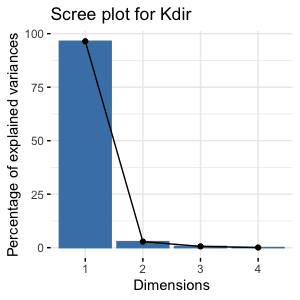}
\includegraphics[width=0.243\linewidth, trim=10 10 10 20]{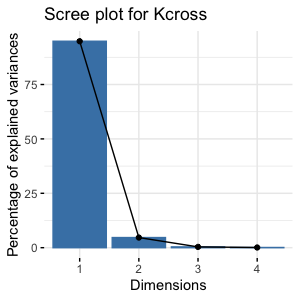} 
\includegraphics[width=0.243\linewidth, trim=10 10 10 20]{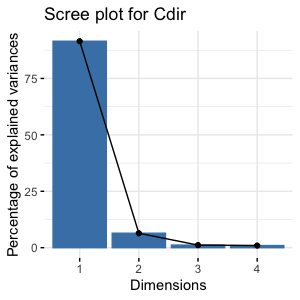}
\includegraphics[width=0.243\linewidth, trim=10 10 10 20]{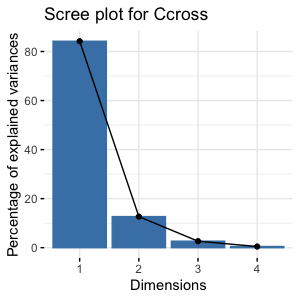}

\vspace{0.25cm}
\begin{tabular}{r | r | r | r | r}
Observed discrepancy  on output & $K_{d}$ & $k_{c}$ & $C_{d}$ & $c_{c}$  \\
\hline
Variation represented in the first-PC & 96.46\% & 94.89\% & 91.49\% & 84.19\% \\
Variation represented in the second-PC& 2.84\% & 4.66\% & 6.41\% & 12.68\% \\
\end{tabular}
\caption{Scree plots and tabulation of PCA variances.}%
\label{fig:scree}%
\end{figure}

For honeycomb, there are $J=4$ properties (blocks):
\{$K_{d}$, $k_{c}$, $C_{d}$, $c_{c}$\}. Thus, we performed four separate PCAs
 in total.  Figure \ref{fig:scree} summarizes these via scree
plots of variance decomposition, accompanied by a table with a numerical
summary via the top variances.
Notice that the first-PCs dominate in all four cases.  Output
$c_c$ which might marginally benefit from the second principal direction.

This suggests that a \citet{Higdon2008}-style basis-based-surrogate
(\ref{eq:pccali}), projecting down to one principal direction for these four
outputs (separately over output property $j$) could be effective here, after
upgrading to accommodate OSSs. Instead of breaking down the simulator and
discrepancies into two separate principal representations  in
(\ref{eq:pccali}), we prefer a single PCA through the matrix of eigenvectors
$\mathbf{W}_j$ trained from  the whole ``on-site'' observed discrepancy in
(\ref{eq:pca}). In particular, we  use the first (column) eigenvector
$\mathbf{w}^1_j$ of $\mathbf{W}_j$ to extract the first-PCs from multiple
frequency outputs of both simulated $\mathbf{y}_j^{M1} =
\mathbf{Y}_j^M
\mathbf{w}_j^1$ and field $\mathbf{y}_j^{F1} = \mathbf{Y}_j^F\mathbf{w}_j^1$.
Once extracted, we can put them together as 
\begin{equation}
\mathbf{y}_j^{F1}(\mathbf{x}) = \mathbf{y}_j^{M1} (\mathbf{x}, \mathbf{u}^{\star1})
 + \mathbf{b}^1_j(\mathbf{x})  + \mathbf{\epsilon}^1_j. 
  \label{eq:pckoh}
\end{equation}
which may be interpreted as ordinary KOH in the first-PC subspace. Observe the
introduction of new notation for model discrepancy
$\mathbf{b}^1_j(\mathbf{x})$, best calibration setting $\mathbf{u}^{\star1}$,
and iid noise $\mathbf{\epsilon}_j^1$, to recognize PC pre-processing to form
$\mathbf{y}_j^{F1}(\mathbf{x})$ and $\mathbf{y}_j^{M1}$. Focus remains on
observed sites only, i.e., $\mathbf{x} \equiv \mathbf{X}_F$ in the on-site setting,
for observed discrepancies, and surrogate(s) trained on $\blunew{N}= 286{,}282$ dense
paired with space-filling $\mathbf{U}_i$'s. After learning,
prediction, etc., may be linearly back-transformed to the
original full space.


\subsection{PC-level OSSs calibration: optimization and full Bayes}
\label{sec:pcacali}

Each first on-site PC can be used in a separate calibration: four this time
instead of sixteen in Section \ref{sec:challenge}.  
Here we illustrate how these may plugged into the OSS calibration framework as an
intermediate step toward a fully joint model in Section
\ref{sec:allcali}.  


\paragraph{Modular calibration via optimization.} Here the goal is to search
for $\hat{\mathbf{u}}_j^1$ by maximizing the posterior probability of
(observed) model discrepancy (\ref{eq:opt1}) in the first principal direction.
Update Eq.~(\ref{eq:opt1}) by swapping the univariate (column vector) observed
discrepancy with that based on first eigenvector $\mathbf{w}^1_j$ obtained in
Eq. (\ref{eq:pca}). I.e., for each of $j=1, \dots, J$,
 \begin{align}
\mathbf{D}^{B}_{\blunew{F}}(\mathbf{u}) =( \mathbf{X}_{\blunew{F}},  
 \mathbf{y}_{\blunew{F}} - \hat{\mathbf{y}}^M_{\blunew{F}}({\mathbf{u}}))
 \rightarrow
 \mathbf{D}^{j1}_{\blunew{F}}(\mathbf{u}_j^1) =( \mathbf{X}_{\blunew{F}},  
(\mathbf{Y}_j^F - \hat{\mathbf{Y}}_j^M(\mathbf{u}_j^1) ) \mathbf{w}^1_j ),  
\label{eq:pcbias}
\end{align}
In (\ref{eq:pcbias}),  $\mathbf{Y}_j^F$ is fixed and
$\hat{\mathbf{Y}}_j^M(\mathbf{u}_j^1)$ are OSS [fitted as in Section
\ref{sec:unioss}] evaluations at $\mathbf{u}_j^1$, which would vary along a
numerical optimizer's search trajectory. The mathematical programs
are identical to Eq.~(\ref{eq:opt1}) but with
$\mathbf{D}^{B}_{\blunew{F}}(\mathbf{u}) \rightarrow
\mathbf{D}^{j1}_{\blunew{F}}(\mathbf{u}_j^1)$ and $\bm{\phi} \rightarrow \bm{\phi}_j^1$
denoting hyperparameters involved in each of $j$ GP-based \blu{(separable
squared exponential)} fitted discrepancies.
 
\paragraph{Bayesian joint inference.} Posterior sampling of $\mathbf{u}^1_j$
requires projecting $\mathbf{Y}_j^F$ (\ref{eq:YF}) and $\mathbf{Y}_j^M$
(\ref{eq:ym}) onto their first principal axis, $\mathbf{y}^{F1}_j$ and
$\mathbf{y}^{M1}_j$ respectively, via Eq.~(\ref{eq:pca}). Then, following
Eq.~(\ref{eq:pckoh}), one may impose a joint MVN (\ref{eq:mvn}), with mean
zero and covariance \blunew{$\mathbf{\Sigma}_j^1(\mathbf{u}^1_j)$}, whose structure is
re-notated below with appropriate indices/first-PC indicators:
\blunew{
\begin{align}
\mathbf{\Sigma}_j^1(\mathbf{u}^1_j)
\equiv \begin{bmatrix} \mathbf{\Sigma}^{j1}_{N}& 
\mathbf{\Sigma}^{j1}_{\blunew{F}, N}(\mathbf{u}^1_j)^\top\\ 
\mathbf{\Sigma}^{j1}_{\blunew{F}, N}(\mathbf{u}^1_j) &
\mathbf{\Sigma}^{j1}_{\blunew{F}}(\mathbf{u}^1_j)  + \mathbf{\Sigma}^{j1}_{b} \end{bmatrix}.    
\label{eq:Sigma}
\end{align}}
Posterior evaluation for $\mathbf{u}^1_j$ in a Metropolis setting
requires decomposing this \blunew{$\mathbf{\Sigma}_j^1(\mathbf{u}^1_j)$} for inverse and
determinant components of the MVN log likelihood.  Note the size of this
matrix is \blunew{$(N
+ F)^2$}, where \blunew{$N + F = 286,574$}  for our honeycomb application.
Consequently, ordinary cubic in \blunew{$(N + F)$} decomposition costs pose a
serious bottleneck.  

Fortunately, \blunew{$\mathbf{\Sigma}_{j}^1(\mathbf{u}^1_j)$} has a convenient
structured-sparsity form under OSSs. 
The largest block  \blunew{$\mathbf{\Sigma}^{j1}_{N}$}, corresponding to the
OSSs themselves, is the most sparse.  It is block-diagonal with $\blunew{F}$
blocks: \blunew{$\mathbf{\Sigma}^{j1}_{N}=\Diag[\mathbf{\Sigma}^{j1}_i(\mathbf{U}_i,
\mathbf{U}_i)]$}.  Each block \blunew{$\mathbf{\Sigma}^{j1}_i(\mathbf{U}_i,
\mathbf{U}_i)$} may be built from the kernel of the $i^\mathrm{th}$ OSS 
conditioned on any fitted hyperparameters from $\mathbf{y}_{ij}^{M1}$.  Since it
does not depend on $\mathbf{u}^1_j$, each may be pre-decomposed separately
at manageable $\mathcal{O}(n_i^3)$ cost.  

The rest of \blunew{$\mathbf{\Sigma}_j^1(\mathbf{u}^1_j)$} requires bespoke construction given
novel $\mathbf{u}^1_j$, but still has convenient block--sparse structure.  The
simulator--field cross covariance piece is block-diagonal, \blunew{$\mathbf{\Sigma}^{j1}_{\blunew{F},
N}(\mathbf{u}^1_j)=\Diag[\mathbf{\Sigma}^{j1}_i(\mathbf{u}^1_j, \mathbf{U}_i)]$},
where \blunew{$\mathbf{\Sigma}^{j1}_i(\mathbf{u}^1_j, \mathbf{U}_i)$} is a row vector of
site-wise covariance between calibration inputs $\mathbf{u}^1_j$ and on-site
design matrix $\mathbf{U}_i$. Finally, although \blunew{
$\mathbf{\Sigma}^{j1}_{\blunew{F}}(\mathbf{u}^1_j) + \mathbf{\Sigma}^{j1}_{b}$} is dense, it is small
($\blunew{F} \times \blunew{F}$) and thus cheap to decompose.  The first component
\blunew{$\mathbf{\Sigma}^{j1}_{\blunew{F}}(\mathbf{u}^1_j)$} is block diagonal, where each block is
like \blunew{$\mathbf{\Sigma}^j_i(\mathbf{U}_i,
\mathbf{U}_i)$} from the $i^\mathrm{th}$ OSS kernel, except with a  stacked
$\mathbf{u}^1_j$ vectors in lieu of the $\mathbf{U}_i$.  The second component
\blunew{$\mathbf{\Sigma}^{j1}_{b}$} is dense and comes from the discrepancy kernel using all
field-data inputs.  Any hyperparameters, e.g., $\hat{\bm{\phi}}_{j}^1$ are most
easily set via maximization-based pre-analysis, e.g., following
Eq.~(\ref{eq:pcbias}), but could also be included in the MCMC.
 \blu{For consistency, zero-mean separable GPs are fitted for $\hat{\bm{\phi}}_{j}^1$.}

Once \blunew{$\mathbf{\Sigma}_j^1(\mathbf{u}^1_j)$} in Eq.~(\ref{eq:Sigma}) is built, for each
proposed $\mathbf{u}^1_j$ setting in a Metropolis scheme, and after its
component parts have been decomposed, a full decomposition -- i.e., combining
from constituent parts -- involves tedious but ultimately straightforward
matrix multiplication  via partition inverse and determinant equations
\citep[e.g.,][]{Petersen:2008}
\blu{ leveraging the multiple-OSSs setup, 
\blunew{\begin{align}
\mathbf{\Sigma}^1_j(\mathbf{u}^1_j)^{-1}  
& = \begin{bmatrix}(\mathbf{\Sigma}^{j1}_{N} )^{-1} +(\mathbf{\Sigma}^{j1}_{N} )^{-1} \mathbf{\Sigma}^{j1}_{F, N}(\mathbf{u}^1_j) ^\top
\mathbf{C}^{-1} (\mathbf{u}^1_j ) \mathbf{\Sigma}^{j1}_{F, N}(\mathbf{u}^1_j)(\mathbf{\Sigma}^{j1}_{N} )^{-1} 
 & (\cdot)^\top \\ 
-\mathbf{C}^{-1} (\mathbf{u}^1_j ) \mathbf{\Sigma}^{j1}_{F, N}(\mathbf{u}^1_j)(\mathbf{\Sigma}^{j1}_{N} )^{-1} 
& \mathbf{C}^{-1} (\mathbf{u}^1_j ) \end{bmatrix}, \nonumber \\
\det[ \mathbf{\Sigma}^1_j(\mathbf{u}^1_j)] 
&= \det(\mathbf{\Sigma}^{j1}_{N} ) \times \det[\mathbf{C} (\mathbf{u}^1_j )], \quad \text{where} \nonumber \\
\mathbf{C} (\mathbf{u}^1_j  ) 
& = \mathbf{\Sigma}^{j1}_{F}(\mathbf{u}^1_j)  + \mathbf{\Sigma}^{j1}_{b}
 - \mathbf{\Sigma}^{j1}_{F, N}(\mathbf{u}^1_j) 
\mathbf{\Sigma}^{j1}_{N} \mathbf{\Sigma}^{j1}_{F, N}(\mathbf{u}^1_j) ^\top.
\end{align}
}}

\subsection{PC-level calibration results}
\label{sec:pcresult}

Isolating each \{$K_{d}$,  $k_{c}$,  $C_{d}$,  $c_{c}$\}, but leveraging
strong linear correlation across frequencies ($4 \times 4 = 16$ total
outputs), eliminates redundant information and brings the unit of analysis
down fourfold to 4.  The computational merits of OSSs, ported to a PC basis,
enables efficient modular optimization and fully Bayes inference on
parameter(s) $\mathbf{u}^1_j$. Here we present the outcome of such analysis
with an eye toward a fully combined setup in Section \ref{sec:allcali}.   We
focus on contrasting to two earlier views, from a fully independent,
separated (16-fold) analysis provided in Section \ref{sec:multioutputs}.

Figure \ref{fig:unibayes} shows 1d posterior marginals for $\mathbf{u}^1_j$,
for each $j=1,\dots,4$, alongside their 28Hz-only analog.  With four
coordinates of $\mathbf{u}$, a total of sixteen comparisons may be made with
this view.  Observe, for example, that $u_1$,  $u_2$, and $u_4$ show
considerable concentration of density under PC for outputs $C_d$ and $c_c$.
There is also a considerable shift in the location of posterior mass (i.e.,
the MAP), with $u_1$ and $u_2$ shifting up and $u_4$ shifting down for $c_c$.
The rest of the PC marginals are similar to their 28Hz counterparts.


\begin{figure}[ht!]
\centering
\includegraphics[width=1\linewidth, trim=0 5 0 0,clip=TRUE]{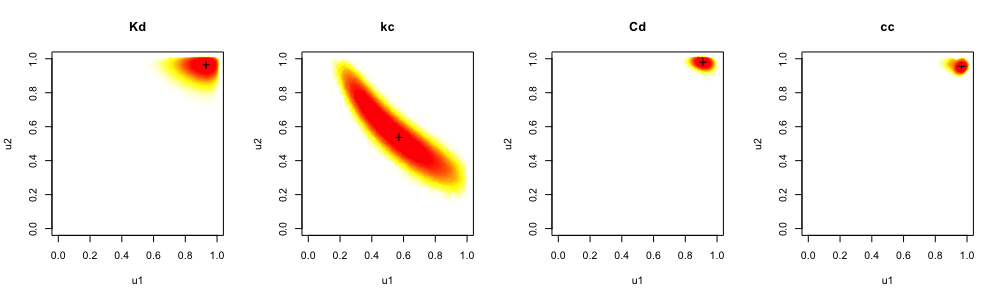}
\includegraphics[width=1\linewidth, trim=0 15 0 50,clip=TRUE]{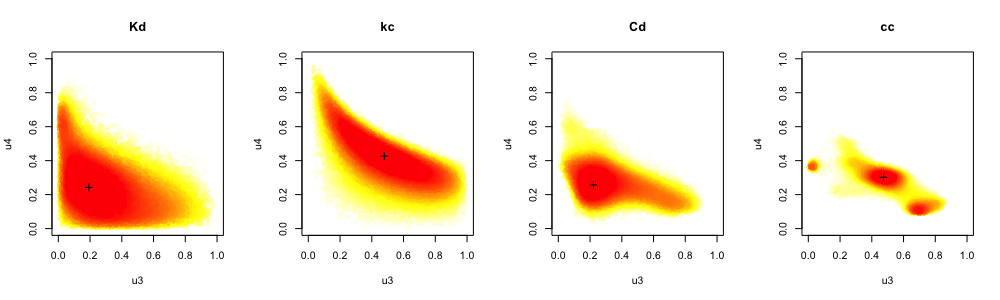}
\caption{Bivariate marginal plots of posterior samples of $\mathbf{u}^1_j$ 
from PC-combined outputs $K_d, k_c, C_d$, and $c_c$. Dots are
 $100{,}000$ MCMC samples of $\mathbf{u}^1_j$, heat colored according to the
 rank of (log) posterior probability. The ``+" indicates the MAP values. }
\label{fig:pcabayes_2}
\end{figure}

Figure \ref{fig:pcabayes_2} offers a similar comparison in 2d when contrasted
with Figure \ref{fig:unibayes_2} in Section \ref{sec:multioutputs}.  As in
that view, only a subset of pairs of outputs are shown.  The rest are in
Supplement \ref{sec:pcap}.  Notice the marked improvement in posterior
concentration in the rightmost four panels.  Two of those panels, $(u_3 \times
u_4)$ for $C_d$ and $(u_1, u_2)$ for $c_c$, also show substantial re-location
of density compared to their 28Hz analog.  The leftmost four panels offer less
stark contrast, a view shared by the 1d analysis in Figure \ref{fig:unibayes}.

These results indicate that we are headed in the right direction.  We embarked
on this analysis knowing well that the diversity of information across
output frequencies would not be substantial.  Nevertheless, representing it in a
parsimonious way, through first-PCs, seems to enhance posterior concentration
even though data is being discarded (i.e., the other three PCs). This is a
hallmark of enhanced learning through dimension reduction.  The last step is
to combine these separate analyses into one, forming a posterior for a single
set of unknowns given the entirety of data on all simulation and field
outputs.

\section{Full integration of outputs}
\label{sec:allcali}

The final step in our analysis is, in some sense, the easiest -- that is, once
all the hard work of building OSSs, forming first-PC posteriors, etc., is
done.  After detailing how outputs $j=1,\dots,J$ may be combined, we provide a
final suite of views into a unified posterior for $\mathbf{u}^1$.  Although
the setup is rather generic in $j$, in what follows we continue to focus the
discussion on the honeycomb application in anticipation of those results.

\subsection{Inferential apparatus}

Section \ref{sec:osspca} demonstrated that a PC basis is effective as a
dimension reduction tool for four pairs of four honeycomb output frequencies.
\blunew{Those four separately calculated bases capture}
around a $4 \times 90\% \approx 360\%$-fold larger amount of multivariate
variability as compared to each singe univariate output.  On the other hand,
each output property $j$ (direct/cross stiffness or damping) has a
distinct physical meaning.  Dependence between them is highly
non-linear, \blunew{despite the linear form of the
differential equation linking them (\ref{eq:trans})}. In Figure
\ref{fig:cor_ym}  we saw no evidence of strong linear dependence across the
$J=4$ output classes, which nudged us toward an independence
assumption into this joint analysis.  We shall return to that in our
discussion in Section \ref{sec:discuss}. 

As a related but practical matter, each output is measured on a
different scale.  PCA helps here. Standardization of the 16d
raw outputs, as a pre-processing step, followed by orthogonal projection onto
the 4-column subspaces spanned by their first eigenvectors $\mathbf{w}^1_j$,
naturally place those quantities on an equal footing. Besides being
represented in the direction of highest variability, they are scale-free
which simplifies joint modeling downstream.

\paragraph{Modular calibration via optimization.} 
Recall that each OSS, i.e., $\blunew{F} \times J \times K = 4,672$ univariate
fits [Section \ref{sec:unioss}] involves a manageable $\mathcal{O}(n^3_i)$
calculation. Then each of $K=4$ output frequencies are combined into their
first-PC (\ref{eq:pca}). Finally,  combine these representations together into
a unified objective to obtain a single $\mathbf{u}^1$ under
Eq.~(\ref{eq:opt1}) for all ($J=4$)  observed discrepancies, with modular
inference for GP hyperparameter $\bm{\phi}^1_j$ in each subspace.  So
basically we wish to \blu{jointly optimize} multiple Eq.~(\ref{eq:opt1})'s for
a single $\mathbf{u}^1$ via \blu{tuning $\mathbf{u}^1_j$ on  observed
discrepancies} $\mathbf{D}^{j1}_{\blunew{F}}(\mathbf{u}^1_j) =(
\mathbf{X}_{\blunew{F}}, (\mathbf{Y}_j^F -
\hat{\mathbf{Y}}_j^M(\mathbf{u}^1_j) ) \mathbf{w}^1_j )$, for $j=1, \dots, J$. 
After imposing conditional independence given common
$\mathbf{u}^1$, the following \blu{joint} objective \blu{across all $J$ outputs} is immediate:
\begin{equation}
\hat{\mathbf{u}}^1 = \mathrm{arg}\max_{\mathbf{u}^1} \left\{ p(\mathbf{u}^1) 
\prod^J_{i = 1} \left[ \max_{\bm{\phi}^1_j} 
p_b(\bm{\phi}^1_j \mid \mathbf{D}^{j1}_{\blunew{F}}(\mathbf{u}^1))\right] \right\}. 
\label{eq:opt2}
\end{equation}
In practice it easiest to solve this in log space. For each value
of $\mathbf{u}^1$ entertained by a numerical optimizer, evaluation requires GP
fitting for $J=4$ observed discrepancies
$\mathbf{D}^{j1}_{\blunew{F}}(\mathbf{u}^1_j)$, each with \blu{optimized GP 
hyperparameter} $\hat{\bm{\phi}}^1_j$
offloaded to a separate, library-facilitated, numerical optimizer.  
\blu{Each of $J$ inner optimizations over hyperparameter $\bm{\phi}_j$ in
$p_b(\bm{\phi}^1_j \mid \mathbf{D}^{j1}_{\blunew{F}}(\mathbf{u}^1))$ is wrapped in an outer optimization over the product-form joint marginal likelihood 
with prior $p(\mathbf{u}^1)$. }
To leverage ubiquitous modern multi-core workstation resources, we fit these in
parallel.  Following \citet{Huang:2018}, we use  \blu{{\tt optim} with BFGS in {\sf R} for the inner GP hyperparameter
optimization(s) and}  \blu{{\tt nloptr}  wrapped in a multi-start scheme} for the  \blu{outer optimization} to find $\hat{\mathbf{u}}^1$.

\paragraph{Bayesian joint inference.}
Posterior sampling $\mathbf{u}^1$ may follow similar \blu{principles.} 
\blunew{Conditioning on the optimized hyperparameters $\hat{\bm{\phi}}^1_j$, }
with the $j^\mathrm{th}$ output in first-PC representation, i.e., $\mathbf{y}^{F1}_j$
and $\mathbf{y}^{M1}_j$ via Eq.~(\ref{eq:pca}), a joint posterior via
conditional independence follows from a likelihood in product form:
\begin{equation}
p(\mathbf{u}^1 \mid \mathbf{y}^{F1}_1, \mathbf{y}^{M1}_1, \dots, 
\mathbf{y}^{F1}_J, \mathbf{y}^{M1}_J )
\propto p(\mathbf{u}^1) \cdot \prod^J_{j = 1} 
p( \mathbf{y}^{F1}_j, \mathbf{y}^{M1}_j \mid \mathbf{u}^1).
\label{eq:lik}
\end{equation}
\blu{To sample from this posterior, a Metropolis-within-Gibbs scheme
can be easily coded up, with each Gibbs step taking a marginal Gaussian 
random walk on $\mathbf{u}^1$.}
Under the multivariate OSSs emulation structure,  Metropolis rejection for
$\mathbf{u}^1$ can be broken down to evaluation of each (log) MVN-likelihood
, $p( \mathbf{y}^{F1}_j, \mathbf{y}^{M1}_j \mid \mathbf{u}^1)$, for $j= 1,
\dots J$. \blu{We see potential for each to be evaluated} in parallel and then put together into the 
joint (log) likelihood, \blu{Still, our serial implementation (with vectored linear algebra subroutines)
was fast enough to furnish thousands of samples/hour.}
A modular optimization solution (\ref{eq:opt2}) was helpful in providing a warm-start to
minimize burn-in efforts.

\subsection{Joint results}
\label{sec:baysresults}

Here we present views into our fully integrated calibration results. Our
modular optimization(s) utilized a 500-random-multi-start BFGS implemented in
parallel. Although more time is required for jointly optimizing four separate
sub-objectives at each iteration, the overall waiting time for joint
optimization turns out to be comparable to univariate and PC-based variations.
A post-optimization analysis summarizes the median number  of {\tt nloptr}
iterations until convergence for each methods  from 500 random initialization:
335 for PC-$K_d$, 114 for PC-$k_c$, 552 for PC-$C_d$, 430 for PC-$c_c$, and
117 for the unified approach.   Speed of the unified approach may owe to the
flatter/smoother surface, shown momentarily.  For full Bayes via MCMC, the
running time is also comparable to its univariate and PC counterparts, thanks
to the multiple, and highly parallelizable, and sparse OSS covariance.

\begin{figure}[ht!]
\centering
\centering
\includegraphics[width=1\linewidth, trim= 0 0 20 80,clip=TRUE]{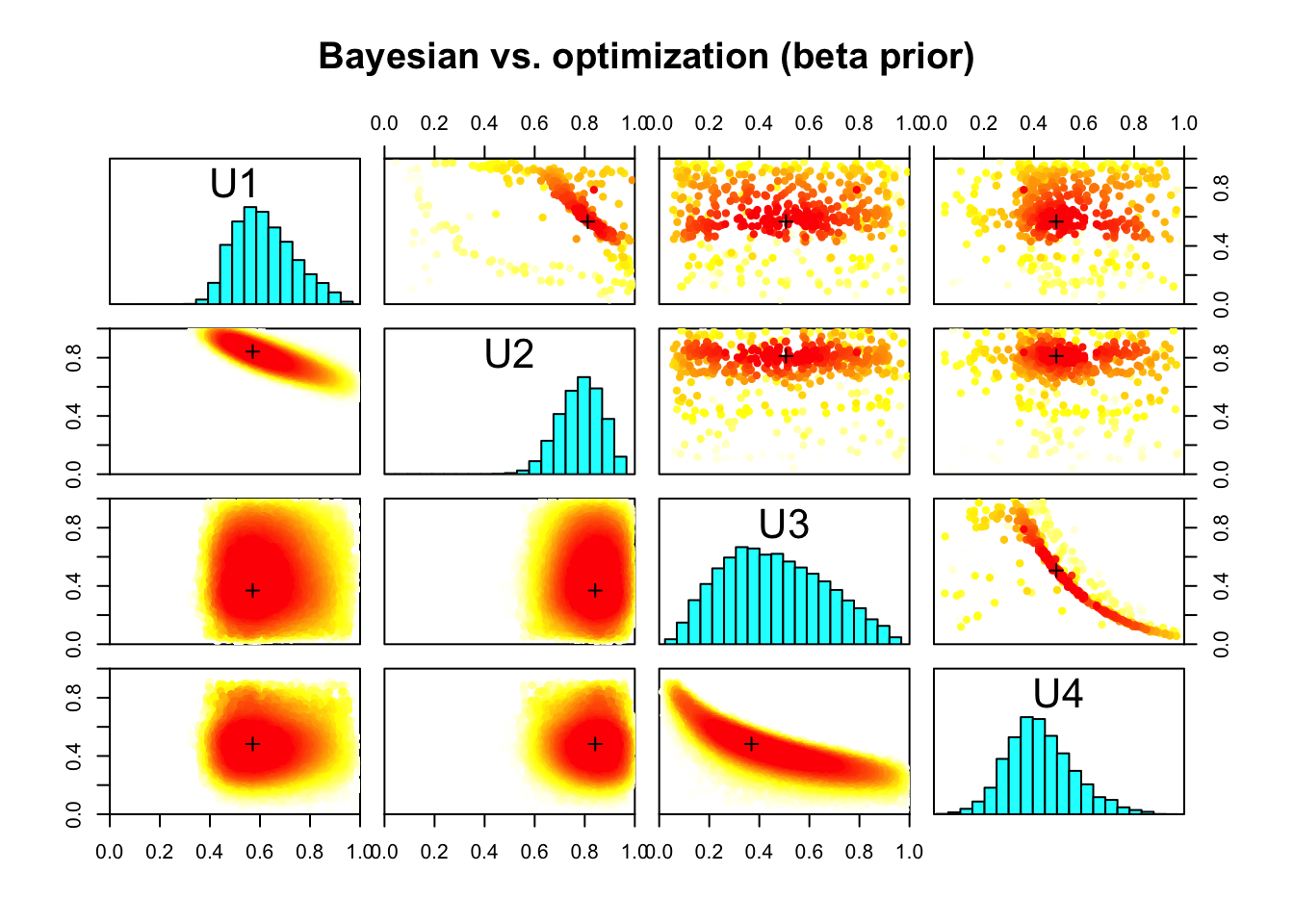}
\includegraphics[width=1\linewidth, trim=50 40 16 55,clip=TRUE]{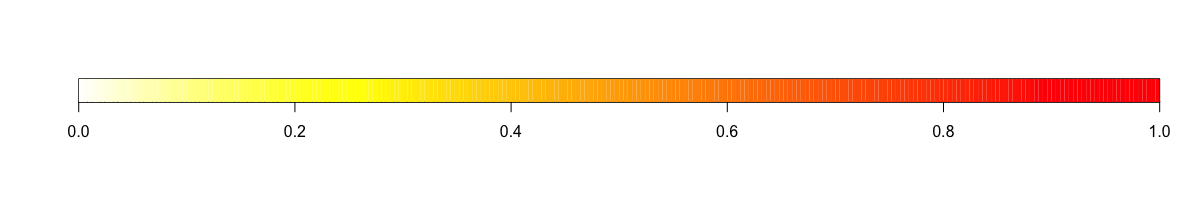}
\caption{Fully Bayesian (lower and diagonal) and modular optimization (upper)
calibration results for  $\mathbf{u}^1$ from the unified approach combining
all 16 outputs. Heat color are derived from rank of (joint) log posterior
probability of these parameter values. Bayesian results are from 100,000 MCMC
samples after burn-in. Optimization results are from  500 converged
optimization under random initialization. ``+" signs indicate the MAP values.}
\label{fig:comb}%
\end{figure}

Figure \ref{fig:comb} provides a 2d look via full posterior (bottom-left
triangle) and multi-start optimization (top-right) and univariate marginals
(diagonal).  This may be contrasted with any of the PC-level analogues in
Supplement \ref{sec:pcap} to reveal how information is synthesized across
outputs in this joint analysis.
While the optimized solutions demonstrate how local dynamics challenge
optimization, its MAP estimation and high density regions are comparable to
the fully Bayesian ones. From both approaches, two dependent structures
between pairs of parameter $(u_1, u_2)$ and $(u_3, u_4)$ can be observed,
displaying an interesting relationship between the friction coefficients
$(u_1, u_2)$ and friction exponents $(u_3, u_4)$. Our BHGE colleagues
concluded that this pattern may be explained by an underlying
turbulent-lubrication friction factor model from bulk-flow theory
\citep{Hirs:1973}.

\begin{figure}[ht!]
\centering
\includegraphics[width=1\linewidth, trim=0 20 0 0,clip=TRUE]{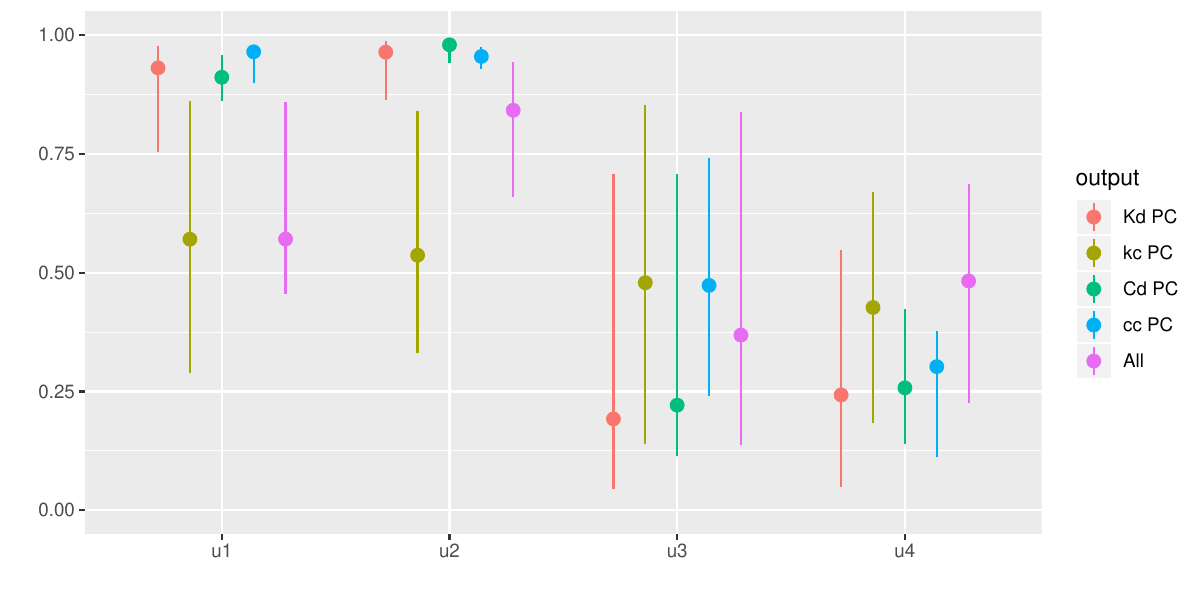}
\caption{Side-by-side comparison between marginal posterior distributions of 
$\mathbf{u}^1_j, j = 1, \dots, 4$  from outputs $K_d, k_c, C_d$,
and $c_c$ (labeled ``PC") and $\mathbf{u}^1$ via unified output (labeled ``All"). Dots
indicate MAP values and error bars form 90\% credible intervals. }
\label{fig:pcandall}
\end{figure}

Figure \ref{fig:pcandall} is lower resolution, providing a 1d look, but allows
a more visually immediate comparison.
Here we see how each output property contributes to the final global solution
$\mathbf{u}^1$, with the output property.  Apparently, cross stiffness ($k_c$) is
the most influential. Notice  the similarity between the $\mathbf{u}^1$
posterior distribution with $k_c$ PC posterior  $\mathbf{u}^1_2$ in each of
the $u$ dimensions, especially in $u_1$. Further comparisons in 2d, via both
the $\mathbf{u}^1$ marginals in Figure \ref{fig:comb}  and PC-calibrated
$\mathbf{u}^1_2$ marginals shown in Figure  \ref{fig:pca_1} from  Supplement
\ref{sec:pcap}, demonstrate dependence in parameters $(u_1, u_2)$.

When comparing the landscape of solutions in upper triangle of Figure
\ref{fig:comb} with those from Figures \ref{fig:pca_1}--\ref{fig:pca_2} one
immediately appreciates the benefit of a smoother posterior surface via full
integration. Notice in those figures that the
optimized solutions from outputs $K_d$, $C_d$ and $c_c$ are quite concentrated
in very small (dense red)  peaks, in contrast to the much flatter surfaces
from $k_c$ [Figure \ref{fig:pca_1}] and the fully integrated [Figure
\ref{fig:comb}]. The full Bayes (lower and diagonal) marginals further exhibit
this pattern: fully integrated and PC-$c_c$ exhibit flatter posteriors, with
two ridges in $(u_1, u_2)$ and $(u_3, u_4)$; the other three PC-based results
produce much narrower peaked posterior surfaces, specially in dimension of
$(u_1, u_2)$. Interestingly,  different outputs could contribute 
unequally to the fully integrated result  [Figure \ref{fig:comb}], 
possibly due to different amount of signal versus noise for 
the output relative to the rest. In the fully integrated honeycomb analysis,  
output $k_c$ seems to pull the parameter posterior  towards itself more than
 the other three outputs, especially in the $(u_1, u_2)$ subspace (See Figure 
 \ref{fig:pcandall}).  

 \blunew{To explore sensitivity to priors and forms of discrepancy, 
 Supplement \ref{sec:sens} demonstrates how these posteriors vary, but are
 ultimately quite robust to prior misspecification.  This extends
 \citeauthor{Huang:2018}'s limited, but ultimately similar analysis in the
 single-output setting.  For example, uniform rather than Beta$(2,2)$ priors
 on $\mathbf{u}^1$ results heavier concentration of posterior mass near the
 boundaries of the study region.  Weak ``prescient'' (greater prior
 concentration on the posterior MAP), and ``weak'' adversarial (prior
 concentration away from the MAP) priors have little effect.  Using L2
 discrepancy, i.e., no GP on the bias, has rather more, and deleterious effect
 on prediction as we show next.}

\section{Prediction}
\label{sec:mpred}

\blu{  
Inference for the calibration parameter is often  the primary interest in a
KOH-style calibration exercise.  
Integrating over the posterior predictive distribution, say at
$\mathbf{x}_{\mathrm{(new)}}$, alongside $\mathbf{u}$ involves many of the same
steps as above, except now with a MVN conditional distribution that has three
components: field observed, simulated, and field unobserved (at
$\mathbf{x}_{\mathrm{(new)}}$).  In the univariate case and with ordinary,
non-OSS surrogates, this is described in textbooks \citep[e.g.,][\blunew{Chapter
8.1.5}]{gramacy2020surrogates}.  With OSSs for
univariate output, \citet{Huang:2018} demonstrate how the same conditioning is
computationally tractable even for larger \blunew{$N$} by extending the block
diagonal structure to the three-component predictive setup.  An updated
variation on those equations are provided here as
Eq.~(\ref{eq:vnew}) momentarily, extending that setup to the
multi-output setting. This is accompanied by predictive results and
comparisons for the honeycomb, beginning in the basis space of the
first-PC (i.e., via $\mathbf{u}^1$), and then back in original output spaces.
}

\subsection{In-basis}
\label{sec:pcpred}

 \blu{
Consider first-PC field outputs
$\hat{\mathbf{y}}^{M1}_j( \mathbf{x}_{\mathrm{(new)}}, \cdot ) +
\hat{\mathbf{b}}^1_j(\mathbf{x}_{\mathrm{(new)}})$ for each of $j=1,\dots, J$, 
where $J=4$ for the honeycomb.
This involves a direct, four-fold independent application of
\citet{Huang:2018}.  No new methodology is being developed here, however we
find this a useful warm-up, thinking ahead to the building blocks required for
a full multi-output setting next in Section \ref{sec:orpred} via
Eq.~(\ref{eq:lik}).  It also enlightening to compare these predictions to the
simulation-only ones $\hat{\mathbf{y}}^{M1}_j( \mathbf{x}_{\mathrm{(new)}},
\cdot)$ as a lens into the nature of the bias
$\hat{\mathbf{b}}^1_j(\mathbf{x}_{\mathrm{(new)}})$. }

\begin{figure}[ht!]
\centering
\includegraphics[width=.49\linewidth, trim=0 0 0 0,clip=TRUE]{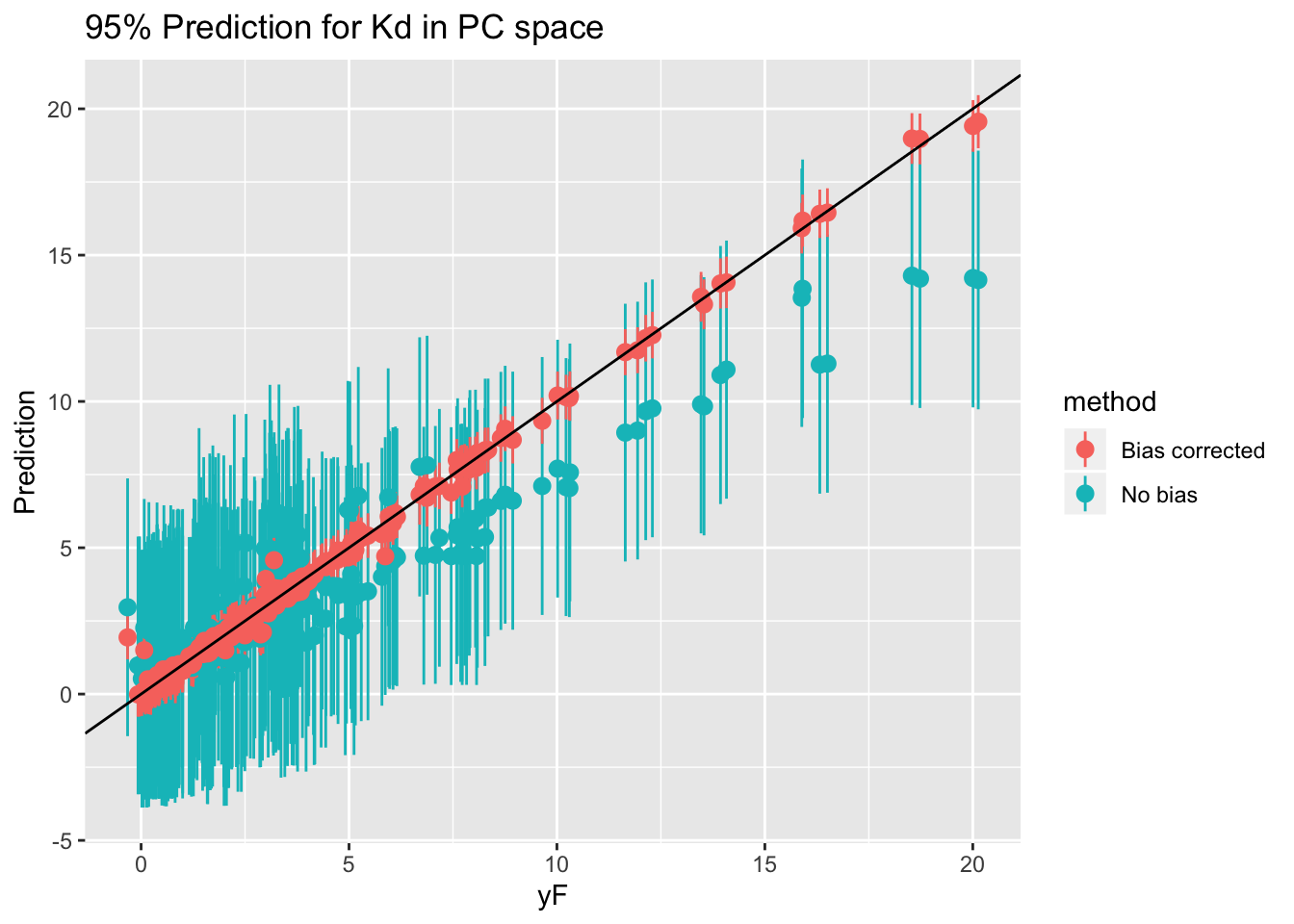}
\includegraphics[width=.49\linewidth, trim=0 0 0 0,clip=TRUE]{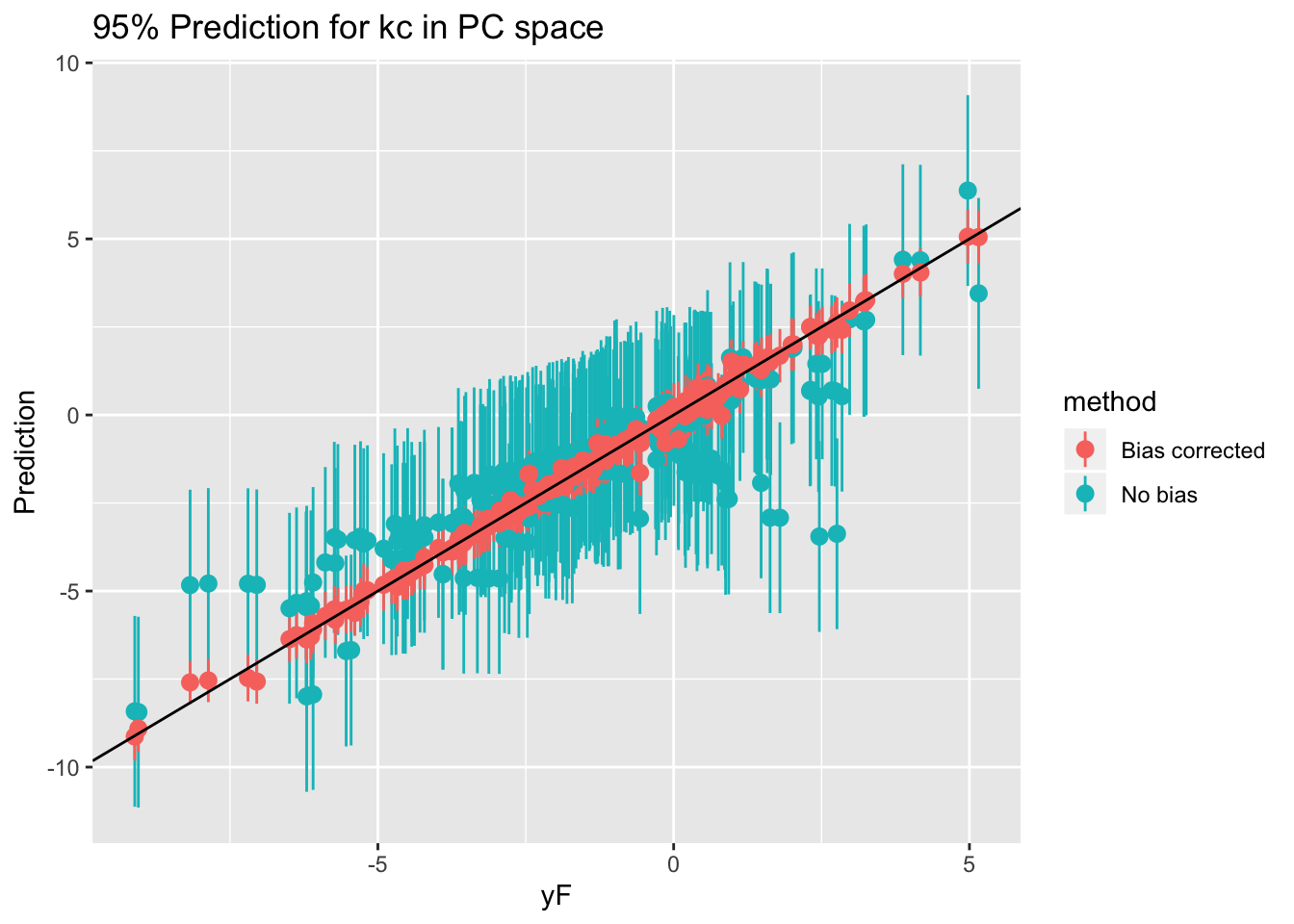}
\caption{LOO-CV first-PC posterior predictive summaries 
for direct stiffness $K_d$ and cross stiffness $k_c$.  
Intervals trace out 95\%; black line has intercept zero, slope one.}
\label{fig:pred_bias}
\end{figure}

 
 \blu{
For example, Figure \ref{fig:pred_bias} shows 
leave-one-out cross-validated (LOO-CV) predictions for field outputs,
  $\hat{\mathbf{y}}^{F1}_j( \mathbf{X}_{\blunew{F}})$, for $K_d$ and $k_c$ ($j=\{1,2\}$), 
 pitting  prediction without bias correction  
$\hat{\mathbf{y}}^{M1}_j( \mathbf{X}_{\blunew{F}}, \hat{\mathbf{u}}^1 )$ against
bias corrected
$\hat{\mathbf{y}}^{M1}_j( \mathbf{X}_{\blunew{F}}, \hat{\mathbf{u}}^1 ) 
+ \hat{\mathbf{b}}^1_j(\mathbf{X}^F_{\blunew{F}})$. 
Here, $\hat{\mathbf{u}}^1$ comes via samples from fully integrated posterior (\ref{eq:lik}). 
Similar views for the other two outputs may be found in Figure \ref{fig:pred_bias2} of Supplement \ref{sec:pred}. 
}
\blu{
Observe in both cases that predictions without bias correction (blue) 
exhibit noticeable deviation from the calibration (diagonal) line
with relatively large credible intervals, suggesting that a considerable portion of dynamics
in the field are not fully captured by simulation.
Bias correction (red) is essential, not only for accuracy but also for confidence:
both have appropriate coverage, but one has much smaller intervals.
Yet these two outputs also exemplify a dual role played by the 
bias correction term $\hat{\mathbf{b}}^1_j(\mathbf{X}_{\blunew{F}})$.
For $K_d$ (left), discrepancies generally widen as the observed output $K_d$ increases.
 For $k_c$ (right), the observed model discrepancy
 demonstrates a more complicated pattern, with relatively larger
 discrepancies at the two extremes.  
 }

\blu{
\subsection{Original outputs}
\label{sec:orpred}
}
 \blu{
Poster predictive sampling of the original $J \times K$ outputs involves
re-centering, re-scaling, and back-rotation from all the PC bases. Once 
$\mathbf{u}^1$ values are sampled from Eq.~(\ref{eq:lik}), we can feed these posteriors into multiple $J \times K$
OSSs  $\hat{\mathbf{y}}_j^{Mk}
 (\mathbf{x}, \mathbf{u}^{1})$  with their own bias corrections:
\begin{equation}
\mathbf{y}_j^{Fk}(\mathbf{x}) = \mathbf{y}_j^{Mk} (\mathbf{x}, \mathbf{u}^{1})
 + \mathbf{b}^k_j(\mathbf{x})  + \mathbf{\epsilon}^k_j,
 \quad j = 1, \dots, J, \; k = 1, \dots K.  
  \label{eq:pckoh2}
\end{equation}
Fitted discrepancies
 $\hat{\mathbf{b}}^k_j(\mathbf{x}) $ are required to recover the complete
 multiple-output prediction on original outputs, while only the
 lower-dimensional first-PC representations are needed for parameter learning
 and sampling at the PC (\ref{eq:pckoh}) and fully
 integrated levels (\ref{eq:lik}). }

\blu{
For the $j^{\mathrm{th}}$ output property, raw predictions at new site location $\mathbf{x}_{\mathrm{(new)}}$
are made by separately applying the OSSs kriging equations from \cite{Huang:2018} $K=4$ times,
but now in their PC represented spaces, where their raw predictive mean and variances 
\blunew{$(\mathbf{\mu}^k_{j\mathrm{(new)}}, \mathbf{\Sigma}^k_{j\mathrm{(new)}} )$}
in the basis spaces are still analytically tractable, 
\blunew{
\begin{align}
\mathbf{\mu}^k_{j\mathrm{(new)}} &= \mathbf{\Sigma}^{jk}_{\blunew{F}', \blunew{F}} 
 \mathbf{C}^{-1}(\mathbf{u}^1)[ \mathbf{y}^{Fk}_j -
\mathbf{\Sigma}^{jk}_{\blunew{F}, N}(\mathbf{u}^1)
(\mathbf{\Sigma}^{jk}_{N}) ^{-1} 
  \mathbf{y}^{Mk}_j]  \nonumber \\
  & \quad\quad\quad
+ \mathbf{\Sigma}^{jk\mathrm{(new)}}_{\blunew{F}, N}(\mathbf{u}^1)
(\mathbf{\Sigma}^{jk\mathrm{(new)}}_{N}) ^{-1} 
 \mathbf{y}^{Mk}_{j\mathrm{(new)}}, \nonumber \\
\mathbf{\Sigma}^k_{j\mathrm{(new)}} &=  \label{eq:vnew}
\mathbf{\Sigma}^{jk\mathrm{(new)}}_{\blunew{F}'}(\mathbf{u}^1)  + \mathbf{\Sigma}^{jk\mathrm{(new)}}_{b}  -
\mathbf{\Sigma}^{jk}_{\blunew{F}', \blunew{F}} \mathbf{C}^{-1}(\mathbf{u}^1)
 (\mathbf{\Sigma}^{jk}_{\blunew{F}', \blunew{F}} )^\top \\
&\quad\quad\quad  -\mathbf{\Sigma}^{jk\mathrm{(new)}}_{\blunew{F}', N'}(\mathbf{u}^1) (\mathbf{\Sigma}^{jk\mathrm{(new)}}_{N'})^{-1}   
  [\mathbf{\Sigma}^{jk\mathrm{(new)}}_{\blunew{F}', N'}(\mathbf{u}^1) ]^\top. \nonumber 
\end{align}}
Details on each new notational element are provided in Supplement \ref{sec:pred}. 
}

\blu{
Seeking insights in an out-of-sample setting, we performed LOO-CV for the
honeycomb like in Section \ref{sec:pcpred}, but this time in the original output space. Using
Eq.~(\ref{eq:vnew}) we first obtain the full-rank raw
predictions   in  the PC spaces, $$(\mathbf{y}^{Fk}_{j}  \mid
\mathbf{y}^{Mk}_{-i},  \mathbf{y}^{Fk}_{-i},
 \mathbf{y}^{Mk}_{i},  \bm{\Phi}, \mathbf{u}^1) 
 \sim \mathcal{N}_{i}(\blunew{\mathbf{\mu}}^k_{ji}, \blunew{\mathbf{\Sigma}}^k_{ji} ), 
 \quad i =1, \dots, \blunew{F},  \; j = 1, \dots, J, \; k =1, \dots, K.$$ 
Once $K$ raw predictions for the $j^{\mathrm{th}}$ output property
$\mathbf{Y}_j^{FK} = (\mathbf{y}_j^{F1}, \dots, \mathbf{y}_j^{Fk})$ in basis
spaces are sampled, they may be rotated back at once using the full matrix
$\mathbf{W}_j$ of the centered and scaled eigenvectors through
$\mathbf{W}^{-1}_j\mathbf{Y}_j^{FK} = \mathbf{Y}_j^F, j=1, \dots, J$, all at once. 
After this fashion, predictions thus obtained synthesize more information than a univariate analysis
could. Since a common  $\mathbf{u}^1$ sample is used for all $k$, predictive
uncertainty may be dramatically reduced.
We take ten thousand MCMC posterior samples $\mathbf{u}^1$ and calculate predictive moments 
using the laws  
 of total expectation and variance:
\begin{align*}
\mathbb{E}(\mathbf{y}^{Fk}_j  \mid \cdot ) = 
\mathbb{E}[\mathbb{E}(\mathbf{y}^{Fk}_j  \mid \cdot, \mathbf{u}^1 )], \quad \mbox{and} \quad
\mathbb{V}(\mathbf{y}^F_{\text{(new)}}  \mid \cdot ) = 
 \mathbb{V}[\mathbb{E}(\mathbf{y}^{Fk}_j   \mid \cdot, \mathbf{u}^1)].
+ \mathbb{E}[\mathbb{V}(\mathbf{y}^{Fk}_j   \mid \cdot, \mathbf{u}^1)].
\end{align*}
These quantities then approximate the full posterior 
(out-of-sample) predictive distribution 
at any $\mathbf{x}_{\mathrm{(new)}}$:  
$(\mathbf{y}^{Fk}_{j}  \mid \mathbf{y}^{Mk}_{-i},  \mathbf{y}^{Fk}_{-i}, 
 \mathbf{y}^{Mk}_{i},  \bm{\Phi})$,  $\forall i,j,k$. }

\begin{figure}[ht!]
\centering
\includegraphics[width=.49\linewidth, trim=0 0 0 0,clip=TRUE]{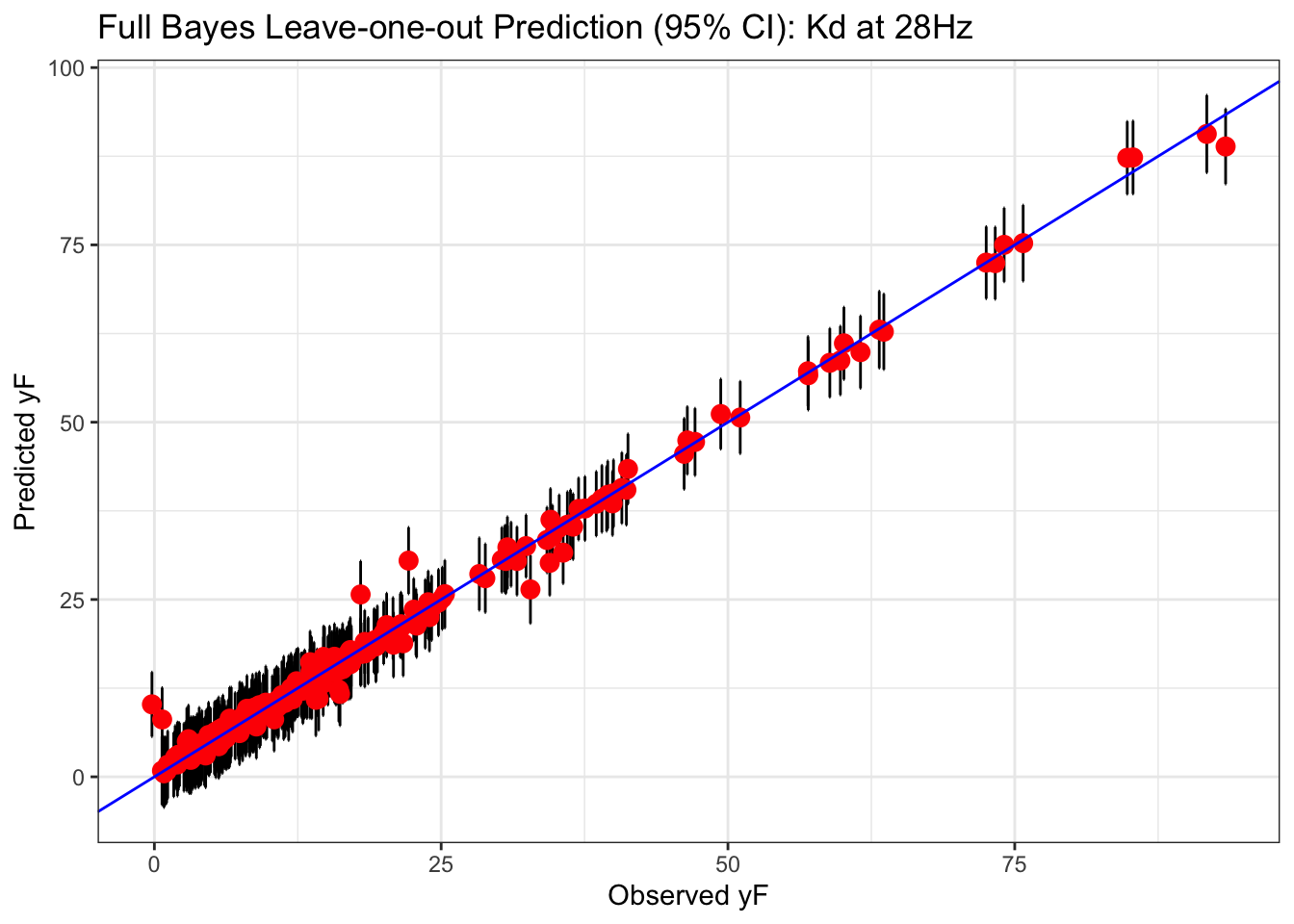}
\includegraphics[width=.49\linewidth, trim=0 0 0 0,clip=TRUE]{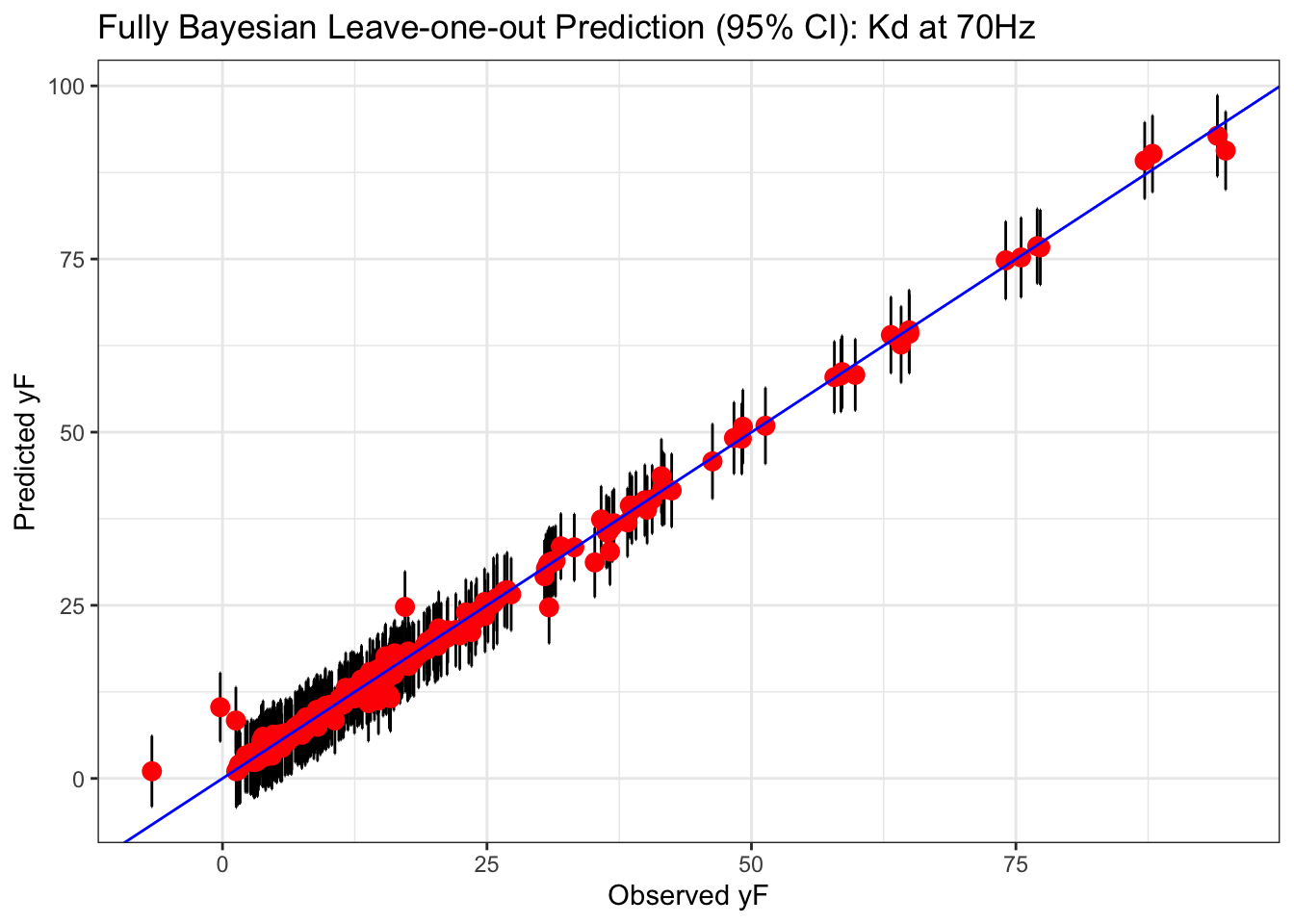}
\includegraphics[width=.49\linewidth, trim=0 0 0 0,clip=TRUE]{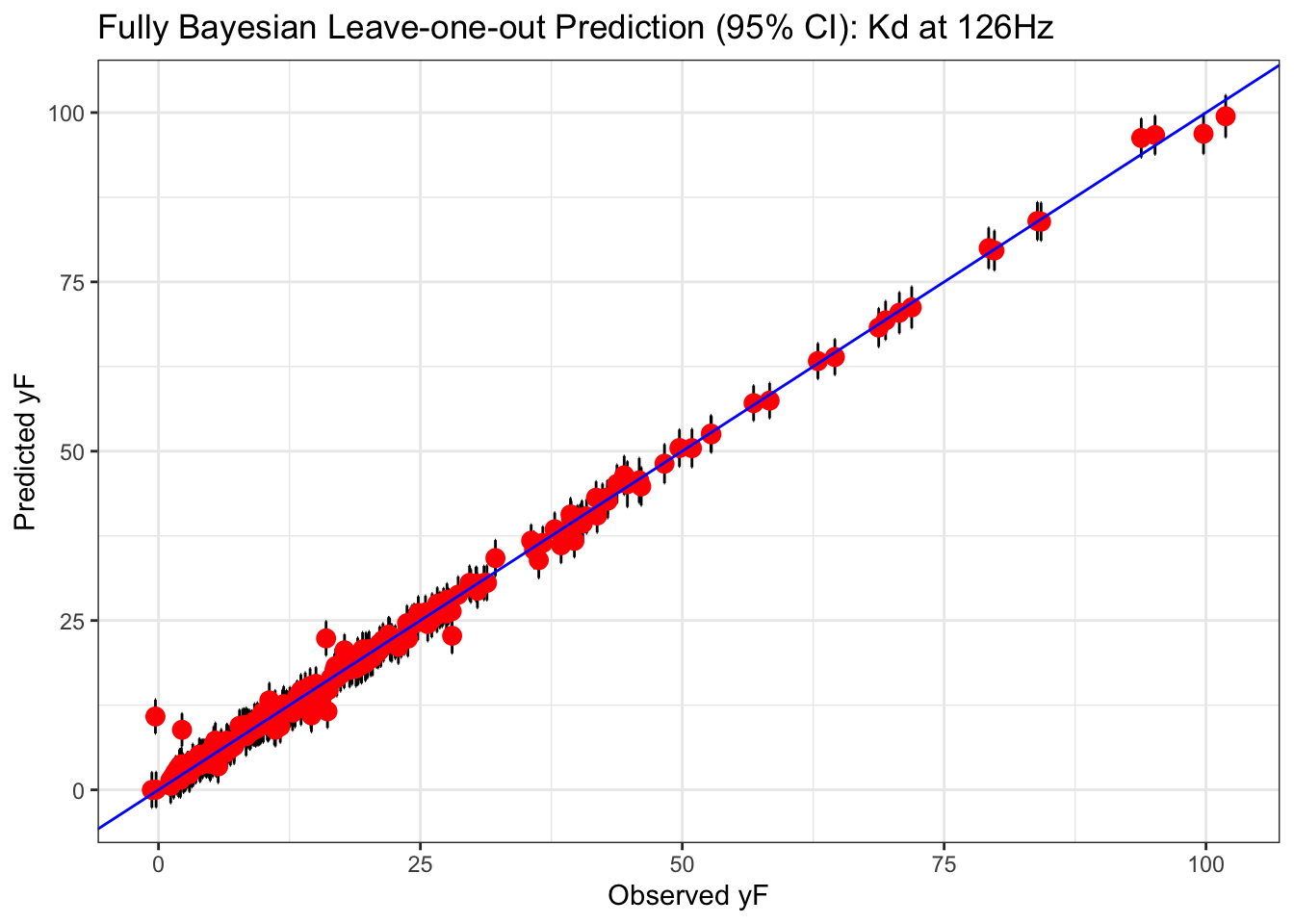}
\includegraphics[width=.49\linewidth, trim=0 0 0 0,clip=TRUE]{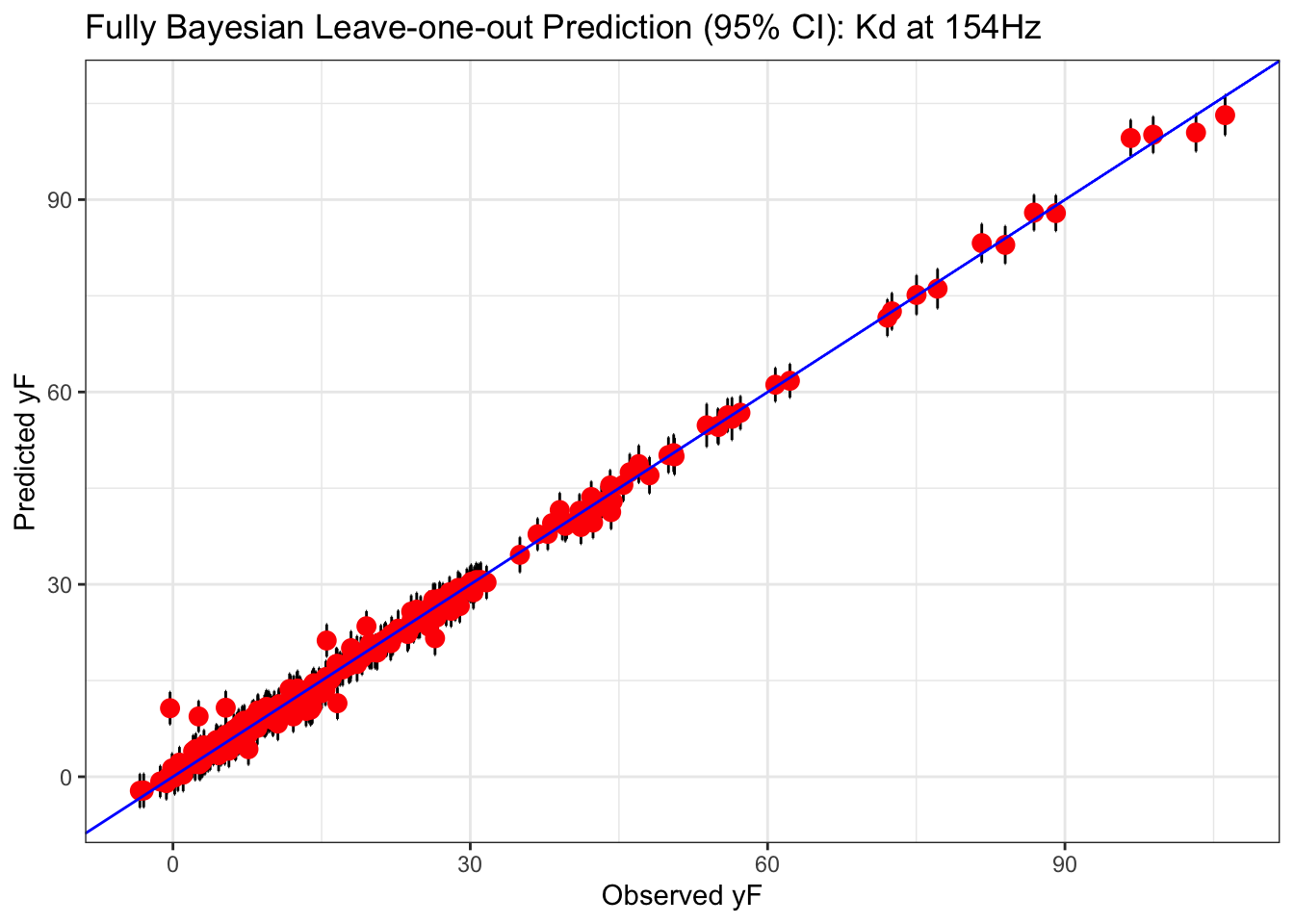}
\caption{Posterior predictive comparison for direct stiffness $K_d$ at 28, 70, 
126, and 154 Hz.  
Red dots are the predicted mean and black bars are 95 \% 
 credible intervals.}
\label{fig:pred_ori}
\end{figure}

\blu{ Table \ref{tab:pred} summarizes the 16 LOO-CV 
RMSEs recovered back in the original $J=4$ output properties at $K=4$ frequencies. 
For example, these fully Bayesian LOO-CV
predictions for direct stiffness $K_d$ at 28 Hz 
across $\blunew{F}=292$ sites have RMSE of 1.460.  This bests its univariate,
separately calibrated counterpart in \citeauthor{Huang:2018} (Table 1) with
RMSE of 1.957.  Although not surprising, 
we speculate this is likely due to better parameter
identification and discrepancy learning through synthesis of a larger corpus
of training data. 
}
\begin{table}[ht!]
\centering
\begin{tabular}{r | r | r | r | r }
Output  & 28 Hz &  70 Hz &  126 Hz & 154 Hz \\
\hline
$K_d$ & 1.460  & 1.407 & 1.313 & 1.442 \\
$k_c$ & 0.869 & 0.901 & 0.880 & 1.120\\
$C_d$ & 2.994 & 3.049 & 2.742 & 2.762 \\
$c_c$ & 3.867 & 2.492 & 2.028 & 1.535 \\
\end{tabular}
\caption{LOO-CV  RMSEs using multiple-output approach. Rounded at 3 digits. }
\label{tab:pred}
\end{table} 
\blu{ For another view of LOO-CV performance in original 
outputs, we plotted posterior predictive mean and 95\% intervals over each
observed field output. Figure \ref{fig:pred_ori} provides these for direct
 stiffness $K_d$ with $K=4$ frequencies; results for the other three outputs are provided
 by Figures \ref{fig:pred_ori2}--\ref{fig:pred_ori4} in Supplement
 \ref{sec:pred}. Comparing $K_d$ at 28 Hz to its univariate
 counterpart in Figure 11 of \cite{Huang:2018}, this multiple-output
 approach yields  more closely aligned out-of-sample predictions, with fewer
 and less substantial deviations from the diagonal (black) line.  Moreover,
 there are no obvious sites with larger level of predictive uncertainty (wider
 error-bars). Notice in Figure \ref{fig:pred_ori} that there are only a few
 sites which are mis-predicted and that the error bars are more uniform than
 those in \citeauthor{Huang:2018} version. Across frequency, predictive
 uncertainty (error-bar width) decreases as the frequency level increases.
 The first three outputs $K_d$, $k_c$, and $C_d$ exhibit stable
 predictive performance.  The last output, $c_c$, reflects high measurement error.
Refer Figure \ref{fig:pred_ori4} to Figure \ref{fig:pred_bias2}.}

\section{Discussion}
\label{sec:discuss}

Motivated by a large-scale industrial multivariate calibration problem
studding (honeycomb) seal flow dynamics from oil and gas research, we
developed a new multivariate calibration method by extending a few features of
a parsimonious univariate strategy based on on-site surrogates
\citep[OSSs;][]{Huang:2018}. Tailored to effectively capture unique simulation
features including high-dimensional input space, local nonstationary, missing
simulations, and model fidelity at scale, this method is practical, but it is
not without drawbacks. Univariate output applications, when replicated across
diverse outputs are not immune to a data-poor Bayesian learning pitfalls for
both high-dimensional calibration parameter and ill-posed model discrepancy.
\blu{Therefore, in lieu of imposing elaborate cross-site correlation structures, 
we opted for a simpler approach.}

Our solution, toward a multivariate OSS-based calibration, gathered together
responses to several features of the honeycomb. 
\blu{Although this multivariate  setup might not be ideal for all applications, especially
when the ground truth posterior distribution of parameter $\mathbf{u}$ differ
strikingly across different outputs, see e.g.,\cite{box},  the focus here is
to leverage all available information through one coherent, and tractable
modeling framework.} A careful exploratory data analysis (EDA) suggested that
a principal component-style basis representation could be effective for
handing honeycombs quadruplet of highly linearly dependent output frequencies.
We then update the OSS framework for both modular/optimized and fully Bayesian
implementations PC-based calibration within the Kennedy \& O'Hagan framework.
Our empirical results indicated improved parameter identification and
posterior concentration for calibration parameters, compared to the (separate)
univariate analog. Then, we designed an independent multi-output PC approach
to gather these models across output property.  This was motivated by a lack
of linear correlation observed in our EDA.  The result is a unified analysis
for the honeycomb, synthesizing millions of simulation runs in a matter of
hours.

It might be possible to entertain nonlinear dependency among output
properties, as suggested by the differential equation (\ref{eq:trans}) known
to govern relationships across outputs of the types studied in honeycomb.  On
the other hand though, our visual inspections (via EDA) did not reveal any
notable patterns.  So at this time, the merits of such an approach are
speculative at best, although extension is certainly possible.  
 \blu{Given posterior draws of the calibration parameter, analytically tractable
prediction is within reach. We evaluated empirical out-of-sample performance
in PC-basis spaces and back on original outputs. This analysis reinforces the
existence of nontrivial model discrepancies for all outputs as well as the
integral role of KOH bias correction. Compared to its univariate, separately
calibrated counterpart, this unified multiple-output solution enjoys
improved out-of-sample accuracy. }

\section*{Supplementary Material}

\blunew{
\textbf{Supplementary Material:} The PDF file contains (\ref{sec:uniap}) 
univariate calibration results: full Bayes for 16 outputs; (\ref{sec:pcap})
PC-level  calibration results: optimization and full Bayes; 
(\ref{sec:pred}) Multiple-output OSSs prediction results; and
(\ref{sec:sens}) Sensitivity analysis.} \\

\noindent \blunew{ \textbf{Computer Code:} {\sf R} code to reproduce 
all the results in this article are available on a private repository on
Bitbucket, available from the authors (many aspects are proprietary to BHGE
and require approval). 
Note that execution requires thousands of core hours even in
high performance computing environments.}

\if0\blind{
\subsubsection*{Acknowledgments}

Authors JH and RBG are grateful for support from National Science Foundation
grants DMS-1521702 and DMS-1821258.  JH and RBG also gratefully acknowledge
funding from a DOE LAB 17-1697 via subaward from Argonne National Laboratory
for SciDAC/DOE Office of Science ASCR and High Energy Physics. We thank Andrea
Panizza (BHGE) for early work on this project, and for initiating the line of
research, Mirko Libraschi (BHGE) for many interesting discussions, and
\blu{thoughtful comments from the Editor V. Roshan Joseph, an AE,  and two referees towards improving this work.}

\fi

\bibliography{multoss}
\bibliographystyle{jasa}

\pagebreak
\appendix

\section{Supplementary Material}
\label{ap:sep}

\noindent \dots for {\bf ``Multi-output calibration of a honeycomb seal
via on-site surrogates''} by Huang and Gramacy.

\subsection{Univariate calibration results: full Bayes for 16 outputs}
\label{sec:uniap}

We provide completed univariate $\mathbf{u}$ posterior
summaries for all 16 outputs: $K_d$, $k_c$, $C_d$,
and $c_c$, each at 28, 70, 126, and 154 Hz, augmenting Figure
\ref{fig:unibayes_5}'s marginal view. Observe greater
differences between output properties (block), than across frequencies (color).
\begin{figure}[ht!]
\centering
\includegraphics[width=1\linewidth, trim=0 20 0 0,clip=TRUE]{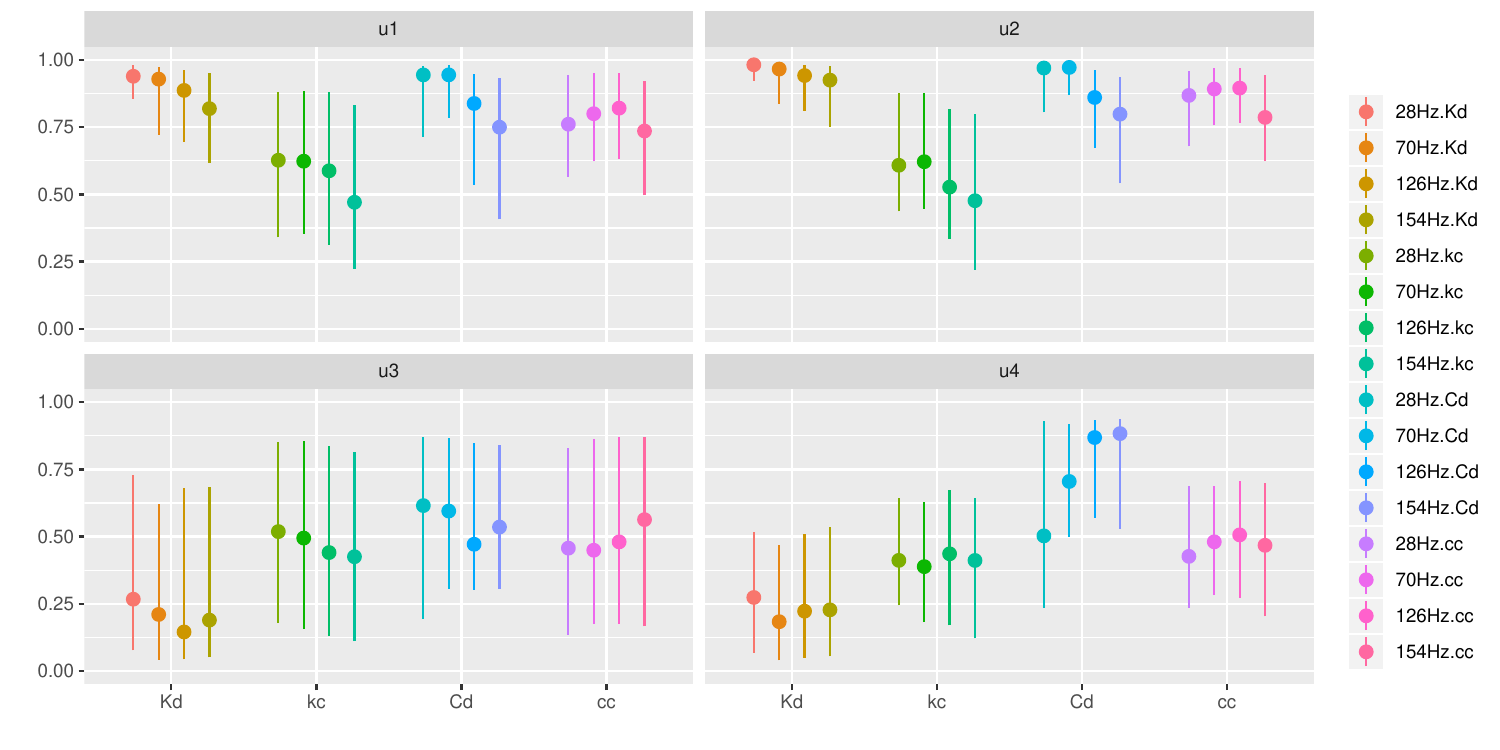}
\caption{16 sets of marginal  $\mathbf{u}$  posteriors 
from outputs $K_d, k_c, C_d$, and $c_c$  at frequencies 28, 70, 126, and 154 Hz.
Dots indicate MAP values and error bars form 90\% intervals.}
\label{fig:unibayes_5}
\end{figure}
Augmenting those 1d marginals, Figures \ref{fig:uni_1} and \ref{fig:uni_3}
project $\mathbf{u}$  onto its six 2d marginals, separately for each
frequency. Each row represents a complete set of 2d projections for one output
property--frequency pair. Cross frequency consistency in MAP estimation and
high density region can be observed from these figures, echoing Figure
\ref{fig:unibayes_5}.
\begin{figure}[ht!]
\centering
\includegraphics[width=0.95\linewidth, trim=0 50 0 20,clip=TRUE]{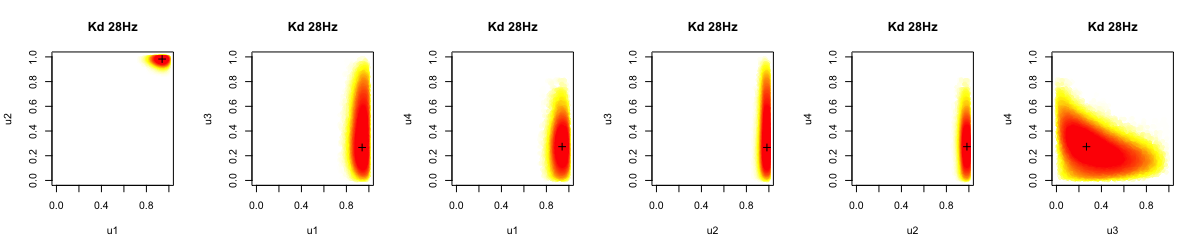}
\includegraphics[width=0.95\linewidth, trim=0 40 0 15,clip=TRUE]{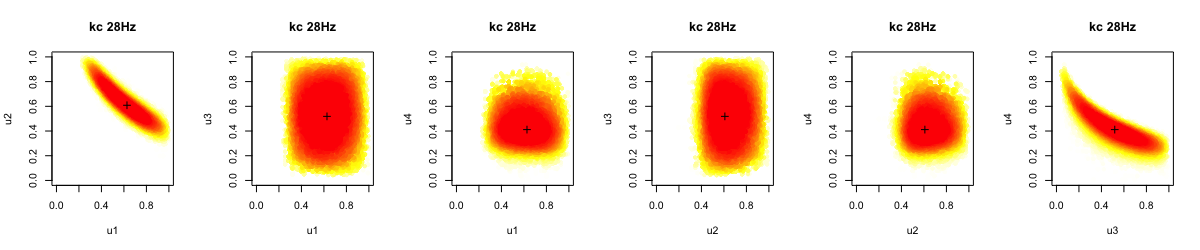}
\includegraphics[width=0.95\linewidth, trim=0 40 0 15,clip=TRUE]{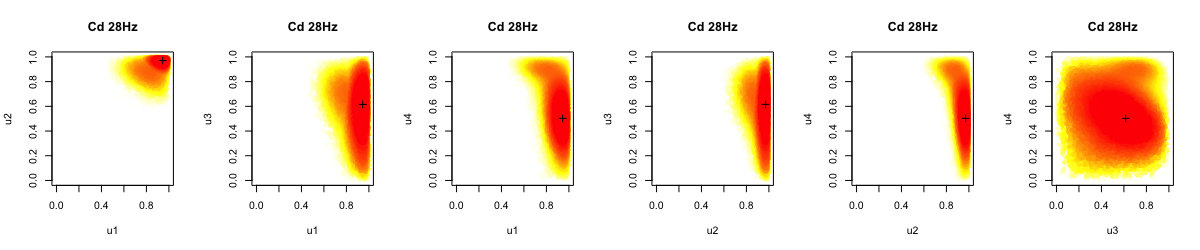}
\includegraphics[width=0.95\linewidth, trim=0 40 0 15,clip=TRUE]{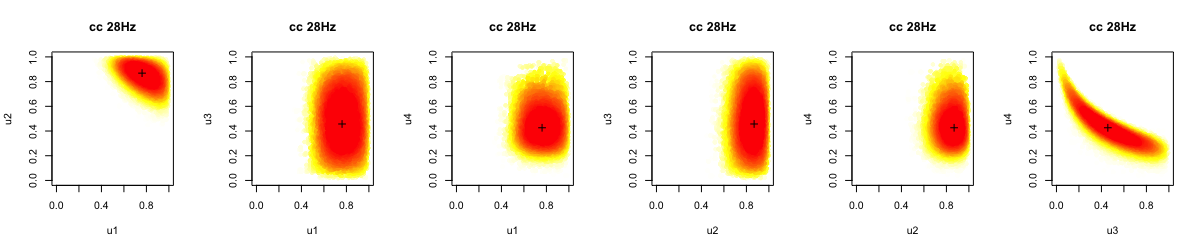}
\includegraphics[width=0.95\linewidth, trim=0 40 0 15,clip=TRUE]{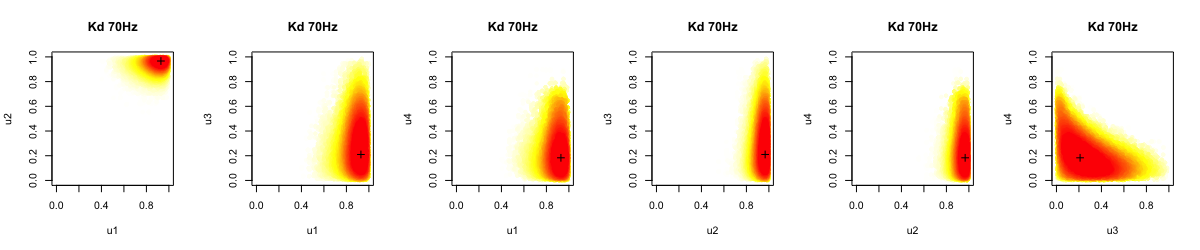}
\includegraphics[width=0.95\linewidth, trim=0 40 0 15,clip=TRUE]{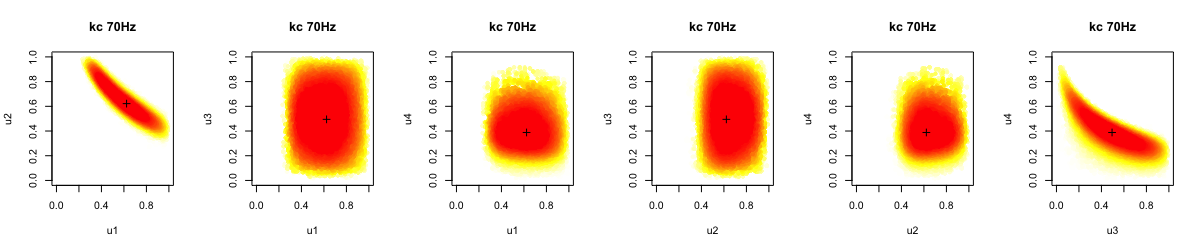}
\includegraphics[width=0.95\linewidth, trim=0 40 0 15,clip=TRUE]{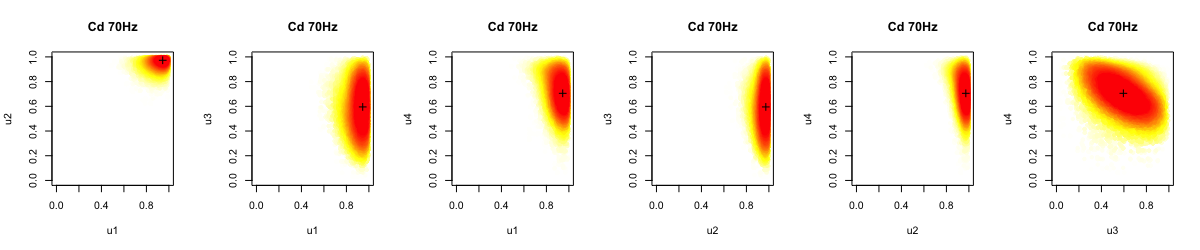}
\includegraphics[width=0.95\linewidth, trim=0 50 0 15,clip=TRUE]{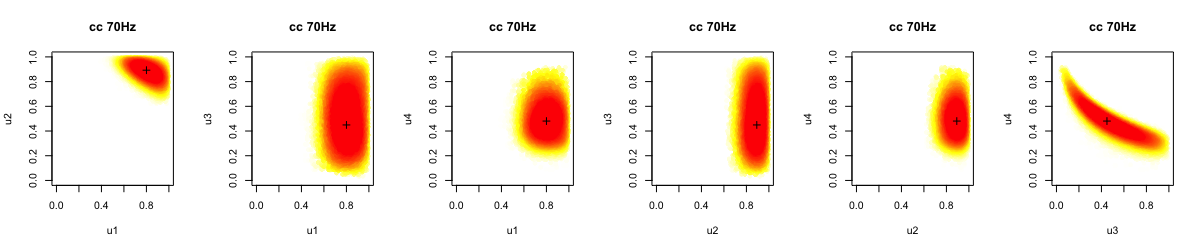}
\caption{Bivariate marginals of single-output posterior samples of $\mathbf{u}$ 
at 28 and 70 Hz.}
\label{fig:uni_1}
\end{figure}
\begin{figure}[ht!]
\centering
\includegraphics[width=0.95\linewidth, trim=0 50 0 15,clip=TRUE]{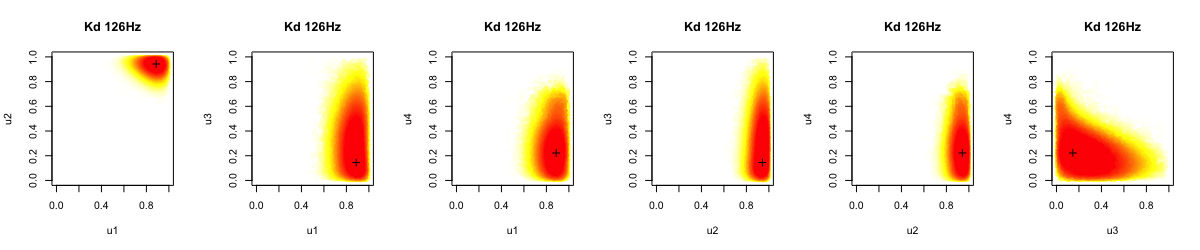}
\includegraphics[width=0.95\linewidth, trim=0 40 0 15,clip=TRUE]{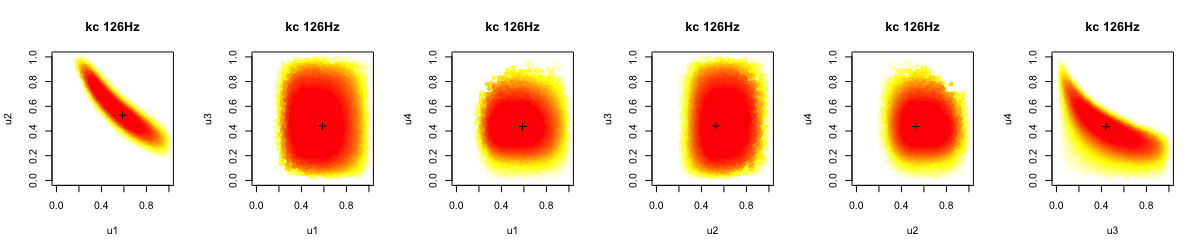}
\includegraphics[width=0.95\linewidth, trim=0 40 0 15,clip=TRUE]{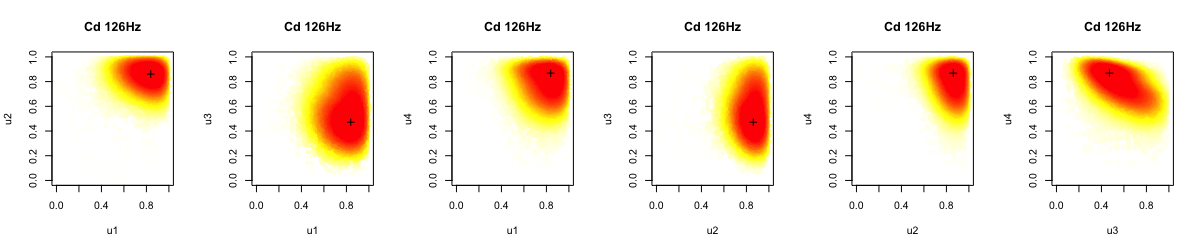}
\includegraphics[width=0.95\linewidth, trim=0 40 0 15,clip=TRUE]{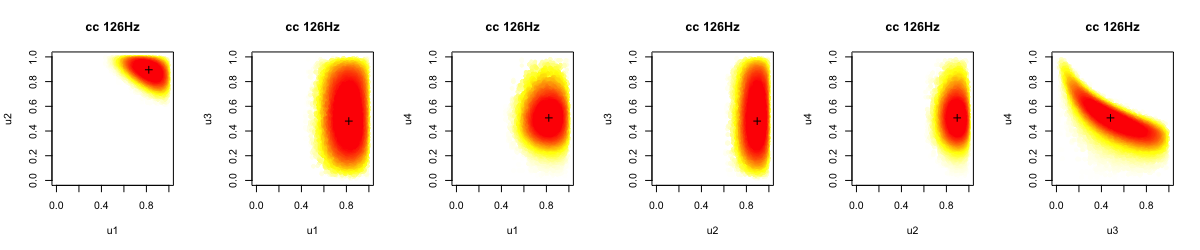}
\includegraphics[width=0.95\linewidth, trim=0 40 0 15,clip=TRUE]{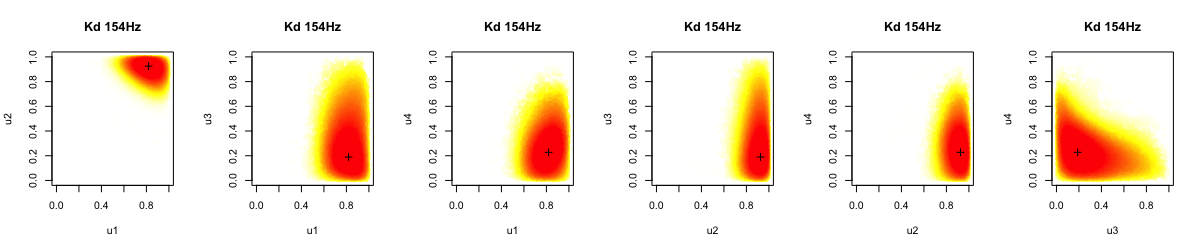}
\includegraphics[width=0.95\linewidth, trim=0 40 0 15,clip=TRUE]{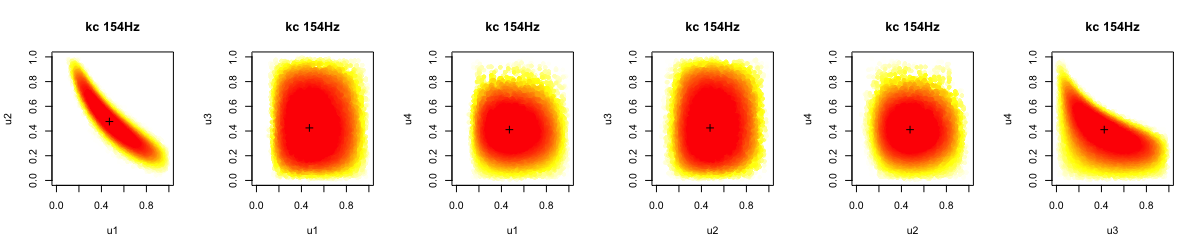}
\includegraphics[width=0.95\linewidth, trim=0 40 0 15,clip=TRUE]{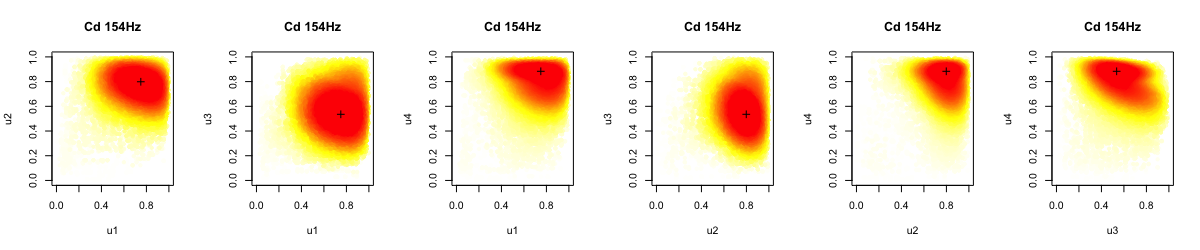}
\includegraphics[width=0.95\linewidth, trim=0 50 0 15,clip=TRUE]{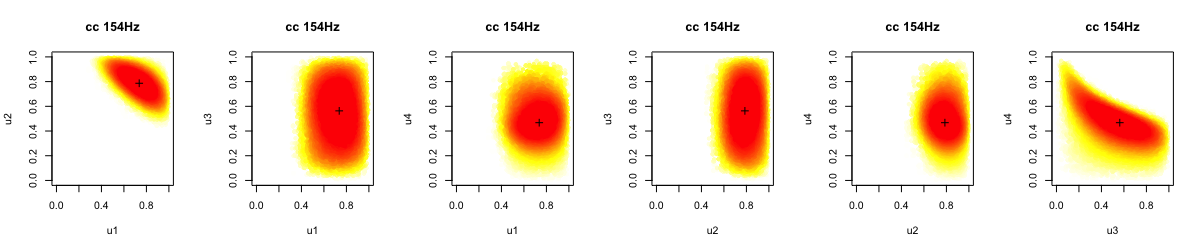}
\caption{Bivariate marginals of single-output posterior samples of $\mathbf{u}$ 
at 126 and 154 Hz.}
\label{fig:uni_3}
\end{figure}

\subsection{PC-level  calibration results: optimization and full Bayes}
\label{sec:pcap}

We  display all 4 sets of PC-calibrated parameters $\mathbf{u}^1_j$, for $j = 1, \dots, J$
using both modular optimization and fully Bayesian approaches developed from 
Section \ref{sec:pcacali} in Figures \ref{fig:pca_1} and \ref{fig:pca_2}. 
These are derived from 100,000 MCMC samples after burn-in. 
Optimization results are from  500 converged random multi-starts; 
``+" signs indicate the MAP values. 
\begin{figure}[ht!]
\centering
\includegraphics[width=.85\linewidth, trim= 0 65 20 80,clip=TRUE]{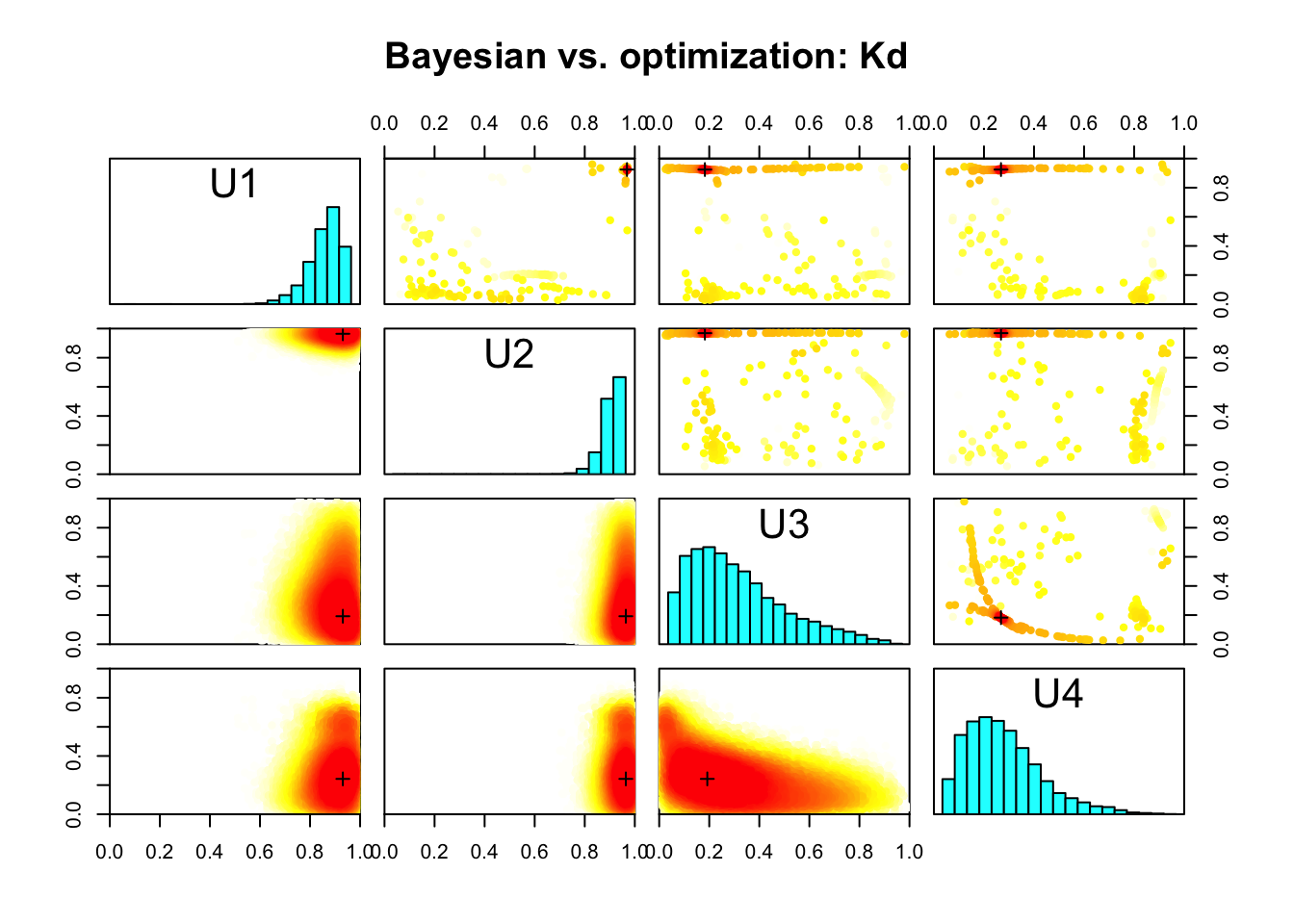}
\includegraphics[width=.85\linewidth, trim= 0 65 20 80,clip=TRUE]{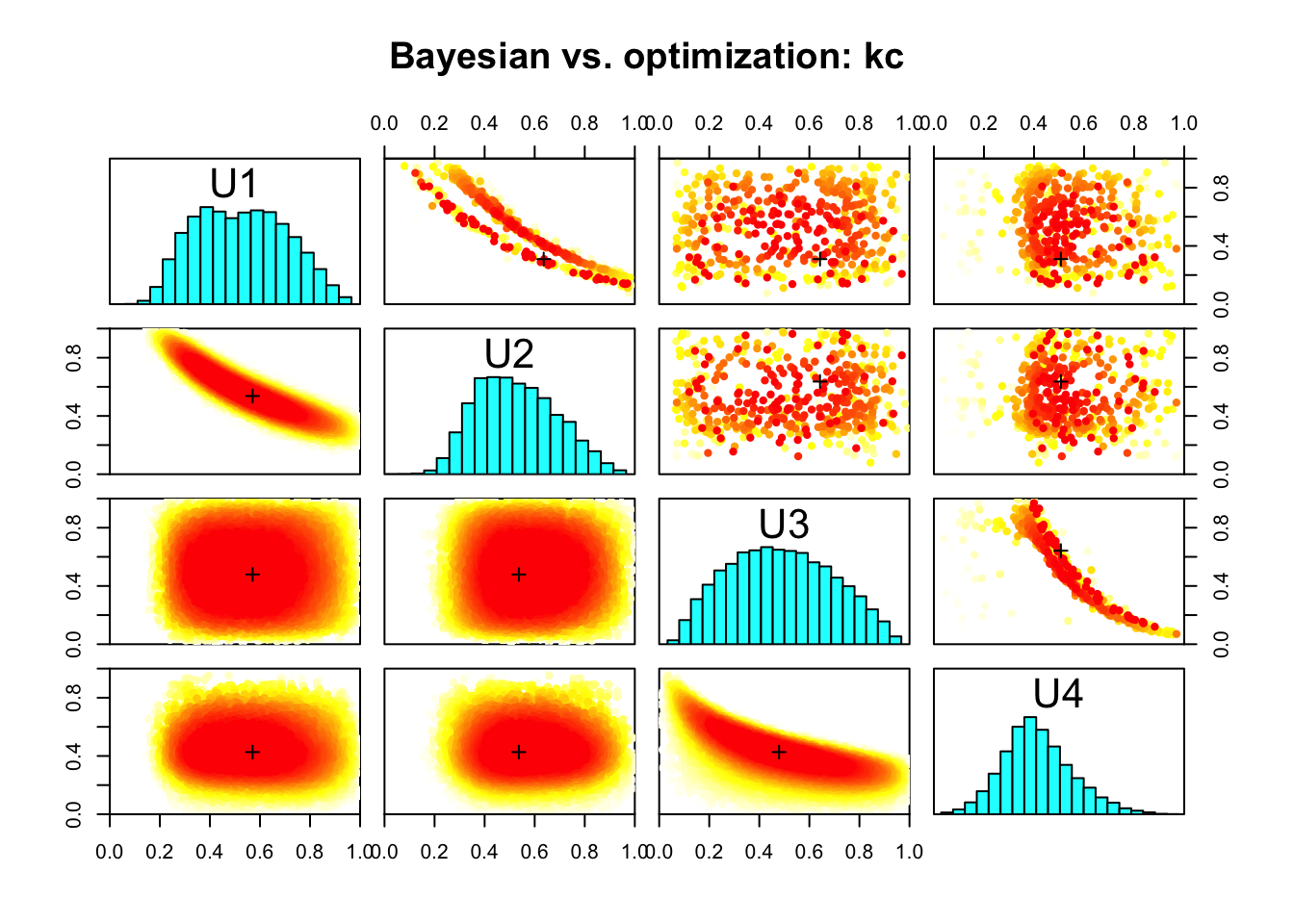}
\includegraphics[width=.85\linewidth, trim=50 40 16 55,clip=TRUE]{heatcolorscale}
\caption{PC fully Bayesian (lower and diagonal) and modular optimization (upper)
calibration results representing $\mathbf{u}^1_1$ for $K_d$ (top) 
and $\mathbf{u}^1_2$ for $k_c$ (bottom) combining 4 frequencies. 
Heat colors derived from rank of log posterior probability of these parameter values.}
\label{fig:pca_1}%
\end{figure}
\begin{figure}[ht!]
\centering
\includegraphics[width=.85\linewidth, trim= 0 65 20 80,clip=TRUE]{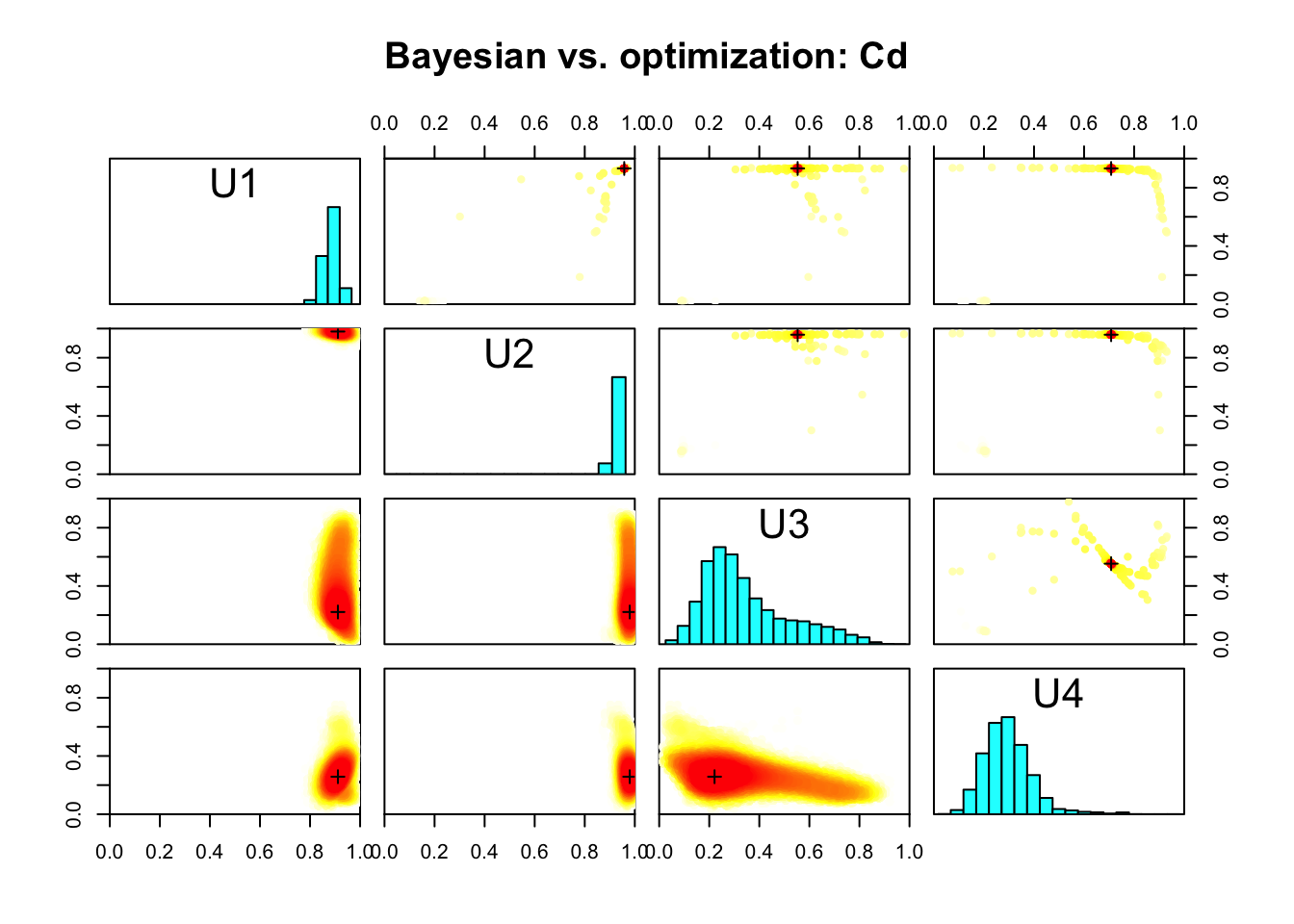}
\includegraphics[width=.85\linewidth, trim= 0 65 20 80,clip=TRUE]{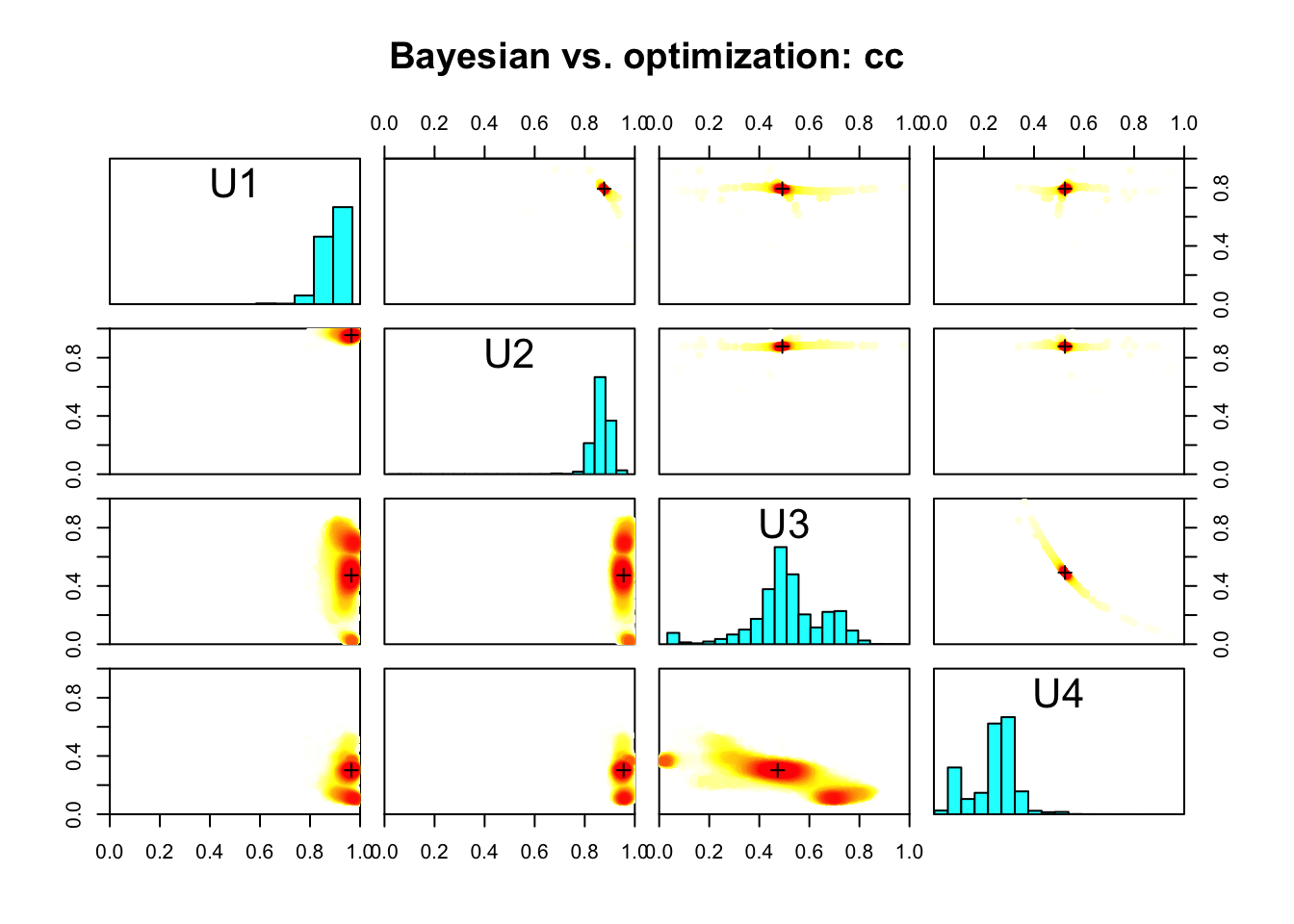}
\includegraphics[width=.85\linewidth, trim=50 40 16 55,clip=TRUE]{heatcolorscale}
\caption{Similar to Figure \ref{fig:pca_1} for $C_d$ (top) 
$c_c$ (bottom) combining 4 frequencies.}
\label{fig:pca_2}%
\end{figure}
To save space, we pair results from stiffness outputs, $K_d$ and $k_c$,
into Figure  \ref{fig:pca_1}  and damping outputs, $C_d$ and $c_c$,
into Figure  \ref{fig:pca_2}. All 4 outputs demonstrate distinct parameter
distributions with varying levels of uncertainty. 
Stronger similarity can be observed between the damping outputs in Figure \ref{fig:pca_2}
than the stiffness ones in Figure \ref{fig:pca_1}. Compared with the 
fully integrated results in Figure \ref{fig:comb}, output cross stiffness $k_c$
appears to be the most similar to the overall distribution.

\subsection{Multiple-output OSSs prediction results}
\label{sec:pred}

\blu{Here we complete multiple-output prediction and provided 
detailed notation and derivation for posterior prediction via 
multiple-output OSSs, described briefly in Section \ref{sec:mpred}.}

\blu{Augmenting Figure \ref{fig:pred_bias} from Section \ref{sec:pcpred}, we showcase
LOO-CV in first-PCs for $C_d$ and $c_c$ in Figure \ref{fig:pred_bias2}.
Contrasting with prediction without bias correction (blue), bias correction
substantially reduces uncertainty in accuracy and enhances precision. Notice
the bias term for $C_d$ demonstrates a similar pattern as $K_d$, increasing
generally as the observed output $C_d$ increases. For output $c_c$, the bias
correction term plays an even more crucial role for prediction when the
computer model becomes noisy and less informative.    }

\begin{figure}[ht!]
\centering
\includegraphics[width=.49\linewidth, trim=0 0 0 0,clip=TRUE]{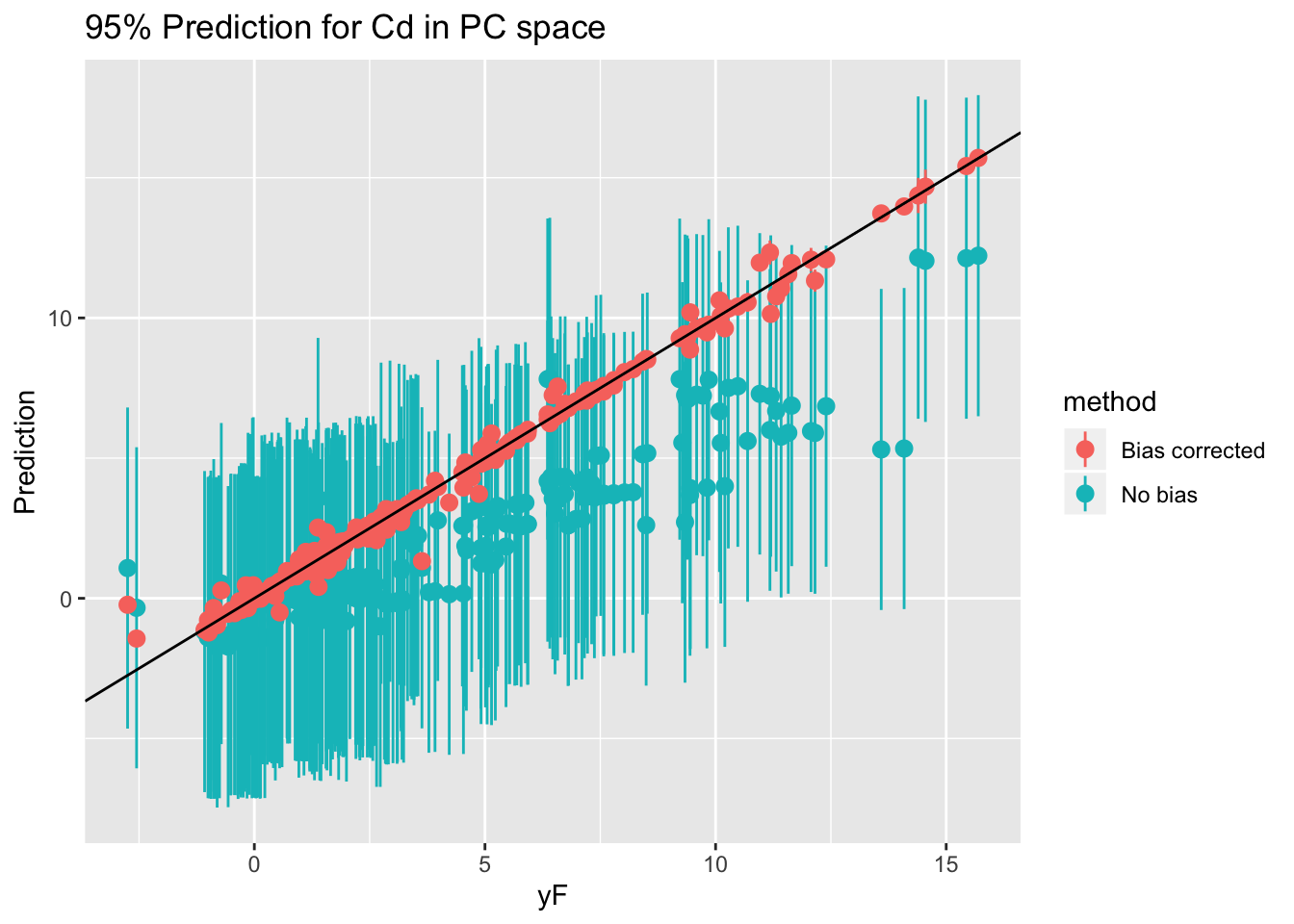}
\includegraphics[width=.49\linewidth, trim=0 0 0 0,clip=TRUE]{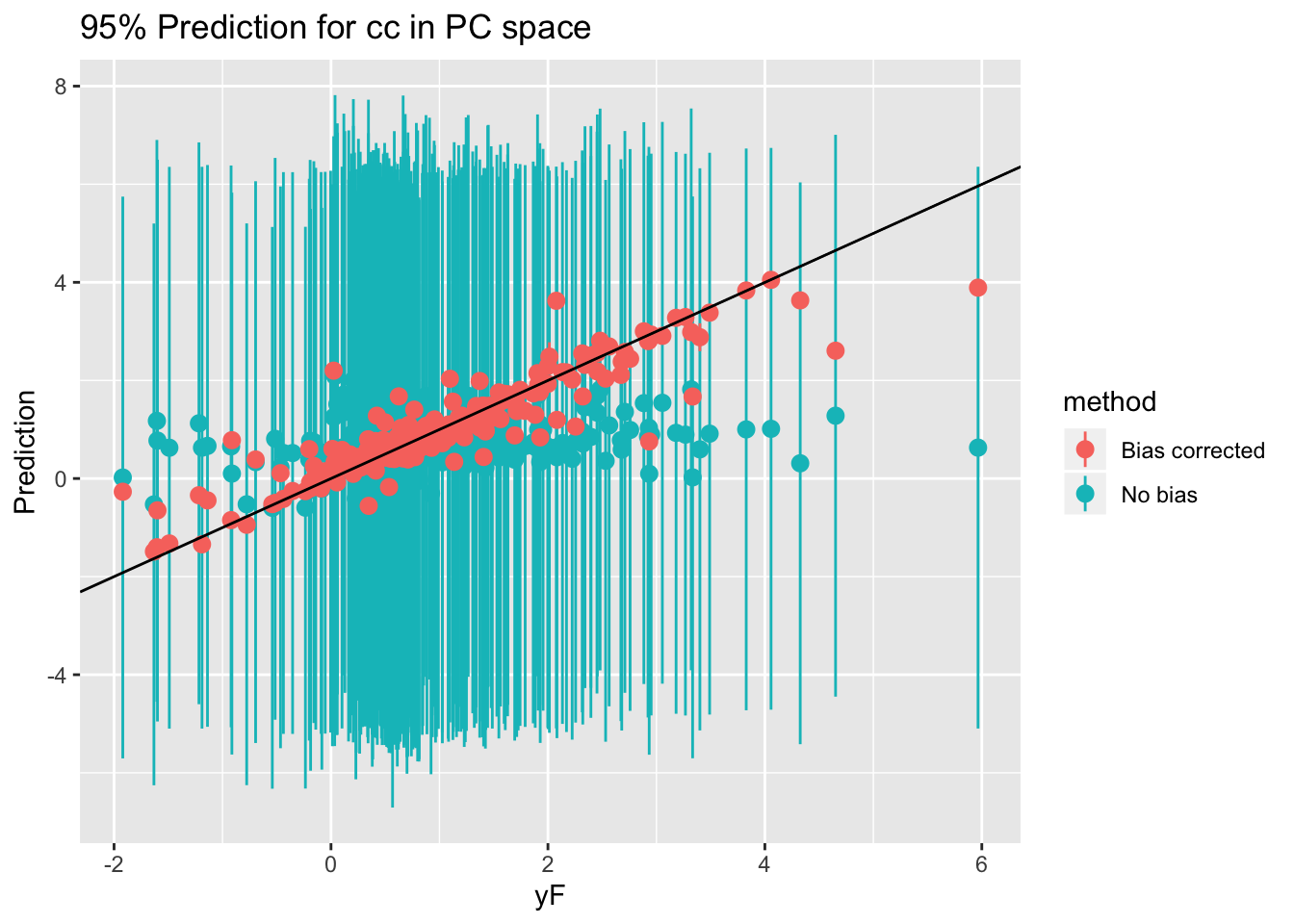}
\caption{LOO-CV first-PC posterior predictive summaries 
for direct damping $C_d$ and cross damping $c_c$.  
Intervals trace out 95\%; black line has intercept zero, slope one.}
\label{fig:pred_bias2}
\end{figure}

\blu{
Figures \ref{fig:pred_ori2}--\ref{fig:pred_ori4} show LOO-CV means and 95\%
credible intervals in original outputs $k_c$,
$C_d$, and $c_c$. These follow procedures described for $K_d$ outputs shown in
Figure \ref{fig:pred_ori} from Section \ref{sec:orpred}, whose RMSEs are
summarized numerically in Table \ref{tab:pred}. }
\begin{figure}[ht!]
\centering
\includegraphics[width=.4\linewidth, trim=0 0 0 0,clip=TRUE]{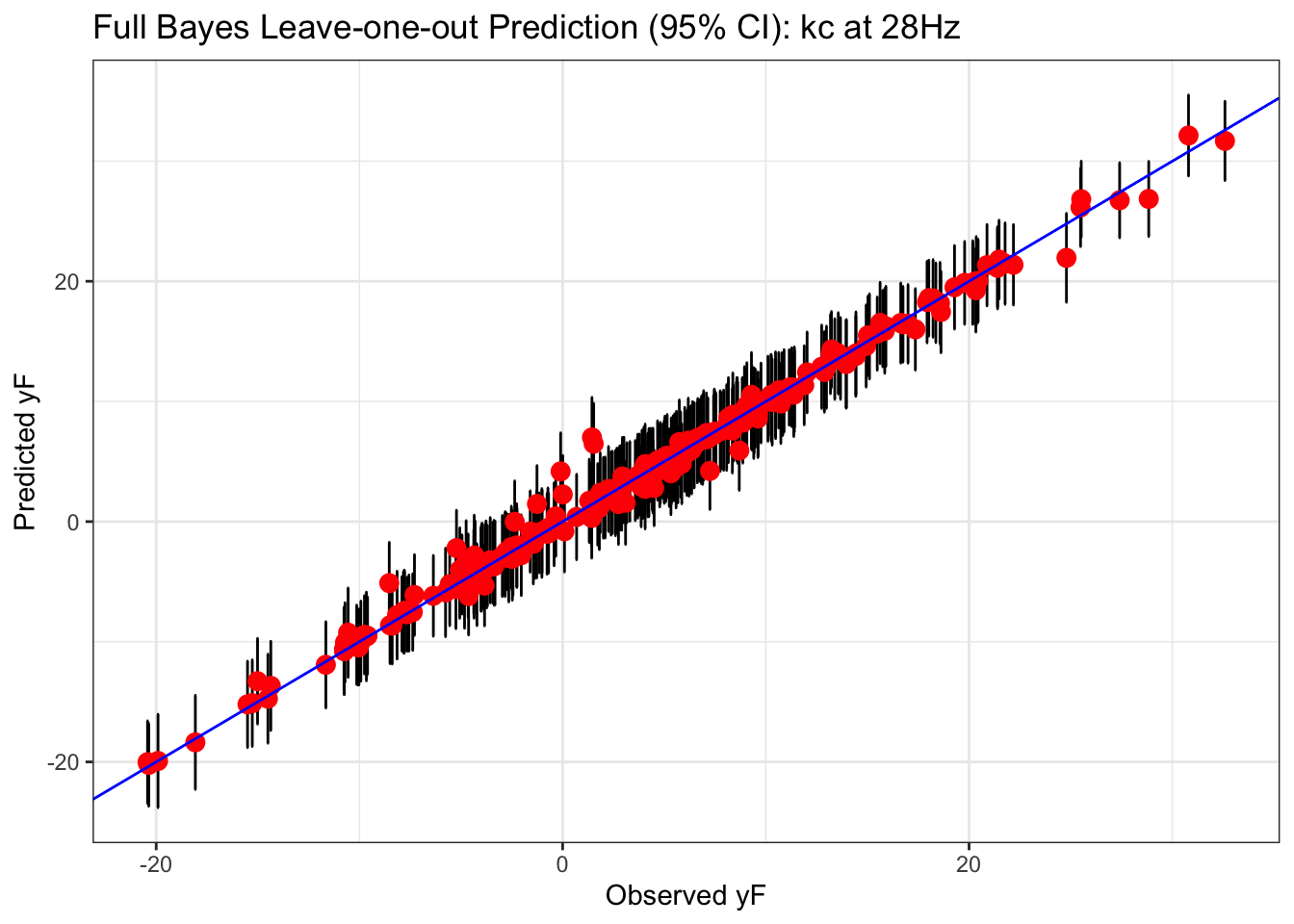}
\includegraphics[width=.4\linewidth, trim=0 0 0 0,clip=TRUE]{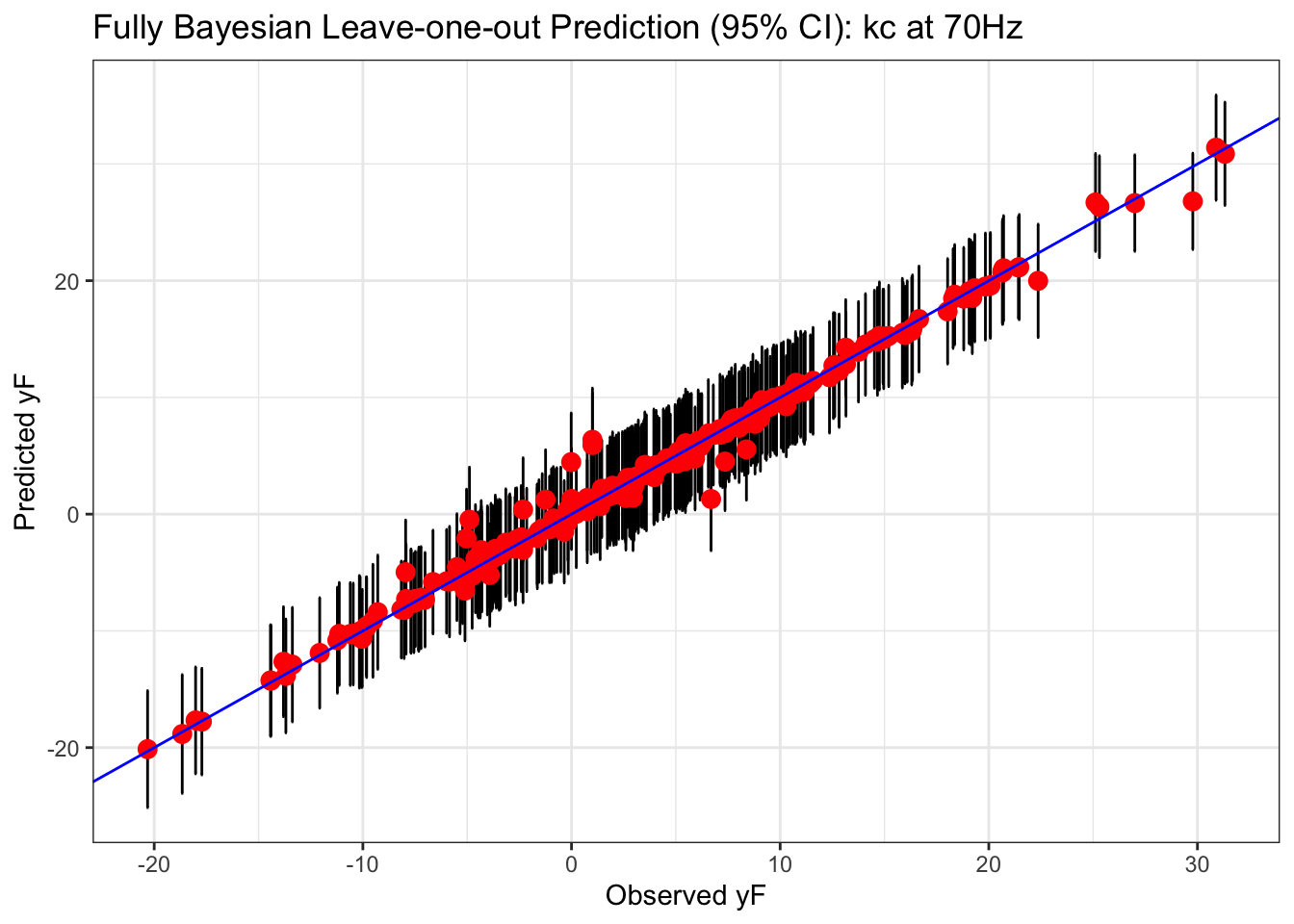}
\includegraphics[width=.4\linewidth, trim=0 0 0 0,clip=TRUE]{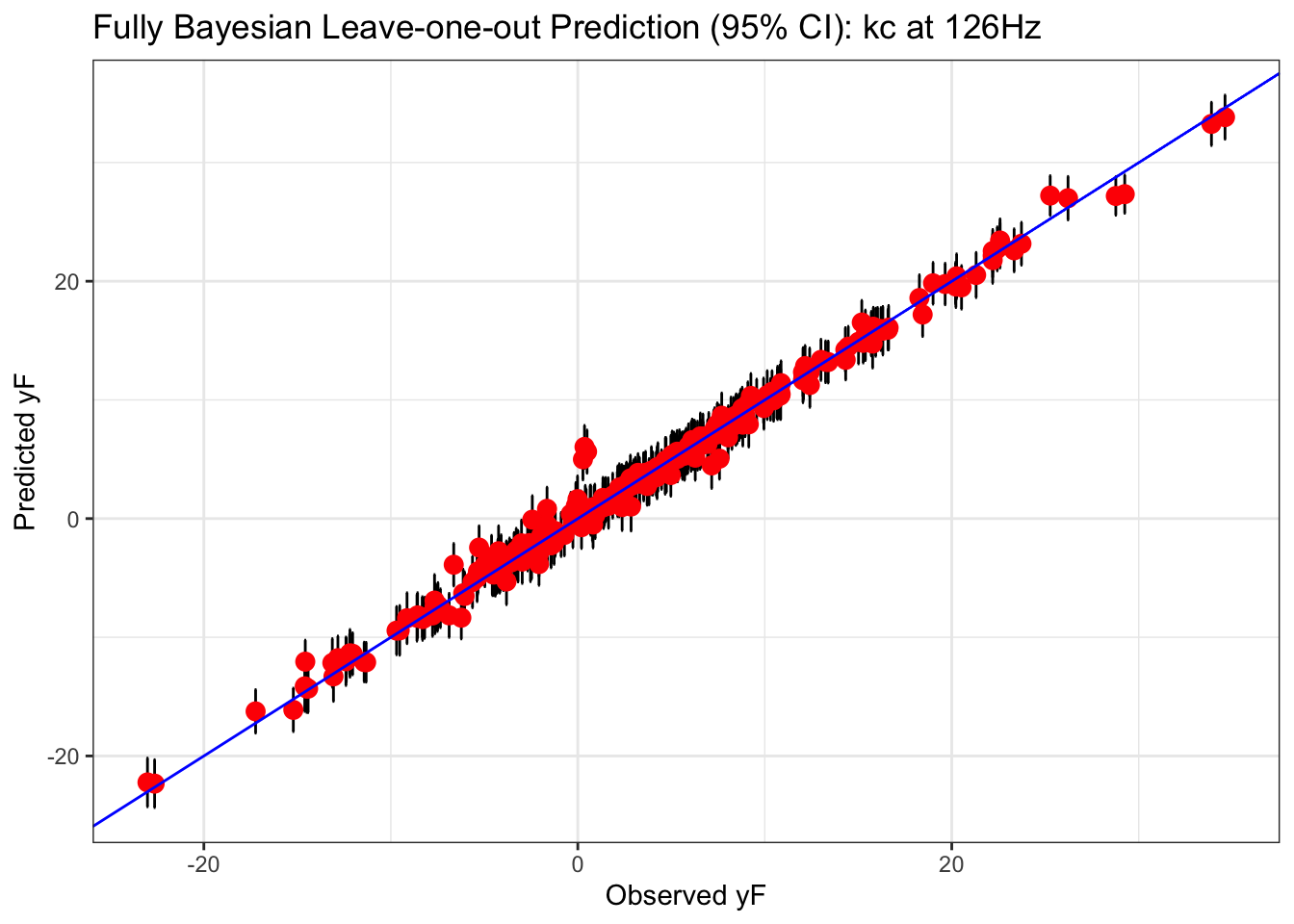}
\includegraphics[width=.4\linewidth, trim=0 0 0 0,clip=TRUE]{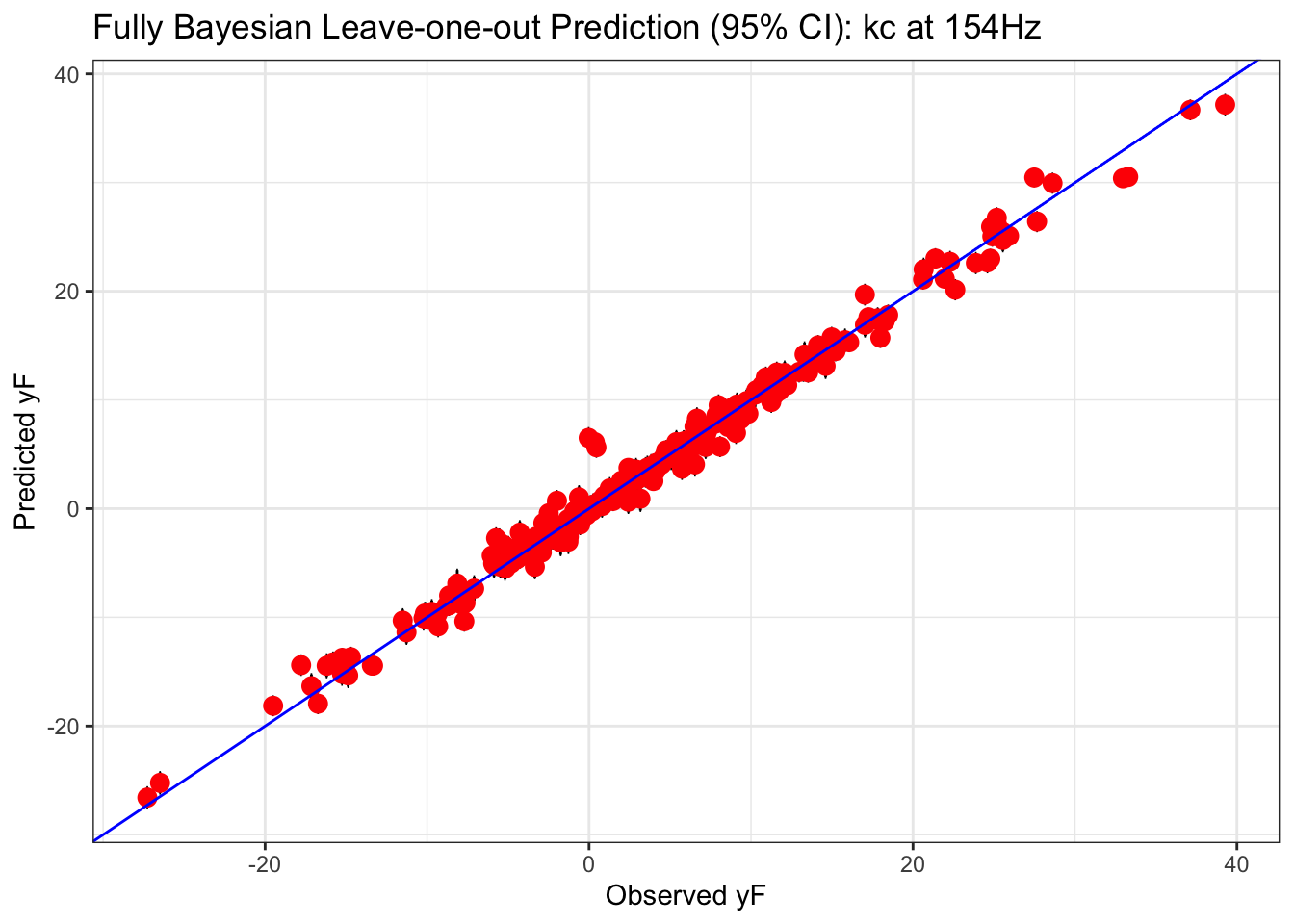}
\caption{Cross stiffness $k_c$ at at 28, 70,
 128, and 154 Hz; similar to Figure \ref{fig:pred_ori}.}
\label{fig:pred_ori2}
\end{figure}
\begin{figure}[ht!]
\centering
\includegraphics[width=.4\linewidth, trim=0 0 0 0,clip=TRUE]{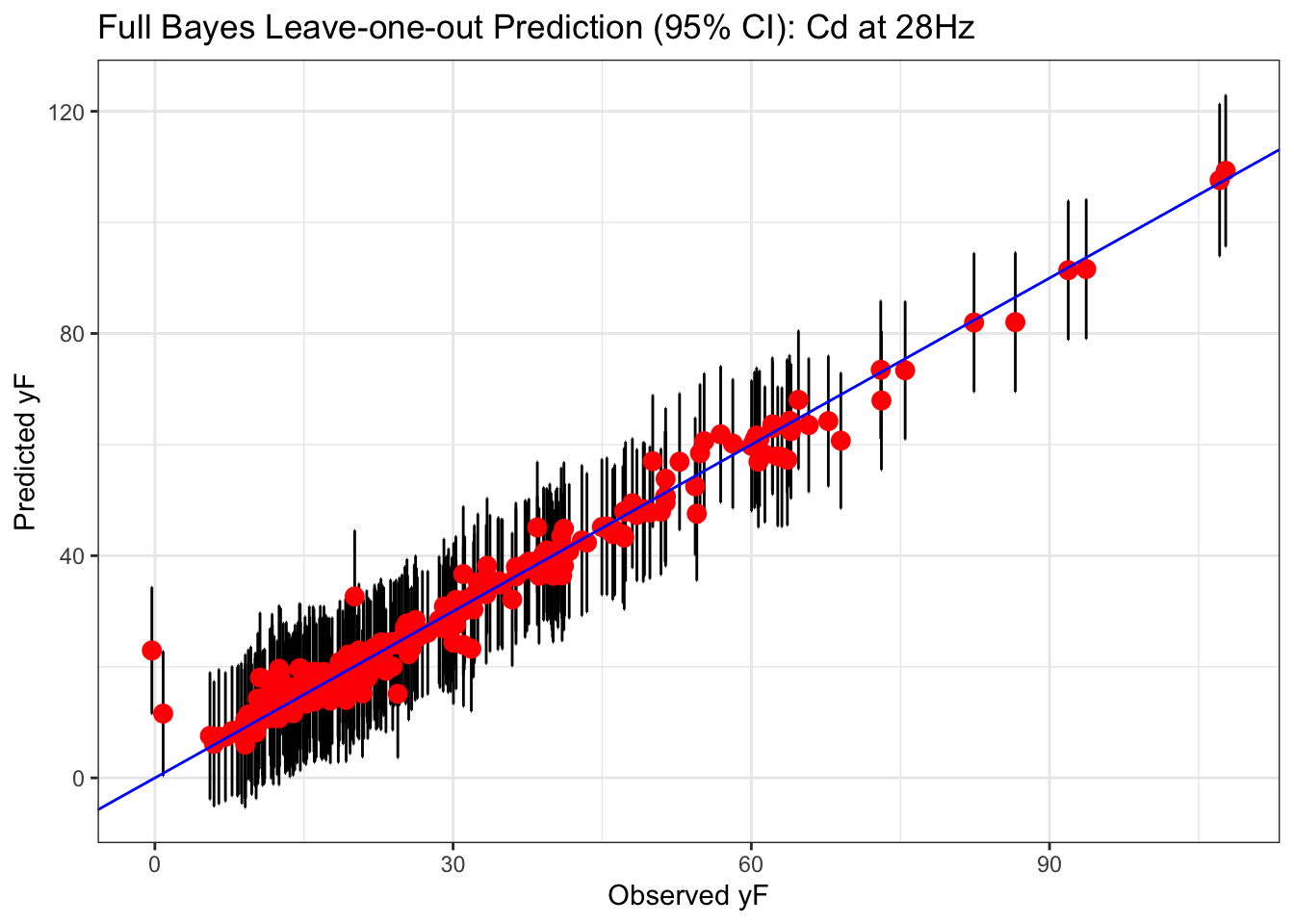}
\includegraphics[width=.4\linewidth, trim=0 0 0 0,clip=TRUE]{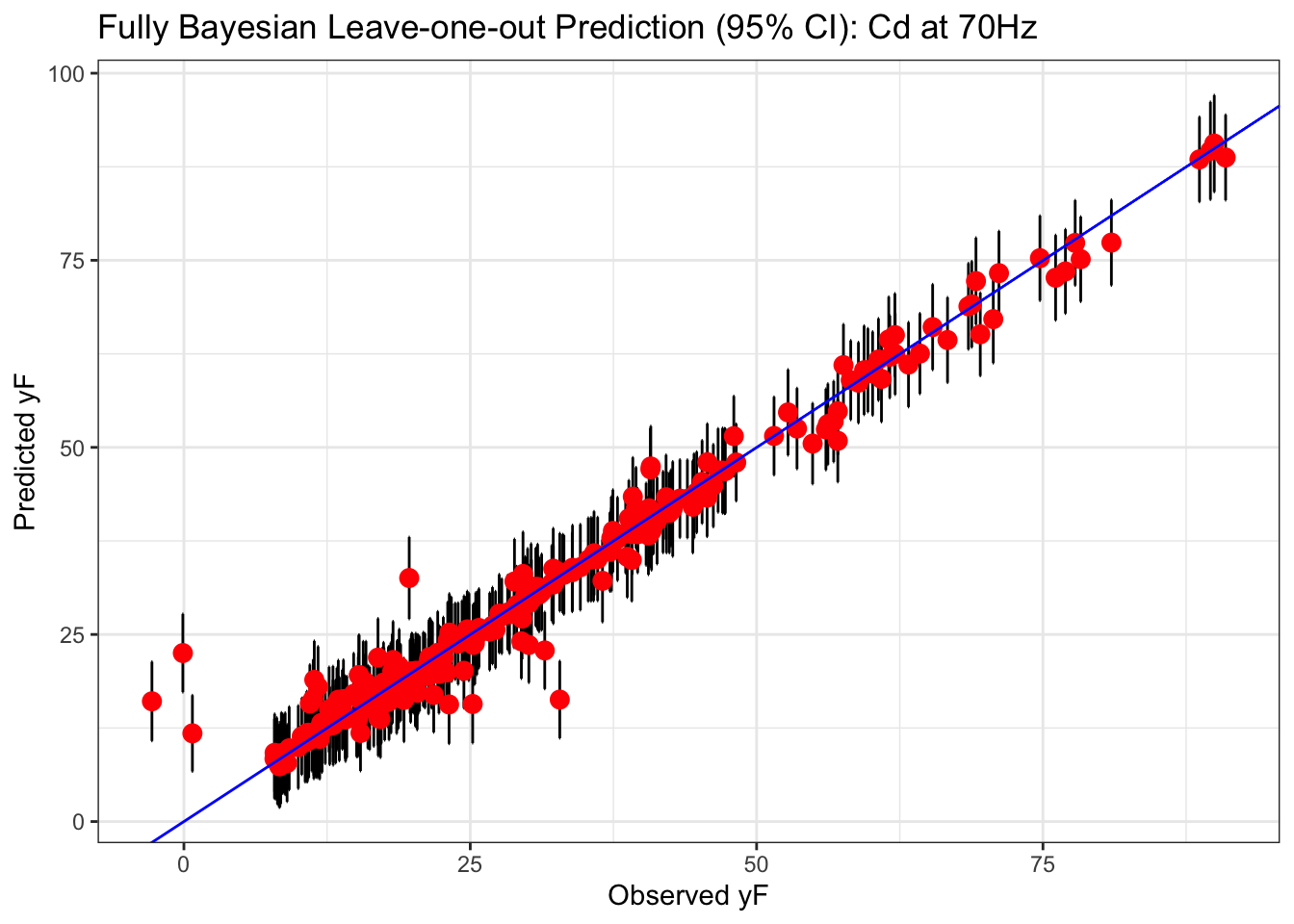}
\includegraphics[width=.4\linewidth, trim=0 0 0 0,clip=TRUE]{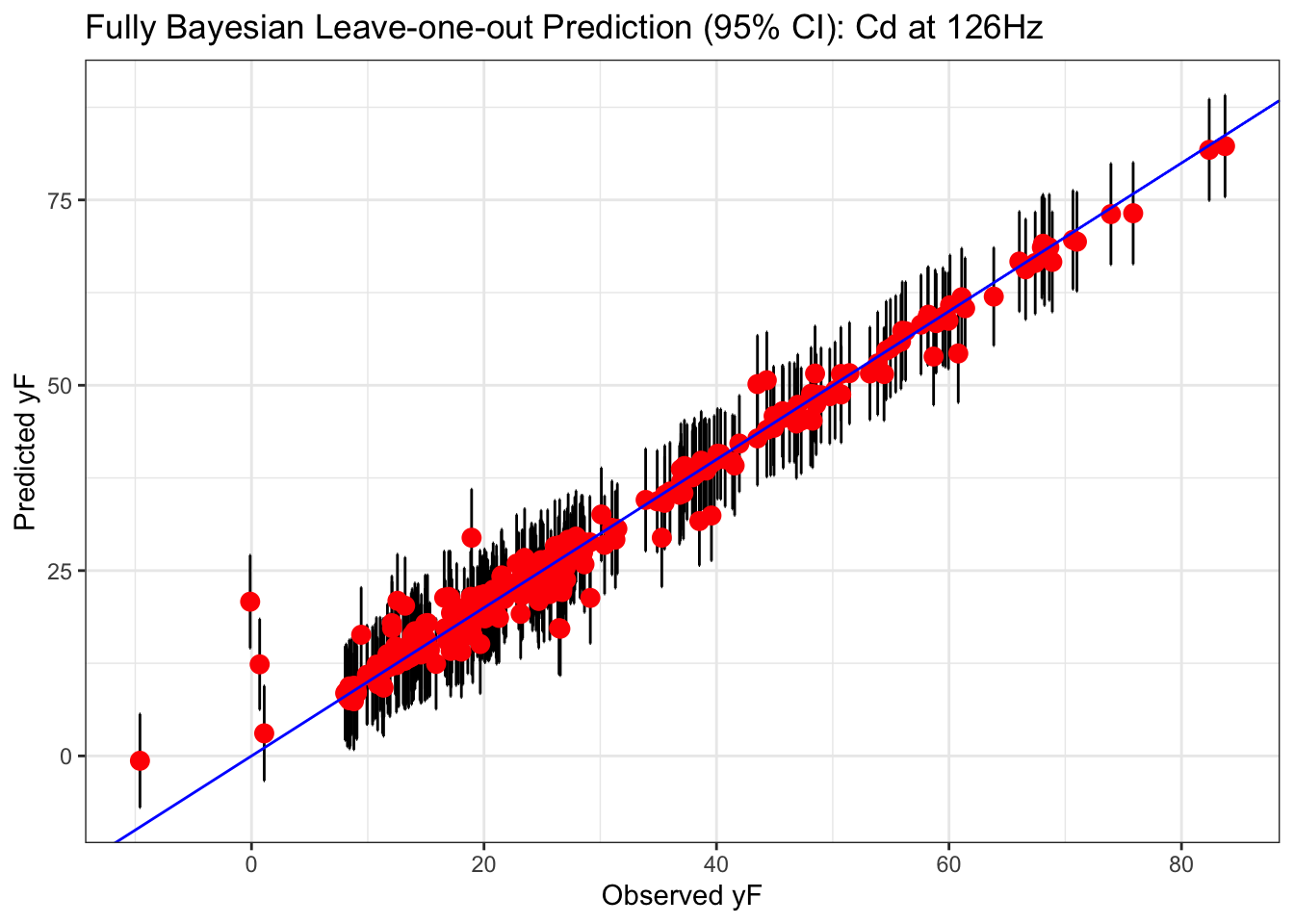}
\includegraphics[width=.4\linewidth, trim=0 0 0 0,clip=TRUE]{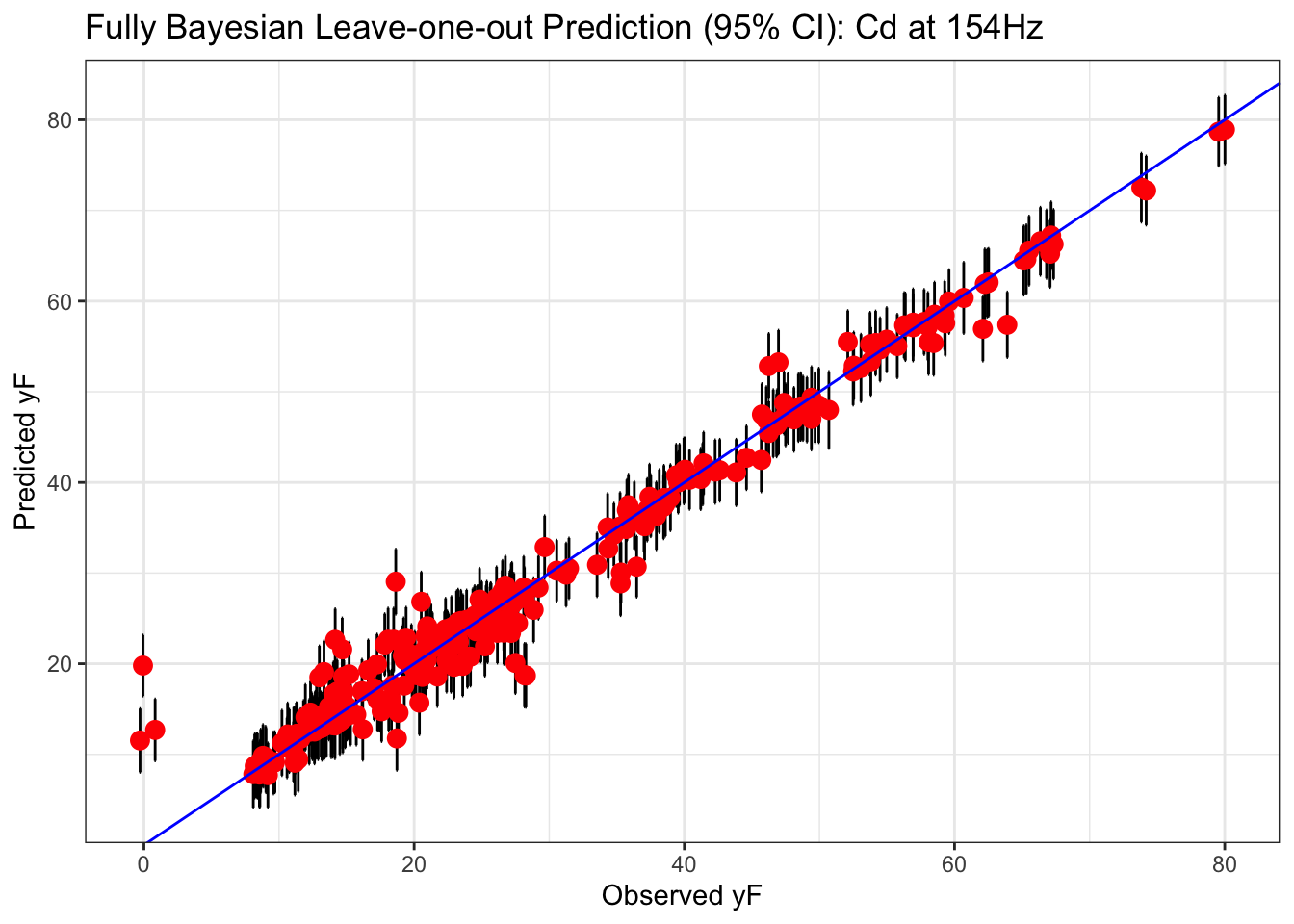}
\caption{Direct damping $C_d$ at at 28, 70,
 128, and 154 Hz; similar to Figure \ref{fig:pred_ori}.}
\label{fig:pred_ori3}
\end{figure}
\begin{figure}[ht!]
\centering
\includegraphics[width=.4\linewidth, trim=0 0 0 0,clip=TRUE]{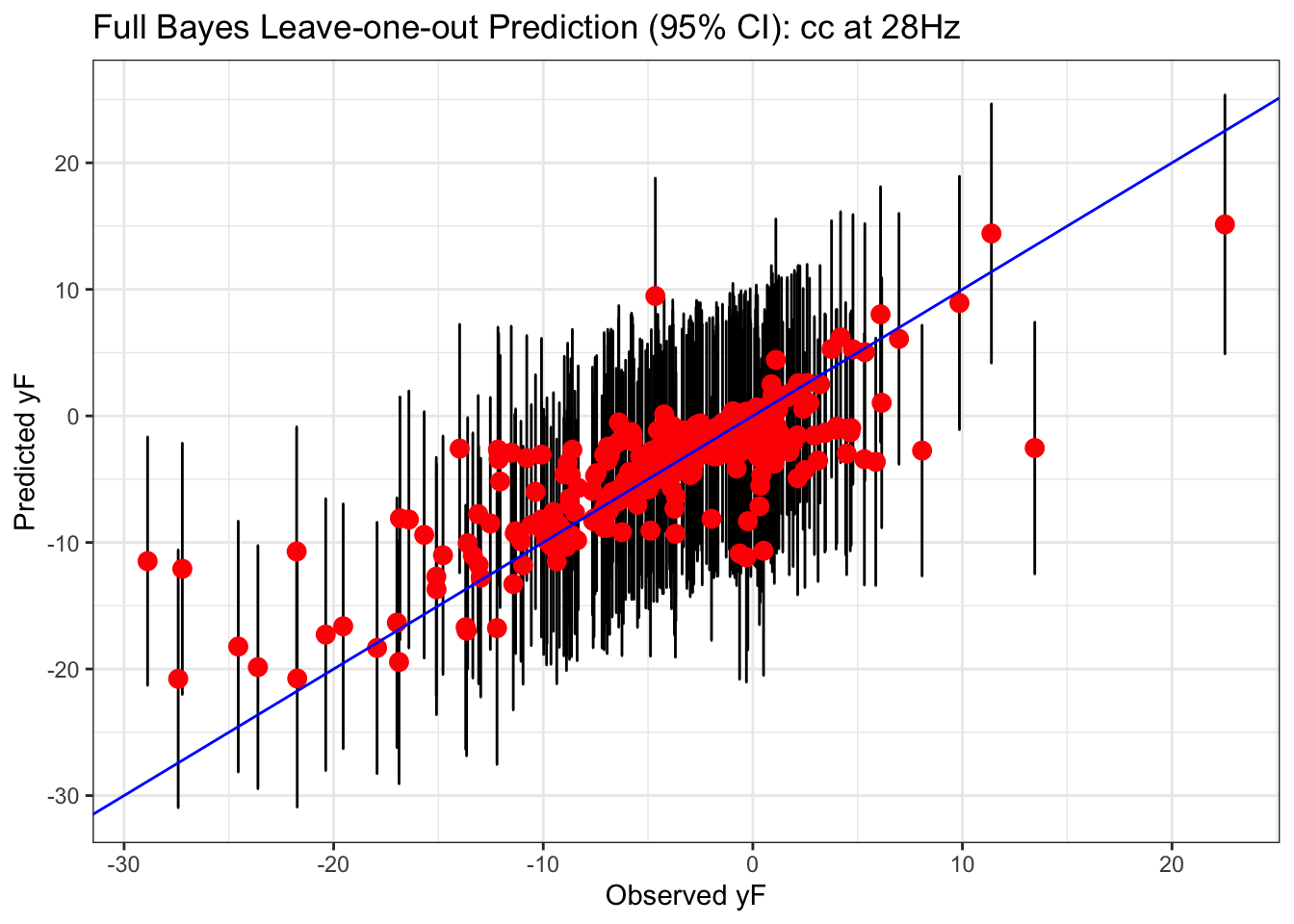}
\includegraphics[width=.4\linewidth, trim=0 0 0 0,clip=TRUE]{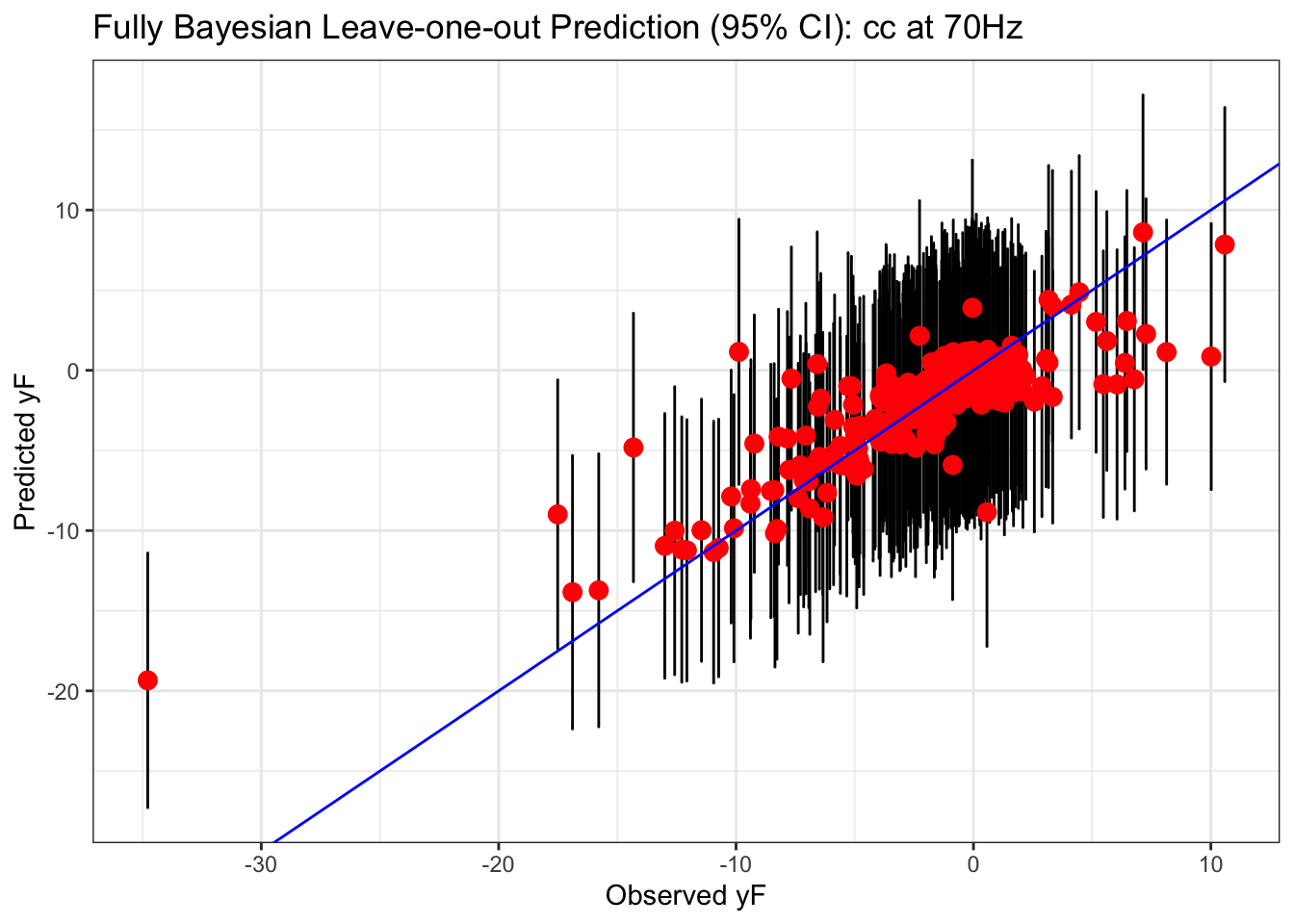}
\includegraphics[width=.4\linewidth, trim=0 0 0 0,clip=TRUE]{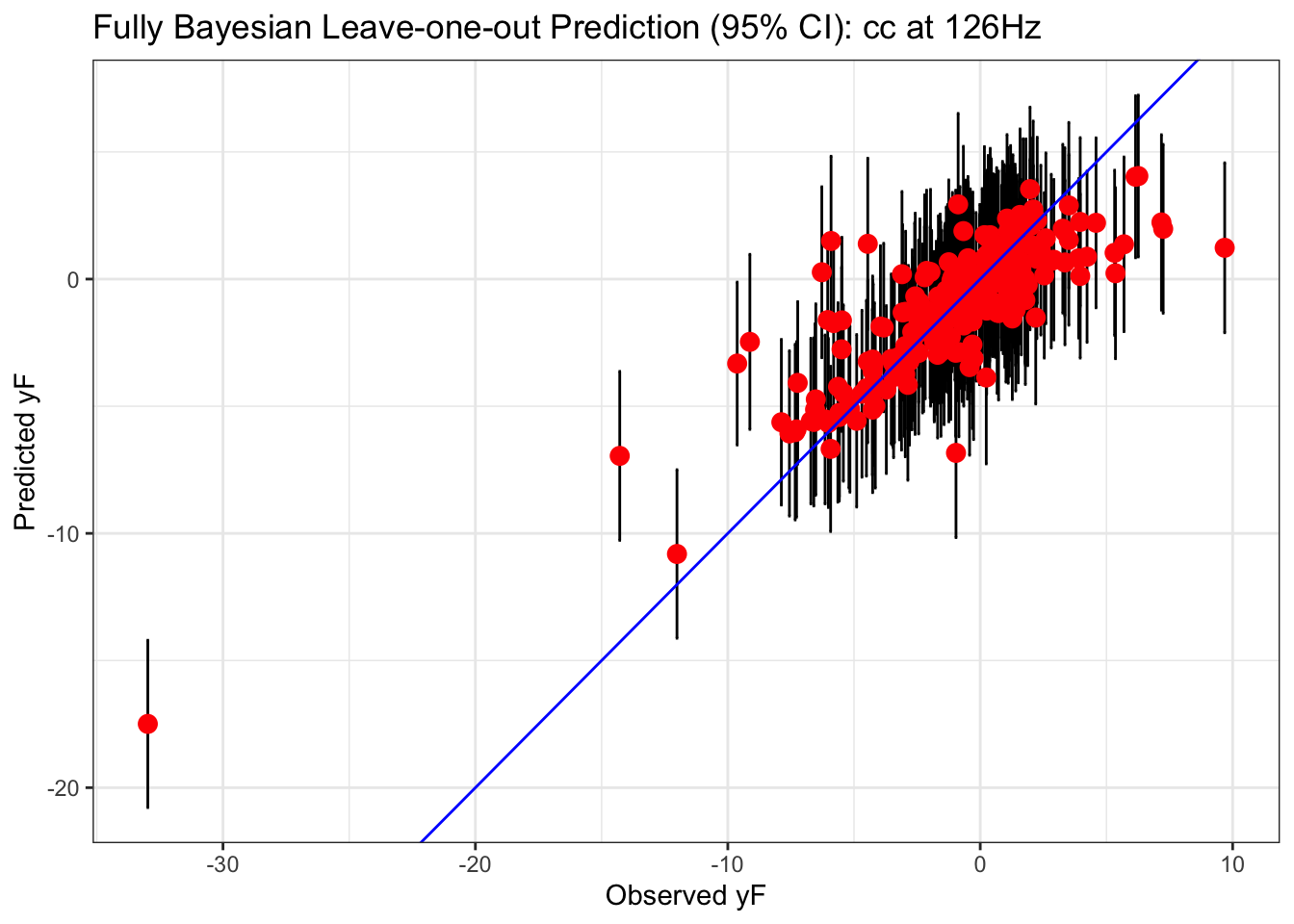}
\includegraphics[width=.4\linewidth, trim=0 0 0 0,clip=TRUE]{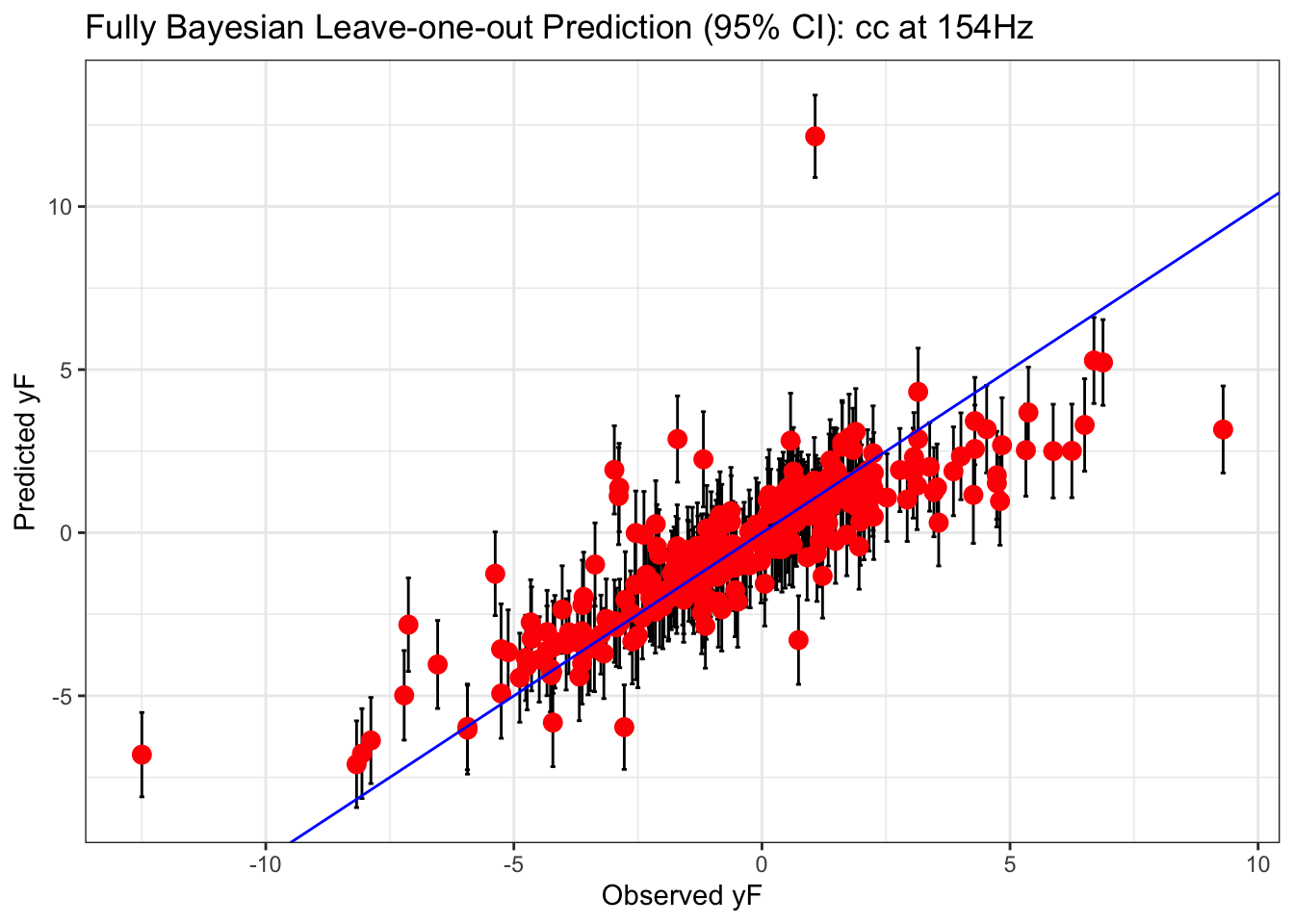}
\caption{Cross damping $c_c$ at at 28, 70,
 128, and 154 Hz; similar to Figure \ref{fig:pred_ori}.}
\label{fig:pred_ori4}
\end{figure}

\blu{
Lastly, we detail the  predictive mean and variance
$(\blunew{\mathbf{\mu}}^k_{j\mathrm{(new)}}, \blunew{\mathbf{\Sigma}}^k_{j\mathrm{(new)}} )$ for
new site location $\mathbf{x}_{\mathrm{(new)}}$
for the $j$th output property at frequency $k$ as 
Eq.~(\ref{eq:vnew}). 
Borrowing notation from Eq.~(\ref{eq:mvn})
but now in PC basis space, 
the joint distribution of the computer model training data
$(\mathbf{X}_{\blunew{N}}, \mathbf{y}^{Mk}_{j})$, field training data
$(\mathbf{X}_{\blunew{F}}, \mathbf{y}^{Fk}_{j})$, 
simulated size \blunew{$N'$} computer model data at new site location 
$(\mathbf{x}_{\mathrm{(new)}}, \mathbf{y}^{Mk}_{j\mathrm{(new)}})$, 
and the unobserved field data of size $\blunew{F}'$ at new location 
$(\mathbf{x}_{\mathrm{(new)}}, \mathbf{y}^{Fk}_{j\mathrm{(new)}})$ 
follows
$ (\mathbf{y}^{Mk}_{j},  \mathbf{y}^{Fk}_{j}, 
\mathbf{y}^{Mk}_{j\mathrm{(new)}} , \mathbf{y}^{Fk}_{j\mathrm{(new)}}) ^\top
\sim \mathcal{N}(\mathbf{0}, \blunew{\mathbf{\Sigma}}_{P}^{jk}(\mathbf{u}^1) )$, where
\blunew{
\begin{align}
\mathbf{\Sigma}_{P}^{jk}(\mathbf{u}^1) 
\equiv 
\begin{bmatrix} 
\mathbf{\Sigma}^{jk}_{N} & \mathbf{\Sigma}^{jk}_{\blunew{F}, N}(\mathbf{u}^1)^\top &
  \mathbf{0}  &  \mathbf{0} \\
\mathbf{\Sigma}_{\blunew{F}, N}^{jk}(\mathbf{u}^1) &   \mathbf{\Sigma}^{jk}_{\blunew{F}}(\mathbf{u}^1)  + \mathbf{\Sigma}^{jk}_{b} 
& \mathbf{0} & 
\mathbf{\Sigma}^{jk\top}_{\blunew{F}', \blunew{F}} \\ 
  \mathbf{0} & \mathbf{0} &  \mathbf{\Sigma}^{jk\mathrm{(new)}}_{N'}  
  & \mathbf{\Sigma}^{jk\mathrm{(new)}}_{\blunew{F}', N'}(\mathbf{u}^1)^\top\\
  \mathbf{0} & \mathbf{\Sigma}^{jk}_{\blunew{F}', \blunew{F}}
  &  \mathbf{\Sigma}^{jk\mathrm{(new)}}_{\blunew{F}', N'}(\mathbf{u}^1) 
  &  \mathbf{\Sigma}^{jk\mathrm{(new)}}_{\blunew{F}'}(\mathbf{u}^1)  + \mathbf{\Sigma}^{jk\mathrm{(new)}}_{b} 
   \end{bmatrix}.
\label{eq:mvn_p}
\end{align}
}}
\blu{In Eq.~(\ref{eq:mvn_p}), multiple block-wise sparse structures 
are retained under multiple OSSs in all PC basis spaces. 
Following first-PC notation in from Eq.~(\ref{eq:Sigma}), 
\blunew{$\mathbf{\Sigma}^{jk}_{N} \equiv \Diag[\mathbf{\Sigma}^{jk}_i(\mathbf{U}_i,
\mathbf{U}_i)]$}, for $i = 1, \dots, \blunew{F}$, an upper-left block
diagonal submatrix.  Similarly, the off-diganal \blunew{$\mathbf{\Sigma}^{jk}_{\blunew{F},
N}(\mathbf{u}^1)$} and field covariance \blunew{$\mathbf{\Sigma}^{jk}_{\blunew{F}}(\mathbf{u}^1)  +
\mathbf{\Sigma}^{jk}_{b}$} follow. The covariance of the newly simulated computer model
data are still block-diagonal, \blunew{$\mathbf{\Sigma}^{jk\mathrm{(new)}}_{N} \equiv
\Diag[\mathbf{\Sigma}^{jk\mathrm{(new)}}_i(\mathbf{U}_i,
\mathbf{U}_i)]$}, for $i = \blunew{F}+1, \dots, \blunew{F}'$.  
Off-diagonal
\blunew{$\mathbf{\Sigma}^{jk\mathrm{(new)}}_{\blunew{F}, N}(\mathbf{u}^1)$} and 
field covariance \blunew{$\mathbf{\Sigma}^{jk\mathrm{(new)}}_{\blunew{F}}(\mathbf{u}^1)  + \mathbf{\Sigma}^{jk\mathrm{(new)}}_{b}$} 
are analogous to training. Now,  condition each new field observation $\mathbf{y}^{Fk}_{j\mathrm{(new)}}$ 
on the rest of all observed computer model data and field data in each of  the PC bases,
\[
(\mathbf{y}^{Fk}_{j\mathrm{(new)}}  \mid \mathbf{y}^{Mk}_{j},  \mathbf{y}^{Fk}_{j}, 
\mathbf{y}^{Mk}_{j\mathrm{(new)}}, \bm{\Phi}, \mathbf{u}^1) 
 \sim \mathcal{N}(\blunew{\mathbf{\mu}}^k_{j\mathrm{(new)}}, \blunew{\mathbf{\Sigma}}^k_{j\mathrm{(new)}} ), 
 \quad \; j = 1, \dots, J, \; k =1, \dots, K.
\]
Following the joint covariance notation
from Eq.~(\ref{eq:mvn_p}), 
closed form predictive mean and variance 
$(\blunew{\mathbf{\mu}}^k_{j\mathrm{(new)}}, \blunew{\mathbf{\Sigma}}^k_{j\mathrm{(new)}} )$ for 
new site location $\mathbf{x}_{\mathrm{(new)}}$ are described in
 Eq.~(\ref{eq:vnew}),  
in which \blunew{$\mathbf{C}(\mathbf{u}^1)=  
\mathbf{\Sigma}^{jk}_{\blunew{F}}(\mathbf{u}^1)  + \mathbf{\Sigma}^{jk}_{b}  
 -  \mathbf{\Sigma}^{jk}_{\blunew{F}, N}(\mathbf{u}^1)
 (\mathbf{\Sigma}^{jk}_{N}) ^{-1} 
 \mathbf{\Sigma}^{jk}_{\blunew{F}, N}(\mathbf{u}^1)^\top$}.}

\blunew{
\subsection{Sensitivity analysis}
\label{sec:sens}
To better understand the subtle relationship between calibration parameter
prior settings, posterior and choices for discrepancy correction, we 
conducted a sensitivity analysis.   For our motivating honeycomb example, we
entertained three additional priors for calibration parameter $\mathbf{u}$
beyond $(u_1, u_2, u_3, u_4) \overset{iid}{\sim} \text{Beta}(2, 2)$, which we
refer to as the ``regular'' prior below.  These new alternatives include: a
``weak'' adversarial prior  $(u_1, u_2)
\overset{\mathrm{iid}}{\sim} \text{Beta}(1.5, 2.5)$ and $ (u_3, u_4)
\overset{\mathrm{iid}}{\sim} \text{Beta}(2.5, 1.5)$, and a ``prescient" prior with
$(u_1, u_2) \overset{\mathrm{iid}}{\sim} \text{Beta}(2.5, 1.5)$ and $ (u_3, u_4)
\overset{\mathrm{iid}}{\sim} \text{Beta}(1.5, 2.5)$. Observe that the ``weak''
prior puts prior mass away from the MAP shown in Figure \ref{fig:comb}
where as the ``precient'' one reinforces this setting. In addition to
exploring how $\mathbf{u}$ posteiors vary under these different prior
settings,  we consider a version without bias correction, ``no bias" to contrast with
``regular" prior. Two-dimensional contour plots from all four scenarios have
been overlaid  in Figure \ref{fig:sens}.}

\begin{figure}[ht!]
\centering
\includegraphics[width=.28\linewidth, trim=0 0 80 0,clip=TRUE]{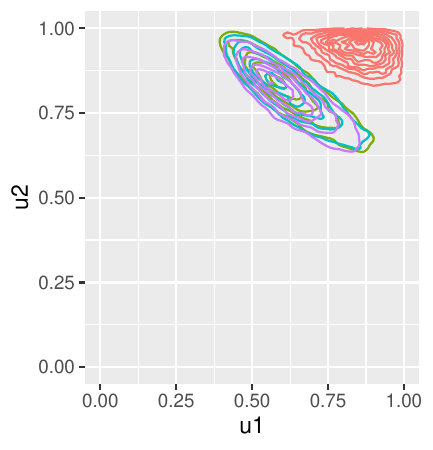}
\includegraphics[width=.28\linewidth, trim=0 0 80 0,clip=TRUE]{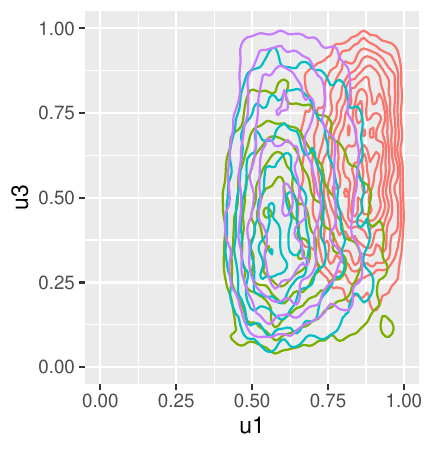}
\includegraphics[width=.39\linewidth, trim=0 0 00 0,clip=TRUE]{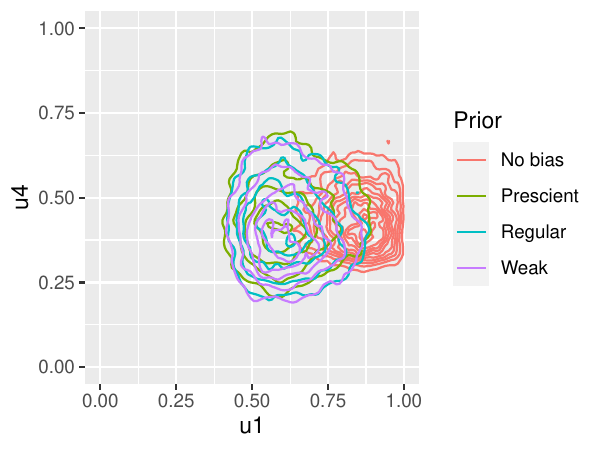}
\includegraphics[width=.28\linewidth, trim=0 0 80 0,clip=TRUE]{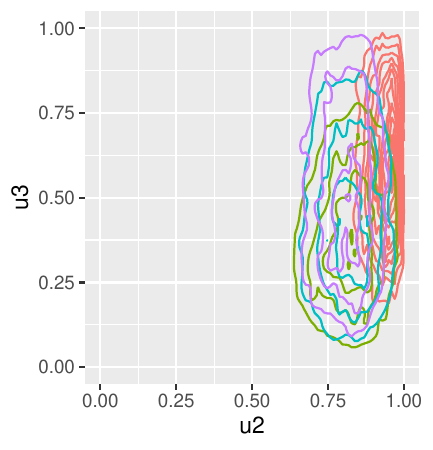}
\includegraphics[width=.28\linewidth, trim=0 0 80 0,clip=TRUE]{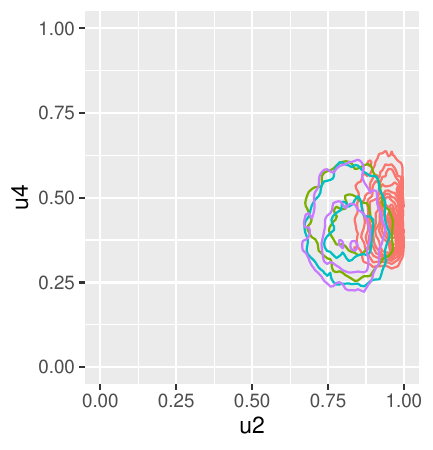}
\includegraphics[width=.39\linewidth, trim=0 0 00 0,clip=TRUE]{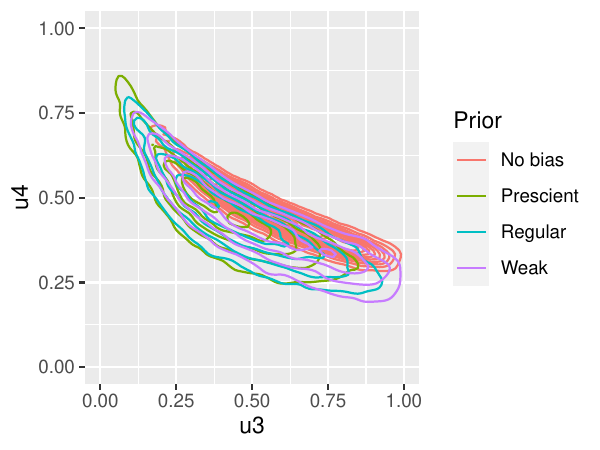}
\caption{Bivariate marginal contour plots of $100{,}000$ MCMC posterior samples of
 $\mathbf{u}$ under: regular (cyan),  weak (purple), and prescient (green)
 priors, and no bias correction(red). }
\label{fig:sens}
\end{figure}

\blunew{
Among all the three ``bias corrected" prior settings, including regular
(cyan), weak (purple),  and prescient (green),  observe that the posterior
contours for  $\mathbf{u}$ largely overlap, suggesting that prior influence is
weak relative to the likelihood (data influence), and that consequently our
preferred analysis (``regular'') is robust in that sense. Taking a closer look
at  the middle plot in the first row and the left plot in the second row, the
weak posterior (purple) contours encircle those of prescient posterior
(green), demonstrating a meaningful and intuitive trend of impact from
different prior choices on the posterior distributions.  It is worth pointing
out that these posteriors complement an analysis in \citep{Huang:2018} showing
that a ``uniform'' prior on $\mathbf{u}$ also works well, except mass is more
concentrated on the boundaries of the input space where MAP is close to the
end of the study region.
}

\blunew{
Now consider the ``no bias" (red) posterior which is in sharp distinction, but
also not altogether different.  Only parameter pair $(u_1, u_2)$ has shifted
substantially, concentrating on boundaries of their support. We already know
that this setting leads to bad predictions 
(Figures \ref{fig:pred_bias} and \ref{fig:pred_bias2}, but
perhaps this does not have a substantial effect on calibration.)}
\end{document}